\documentclass{sig-alternate-sigmod08}

\begin{document}
\conferenceinfo{ACM SIGMOD}{'08 Vancouver, BC, Canada}


\newtheorem{thm}{Theorem}[section]
\newtheorem{cor}[thm]{Corollary}
\newtheorem{lemma}[thm]{Lemma}
\newtheorem{conjecture}[thm]{Conjecture}
\newtheorem{prop}[thm]{Proposition}
\newtheorem{claim}[thm]{Claim}
\newtheorem{fact}[thm]{Fact}
\newtheorem{definition}[thm]{Definition}
\newtheorem{rem}[thm]{Remark}
\newtheorem{corollary}[thm]{Corollary}
\newtheorem{problem}[thm]{Problem}

%

\def\pri{\mbox{\sc pri}}
\def\ws{\mbox{\sc ws}}
\def\rc{\mbox{\sc RC}}
\def\ssc{\mbox{\sc SC}}
\def\ml{\mbox{\sc ML}}
\def\wsr{\mbox{\sc wsr}}
\def\wsrd{\mbox{\sc wsrd}}
\def\fws{\mbox{\sc f-ws}}
\def\NF{\mbox{\sc NF}}
\def\aNF{\mbox{\sc aNF}}
\def\sNF{\mbox{\sc sNF}}
\def\sSH{\mbox{\sc sSH}}
\def\SH{\mbox{\sc SH}}
\def\aSH{\mbox{\sc aSH}}

\def\orr{\overline{r}}
\def\oww{\overline{w}}
\def\oWW{\overline{W}}
\def\uww{\underline{w}}
\def\uWW{\underline{W}}

\def\inperm{\mbox{\sc inperm}}
\def\permnum{\mbox{\sc permnum}}

\def\HT{\mbox{\sc HT}}
\def\HTp{\mbox{\sc HTp}}

\def\vecs{\mbox{\bf s}}
\def\vecr{\mbox{\bf r}}

\newcommand{\ignore}[1]{}

\def\CondPref{\mbox{\sc CondPref}}
\def\pr{\mbox{\sc pr}}
\def\var{\mbox{\sc var}}
\def\cov{\mbox{\sc cov}}
\def\pref{\mbox{\sc pfx}}

\newcommand{\notinproc}[1]{#1}
\newcommand{\onlyinproc}[1]{}

\def\marrow{{\marginpar[\hfill$\longrightarrow$]{$\longleftarrow$}}}
\def\haim#1{{\sc Haim says: }{\marrow\sf #1}}
\def\edith#1{{\sc Edith says: }{\marrow\sf #1}}


\title{Sketch-Based Estimation of Subpopulation-Weight}

\numberofauthors{2}

\author{
\alignauthor Edith Cohen\\
       \affaddr{AT\&T Labs--Research}\\
       \affaddr{180 Park Avenue}\\
       \affaddr{Florham Park, NJ 07932, USA}\\
       \email{edith@research.att.com}
\alignauthor  Haim Kaplan \\
       \affaddr{School of Computer Science}\\
       \affaddr{Tel Aviv University}\\
       \affaddr{Tel Aviv, Israel}\\
       \email{haimk@cs.tau.ac.il}
}


 \maketitle 

  \begin{abstract}

 Summaries of massive data sets support approximate query processing
over the original data.  A basic aggregate over a set of records is
the weight of subpopulations specified as a predicate over records'
attributes.  {\em Bottom-$k$} sketches are a powerful summarization
format of weighted items that includes priority
sampling~\cite{DLT:sigmetrics04} (\pri) and the classic weighted
sampling without replacement (\ws). They can be
computed efficiently for many representations of the data including
distributed databases and data streams.

We derive novel unbiased estimators and efficient confidence bounds
for subpopulation weight.  Our estimators and bounds are tailored by
distinguishing between applications (such as data streams) where the
total weight of the sketched set can be computed by the summarization
algorithm without a significant use of additional resources, and
applications (such as sketches of network neighborhoods) where this is
not the case.  Our {\em rank conditioning} (\rc) estimator, is
applicable when the total weight is not provided.  This estimator
generalizes the known estimator for \pri{} sketches
\cite{DLT:sigmetrics04} and its derivation is simpler.  When the total
weight is available we suggest another estimator, the {\em subset
  conditioning\/} (\ssc) estimator which is tighter.

 Our rigorous derivations, based on clever applications of the
Horvitz-Thompson estimator (that is not directly applicable to
bottom-$k$ sketches), are complemented by efficient computational
methods.  Performance evaluation using a range of Pareto weight distributions
demonstrate considerable benefits of the \ws\ \ssc\ estimator on
larger subpopulations (over all other estimators); of the \ws\ \rc\
estimator (over existing estimators for this basic sampling method);
and of our confidence bounds (over all previous approaches).  Overall,
we significantly advance the state-of-the-art estimation of
subpopulation weight queries.
\end{abstract}

\ignore{
\vspace{0.7mm}
\noindent
{\bf Categories and Subject Descriptors:} G.3: probabilistic algorithms; C.2.3: network monitoring $\quad$
\noindent
{\bf General Terms:} Algorithms, Measurement, Performance $\quad$
}

\vspace{-0.1in}
\section{Introduction}

  Sketches or statistical summaries of massive data sets are an
extremely useful tool. Sketches are obtained by applying a
probabilistic summarization algorithm to the data set.  The algorithm
returns a sketch that has smaller size than the original data set but
supports approximate query processing on the original data set.

 Consider a set of records $I$ with associates weights $w(i)$ for
$i\in I$.  A basic aggregate over such sets is
{\em subpopulation weight}.  
A {\em subpopulation weight query} 
specifies a subpopulation $J\subset I$ as a predicate
on attributes values of records in $I$.   The result of the
query is $w(J)$, the sum of the weights of records in $J$.
This aggregate can be used
to estimate other aggregates over subpopulations 
such as selectivity ($w(J)/w(I)$), and variance and higher moments
of $\{w(i)|i\in J\}$~\cite{CoKa:moments03}.

As an example consider a set of all IP flows going through a router or
a network during some time period.  Flow records containing this
information are collected at IP routers by tools
such as Cisco's NetFlow~\cite{Netflow} (now emerging as an IETF
standard).  Each flow record contains the number of packets and bytes
in each flow.  Possible subpopulation queries in this example are
numerous.  Some examples are ``the bandwidth used for an application
such as p2p or Web traffic'' or ``the bandwidth destined to a
specified Autonomous System.''  The ability to answer such queries is
critical for network management and monitoring, and for anomaly
detection.

Another example is census database that includes a record for
each households with associated weight equal to 
the household income.
Example queries are to find total income by region or by the
gender of the head of the household.

To support subpopulation selection with arbitrary predicates, 
the summary must retain
content of some individual records.  Two common summarization methods
are {\em $k$-mins} and {\em bottom-$k$} sketches.  Bottom-$k$ sketches
are obtained by assigning a {\em rank value}, $r(i)$, for each $i\in
I$ that is independently drawn from a distribution that depends on
$w(i)\geq 0$.  The {\em bottom-$k$ sketch\/} contains the $k$ records
with smallest rank values \cite{ECohen6f,associations01}.  The
distribution of the sketches is determined by the family of
distributions that is used to draw the rank values:  By appropriately
selecting this family, we can obtain sketches that are distributed as
if we draw records without replacement with probability proportional
to their weights (\ws), which is a classic sampling method with a
special structure that allows sketches to be computed more efficiently
than other bottom-$k$ sketches. A different selection corresponds to
the recently proposed {\em priority sketches} (\pri)
\cite{DLT:sigmetrics04,dlt:pods05}, which have estimators that
minimize the sum of per-record variances~\cite{Szegedy:stoc06}.
 {\em $k$-mins\/}
sketches~\cite{ECohen6f} are obtained by assigning independent random
ranks to records (again, the distribution used for each record depends on
the weight of the record).  The record of smallest rank is selected, and
this is repeated $k$ times, using $k$ independent rank assignments.
$k$-mins sketches include weighted sampling with replacement (\wsr).
  Bottom-$k$ sketches are more informative than respective $k$-mins
sketches (\ws\ bottom-$k$ sketches can mimic \wsr\ $k$-mins
sketches~\cite{bottomk07:ds}) and in most cases can be derived much
more efficiently.

 Before delving into the focus of this paper, which is estimators and
confidence bounds for subpopulation weight, 
we overview classes of applications where these
sketches are produced, and which benefit from our results.

Bottom-$k$ and $k$-mins sketches are used as summaries of
 a single weighted set or as
summaries of multiple subsets that are defined over the same ground
set.  In the latter case, the sketches of different subsets are
``coordinated'' in the sense that each record  obtains a consistent rank
value across all the subsets it is included in.  These coordinated
sketches support subpopulation selection based on subsets' memberships
(such as set union and intersection).

  We distinguish between {\em explicit\/} or {\em implicit\/}
representations of the data~\cite{bottomk07:ds}.  Explicit
representations list the occurrence of each record in each subset.  They
include a representation of a single weighted set (for example, a
distributed data set or a data stream~\cite{CoSt:pods03,dlt:pods05})
or when there are multiple subsets that are represented as item-subset
pairs (for example, item-basket associations in a market basket data,
links in web pages, features in
documents~\cite{Broder:CPM00,BharatBroder:www99,associations01,SPWH:sigcomm00,BHRSG:sigmod07}).
Bottom-$k$ sketches can be computed much more efficiently than
$k$-mins sketches when the data is represented
explicitly\notinproc{~\cite{BRODER:sequences97,associations01,bottomk:sigmetrics07,bottomk07:ds}).}\onlyinproc{~\cite{BRODER:sequences97,associations01,bottomk07:ds}}).

Implicit representations are those where the multiple subsets are
specified compactly and implicitly (for example, as neighborhoods in a
metric
space~\cite{ECohen6f,CoWaSu,CoSt:pods03,MS:PODC06,KS06,CoKa:jcss07}.)
In these applications, the summarization algorithm is applied to the
compact representation.  Beyond computation issues, the distinction
between data representations is also important for estimation: In
applications with explicit representation, the summarization algorithm
can provide the total weight of the records without a significant
processing or communication overhead.  In applications with implicitly
represented data, and for sketches computed for subset relations, the
total weight is not readily available.

  An important variant uses {\em hash\/}  values of the identifiers 
of the records instead of random ranks.
For $k$-mins sketches, families of min-wise independent hash functions
or $\epsilon$-min-wise functions have the desirable
properties~\cite{Broder:CPM00,broder00minwise,DasuJMS:sigmod02}.
Hashing had also been used with bottom-$k$
sketches~\cite{BRODER:sequences97,associations01,BHRSG:sigmod07}.
This variant has the property that all copies of
the same record obtain the same rank value across subsets
without the need for
coordination between copies or additional book keeping. 
Therefore hashing
allows to perform aggregations over {\em distinct occurrences}
(see~\cite{FlajoletMartin85}).

  For records associated with points
in some metric space such as a graph, the
Euclidean plane, a network, or the time axis (data streams)
\cite{ECohen6f,CoKa:sigmod04,KS06}, sketches are produced for
neighborhoods of locations of interest.  For example, all records that
lie within some distance from a location or happened within some
elapsed time from the current time.  For such metric applications, we
do not want to explicitly store a separate sketch for each possible distance.
This is addressed by {\em all-distances sketches}.  The all-distances
sketch of a location is a 
succinct representation of the sketches of 
neighborhoods of all distances from the location.
All-distances $k$-mins sketch were introduced
in~\cite{ECohen6f,CoKa:sigmod04}.  All-distances bottom-$k$ sketches
were proposed and analyzed in~\cite{bottomk07:ds}.  All-distances
sketches also support {\em spatially or temporally decaying
aggregation}~\cite{KS06,CoKa:sigmod04}. One application of decaying
aggregates is kernel density estimators~\cite{ScottDW:Book} and
typicality estimation~\cite{HPFLL:VLDB2007} -- The estimated density
is a linear combination of the subpopulation weight over neighborhoods.


\vspace{-0.1in}
\paragraph*{Overview}
Section~\ref{sketches:sec} contains some background and definitions.
In Section~\ref{ML:sec} we apply the Maximum
Likelihood principle to derive \ws\ \ml\ estimators.\ignore{
for weight, subpopulation
weight, and a tighter estimator for subpopulation weight when the
total weight is provided.}  These estimators are applicable to \ws\
sketches as our derivation exploits special properties of the
exponential distribution used to produce these sketches.  While
biased, \ws\ \ml\ estimators can be computed efficiently and perform well
in practice.

Section~\ref{AW:sec}  introduces a variant of the Horvitz-Thompson
(\HT) estimator \cite{HT52}.  The \HT\ estimators assign a
positive {\em adjusted
weight} to each record that is included in the sketch.
Records not included in the sketch have zero adjusted weight.
  The adjusted
weight has the property that for each record, the expectation of its
adjusted weight over sketches is equal to its actual weight.
  The
adjusted weight is therefore an unbiased estimator of the weight of
the record. 
From linearity of expectation, the sum of the adjusted weights of
records in the sketch that are members of a subpopulation constitutes an
unbiased estimate of the weight of the subpopulation.  

The \HT\ estimator
assigns to each included record an adjusted weight equal to its actual
weight divided by the probability that it is included in a
sketch.  This estimator minimizes the per-record variance of the
adjusted weight for the particular distribution over sketches.  The
\HT\ estimator, however,
cannot be computed for bottom-$k$ sketches, since the
probability that a record is included in a sketch cannot be determined
from the information available in the sketch 
alone~\cite{Sampath:book,surveysampling:book}.  
Our variant, which we refer to as {\em \HT\ on a partitioned sample space}
(\HTp), 
overcomes this hurdle by applying the \HT\ estimator on a set of partitions of
the sample space such that this probablity can be computed in each
subspace.

 We apply \HTp\ to derive {\em Rank Conditioning estimators} (\rc) for
general bottom-$k$ sketches (that is, sketches produced with {\em
arbitrary}\/ families of rank distributions).  Our derivation
generalizes and simplifies one for \pri{} sketches 
(\pri\ \rc\ estimator)~\cite{DLT:sigmetrics04}
and reveals general principles.  It  provides tighter and simpler 
estimators for
\ws\ sketches than previously
known. We  show that the covariance between
adjusted weights of different records is zero and therefore the variance
of the subpopulation weight estimator is equal to the sum of the
variances of the records.\ignore{This property establishes the sum of per-record variances (used in~\cite{dlt:pods05,Szegedy:stoc06})
as a performance metric for all \rc\ estimators.}

In Section~\ref{WI:sec} we again apply \HTp\ and derive {\em
subset conditioning} estimators for \ws\ sketches (\ws\ \ssc).  These
estimators use the total weight $w(I)$ in the computation of the
adjusted weights.  The \ws\ \ssc\ estimator is superior to the \ws\
\rc\ estimator, with lower variance on {\em any} subpopulation: The
variance for each record is at most that of the \ws\ \rc\ estimator,
covariances of different records are {\em negative}, and the sum of all
covariances is zero. These properties give
the \ws\ \ssc\ estimator
 a distinct advantage
as the relative variance decreases for larger subpopulations.
The \ssc\ derivation exploits special properties of
\ws\ sketches  -- there is no known \pri\ estimator with negative
covariances.  Moreover, the \ws\ \ssc\ estimator is strictly better than
any \wsr\ estimator: it has a lower sum of per-record variances than
the \HT\ \wsr\ estimator (that minimizes the sum of per-record variances
for \wsr\ but covariances do not cancel out) 
and is also better than the \wsr\ ``ratio'' estimator based on the sum of 
multiplicities in the sample 
of records that are members of the subpopulation (which does 
has negative covariances that cancel out but a much higher sum of per-record
variances on skewed distributions).

The \ws\ \ssc\ estimator is expressed as a definite integral.  We
provide an efficient approximation method that is based on a Markov
chain that converges to this estimator.  After any fixed number of
steps of the Markov chain we get an unbiased estimate that is at least
as good as \ws\ \rc.  We implemented and compared the performance of a
$k$-mins estimator (\wsr), \ws\ \ml, \pri\ \rc, \ws\ \rc, and the
approximate \ws\ \ssc\ estimators on Pareto weight distributions with
a range of skew parameters (see Section~\ref{simul:sec}).  When the
total weight is unknown or is not used, the performances of \ws\ \ml,
\ws\ \rc, and \pri\ \rc\ are almost indistinguishable.  They
outperform \wsr\ and the performance gain grows with the skew of the
data.  Therefore, our estimator for \ws\ sketches nearly match
the best estimators on an optimal sketch distribution.

When the total weight is provided, the \ws\ \ssc\ estimator has
a significant advantage (smaller variance) on larger subpopulations
and emerges as the best estimator.  The simulations also show that the
approximate \ws\ \ssc\ estimator is very effective even with a small
number of steps.

  Confidence intervals are critical for many applications.  In
 Section~\ref{conf:sec} we derive confidence intervals (tailored to
 applications where the total weight is or is not provided) and
 develop methods to efficiently compute these bounds.
In Section~\ref{simul:sec} we compare our confidence bounds with
previous approaches (a bound for \pri{}
sketches~\cite{Thorup:sigmetrics06} and known \wsr\ estimators) 
using a range of Pareto
distributions with different skew parameters.  Our bounds for \ws\
sketches are significantly tighter than the \pri\ bounds, even when
the total weight is not used. This may seem surprising since combined
with our results, the \pri\ \rc\ estimator has nearly optimal variance
\cite{Szegedy:stoc06} among all \rc{} estimators.  The explanation is
that the confidence intervals do not reflect this near optimality.
Our \ws\ confidence bounds derivation, based on some special
properties of \ws\ sketches, exploits the information available in the
sketch.  We point on the sources of slack in the \pri\ confidence
bounds of~\cite{Thorup:sigmetrics06} that explain its inferior
behavior. We propose approaches to address some non-inherent sources
of slack.  Our \ws{} bounds that use the total weight are tighter, in
particular for large subpopulations, than those that do not use the
total weight.

\notinproc{
 A short summary of some of the results in this paper appeared in~\cite{bottomk:sigmetrics07}.
  }


\section{Preliminaries} \label{sketches:sec}

Let $(I,w)$ be
a weighted set.
A {\em rank assignment}\/ maps each item $i$ to a random
rank $r(i)$.  The ranks of items are drawn independently using a
family of distributions $\bf{ f}_w$ ($w\geq 0$), where the rank of an
item with weight $w(i)$ is drawn from $\bf{ f}_{w(i)}$.
For a subset $J$ of items and a rank assignment $r()$ we define
$B_i(r(),J)$ to be the item in $J$ with $i$th smallest rank according
to $r()$ and $r_i(J)\equiv r(B_i(r(),J))$ to be the $i$th smallest rank
value of an item in $J$.

\begin{definition}
  {\em $k$-mins sketches} are produced from $k$ independent rank
  assignments, $r^{(1)}(),\ldots,r^{(k)}()$.  The sketch of a subset $J$ is
  the $k$-vector $(r_1^{(1)}(J),r_1^{(2)}(J),\ldots,r_1^{(k)}(J))$.
 For some
  applications, we use a sketch that includes with each entry an identifier
  or some other attributes such as the
  weight of the items $B_1(r^{(j)}(),J)$ ($j=1,\ldots,k$). 
\end{definition}

\begin{definition} \label{def:botk} {\em Bottom-$k$ sketches}\/ are produced from a
  single rank assignment $r()$.    The bottom-$k$
  sketch $s(r(),J)$ of the subset $J$ is a list of 
  entries $(r_i(J),w(B_i(r(),J)))$ for
  $i=1,\ldots,k$. (If $|J| < k$ then the list contains only 
$|J|$ items.)
The list is ordered by
  rank, from smallest to largest.  In addition to the weight, the sketch
  may include an identifier and attribute values of items $B_i(r(),J)$
  ($i=1,\ldots,k$). We also include with
the sketch the $(k+1)$st smallest rank value $r_{k+1}(J)$ (without additional
attributes of the item with this rank value).
\end{definition}


In fact, bottom-$k$ sketches must
 include the items' weights but do 
not need to store all rank values and it suffices to 
store $r_{k+1}$. 
Using the weights of the items
with $k$ smallest ranks and $r_{k+1}$, we can {\em redraw}\/ 
rank values to items in $s$ using the
density function $\bf{f}_w(x)/\bf{F}_w(r_{k+1})$ for $0\leq x\leq r_{k+1}$ and 
$0$ elsewhere, for item with weight $w$~\cite{bottomk07:ds}.
\notinproc{
\begin{lemma} \label{newranks}
This process of re-assigning ranks
is equivalent to drawing a random rank assignment $r()$ and
taking $s(r'(),J)$ from the probability subspace where
$$\{B_1(r'(),J),\ldots,B_k(r'(),J)\}=\{B_1(r(),J),\ldots,B_k(r(),J)\}$$
(the same subset of items with 
$k$ smallest ranks, not necessarily in the same order) and 
$r_{k+1}(J)=r'_{k+1}(J)$.\footnote{As we shall see in Section~\ref{subscond}, 
if $w(J)$ is provided and we use \ws{} sketches,
we can redraw all rank values, effectively obtaining a rank assignment
from the probability subspace where the subset of items with $k$
smallest ranks is the same.}
\end{lemma}

Bottom-$k$ and $k$-mins sketches have the following useful property: 
The sketch of a union of
two sets can be generated from the sketches of the two sets.  Let
$J,H$ be two subsets.
For any rank assignment $r()$, $r(J\cup H)=\min\{ r(J), r(H)\}$.
Therefore, for $k$-mins sketches we have $(r_1(J\cup
H),\ldots,r_k(J\cup H))=$ \\ $(\min\{r_1(J), r_1(H)\},\ldots,\min\{r_k(J),
r_k(H)\})\ .$
This property also holds for bottom-$k$ sketches.  The $k$ smallest
ranks in the union $J\cup H$ are contained in the union of the sets of
the $k$-smallest ranks in each of $J$ and $H$.  That is, $B_k(r(),J\cup
H)\subset B_k(r(),J)\cup B_k(r(),H)$.  Therefore, the bottom-$k$ sketch of
$J\cup H$ can be computed by taking the pairs with $k$ smallest ranks
in the combined sketches of $J$ and $H$.  To support
subset relation queries and subset unions, the sketches must preserve
all rank values. 
} 

\vspace{-0.1in}
\paragraph*{ws sketches}
 The choice of which family of random rank functions to use matters
only when items are weighted.  Otherwise, 
we can map (bijectively) the ranks of one rank function to ranks
of another rank function in a way that preserves  the bottom-$k$
sketch.\footnote{We map $r$ such that $F_1(r) = \alpha$
to $r'$ such that $F_2(r') = \alpha$, where 
$F_1$ is the CDF of the first rank function and $F_2$ 
is the CDF of the other (assuming the CDFs are continuous).}
Rank functions $\bf{f}_w$
with some convenient properties
are exponential 
distributions with parameter $w$~\cite{ECohen6f}. 
The density function of this distribution is
$\bf{ f}_w(x) = w e^{-wx}$, and its cumulative
distribution function is $\bf{ F}_w(x) = 1- e^{-wx}$.
 The minimum rank $r(J)=\min_{i\in J} r(i)$ of an
item in a subset $J\subset I$ is exponentially distributed with
parameter $w(J)=\sum_{i\in J} w(i)$ (the minimum of independent 
exponentially distributed random variables 
is exponentially
distributed with parameter equal to the sum of the
parameters of these distributions).
Cohen \cite{ECohen6f} used this property
to obtain unbiased low-variance estimators
for both the weight and the inverse weight of the subset.\footnote{Estimators 
for the inverse-weight are useful for obtaining unbiased estimates for
quantities where the weight appears in the denominator such as
the weight ratio of two different subsets.}

 With exponential ranks the item with the minimum rank
$r(J)$ is a {\em weighted random sample}\/ from $J$: The probability
that an item $i\in J$ is the  item of minimum rank is $w(i)/w(J)$.  
Therefore, a
$k$-mins sketch of a subset $J$ corresponds to a weighted random
sample of size $k$, drawn {\bf with replacement\/} from $J$.
We call $k$-mins sketch using exponential ranks a \wsr\ sketch.
On the other hand,
a bottom-$k$ sketch of a subset $J$ 
with exponential ranks corresponds to
a weighted $k$-sample  drawn {\bf without replacement\/} from $J$~\cite{bottomk07:ds}. We call such a sketch a \ws\ sketch.
\ignore{
 \begin{lemma}
A \ws\ bottom-$k$
sketch of a subset $J$ is a sample of size
$k$ drawn without replacement from $J$.
 \end{lemma}
 \begin{proof}
The probability that the smallest item in $I$ is $i\in I$ is 
$w(i)/w(I)$.  Given that the bottom-$j$ items in $I$ are $i_1,\ldots,i_{j}$, the
   $(j+1)$st item is $i\in I\setminus\{i_1,\ldots,i_{j}\}$ with
   probability $w(i)/(w(I)-\sum_{h=1}^j w(i_h))$.
The first $k$ steps in this  process are
   identical to sampling $k$ items from the set $I$ without replacement.
 \end{proof}
}

The following property of exponentially-distributed ranks is a
consequence of the memoryless nature of the exponential distribution.
\ignore{
\begin{lemma} \label{memoryless}~\cite{bottomk07:ds}
Consider a subspace of the rank assignments
where the order of the items according to rank values is fixed,
say $i_1,\ldots,i_n$, and
the rank values of the first $j$ items are fixed.
Let $r(i_{j+1})$ be the random variable that is the $(j+1)$st smallest 
rank.
  The conditional 
distribution of $r(i_{j+1})-r(i_j)$ is exponential with parameter 
$\sum_{\ell=j+1}^n w(i_\ell)$.
\end{lemma}
By relaxing the conditioning of Lemma~\ref{memoryless}, we obtain
the following:
}
  \begin{lemma} \label{nextdistk:cor}~\cite{bottomk07:ds}
 Consider a probability
    subspace of rank assignments
 over $J$
where the 
$k$ items of
smallest ranks are 
$i_1,\ldots,i_k$ in increasing rank order.
   The rank differences $r_1(J), r_2(J)-r_1(J), \ldots, r_{k+1}(J)-r_k(J)$ are
    independent random variables, where $r_j(J)-r_{j-1}(J)$
    ($j=1,\ldots,k+1$) is exponentially distributed with parameter
    $w(J)-\sum_{\ell=1}^{j-1} w(i_\ell)$. (we formally define
    $r_0(J)\equiv 0$.)
  \end{lemma}

 \ws\ sketches can be computed more efficiently than other bottom-$k$
sketches in some important settings.  One such example is unaggregated
data (each item appears in multiple ``pieces'')
\cite{CDKLT:pods07,CDKLT:IMC07} that is distributed or resides in
external memory.  Computing a bottom-$k$ sketch generally requires
{\em pre-aggregating} the data, so that we have a list of all items and
their weight, which is a costly operation.  A key property of
exponential ranks is that we can obtain a rank value for an item by
computing {\em independently} a rank value for each piece, based on
the weight of the piece.  The rank value of the item is the minimum
rank value of its pieces. 

 The \ws\ sketch contains the items of the
$k$ pieces of distinct items with smallest ranks  and can be
computed in two communication rounds over distributed data or in two
linear passes: The first pass identifies the $k$ items with smallest
rank values.  The second pass is used to add up the weights of the
pieces of each of these $k$ items. 

 Another example is when items are
partitioned such that we have the weight of each part.  In this case,
a \ws\ sketch can be computed while processing only a fraction of the
items.  A key property is that the minimum rank value over a set of
items depends only on the sum of the weights of the items.  Using this
property, we can quickly determine which parts contribute to the
sketch and eliminate chunks of items that belong to other parts. 

The
same property is also useful when sketches are computed online over a
stream.  Bottom-$k$ sketches are produced using a priority queue that
maintains the $k+1$ items with smallest ranks.  We
 draw a rank for each item and update the queue if this rank
is smaller than the largest rank in the queue.
  With
\ws\ sketches, we can simply draw directly from a distribution the
accumulated weight of items that can be ``skipped'' before we obtain
an item with a smaller rank value than the largest rank in the queue.
The stream algorithm simply adds up the weight of items until it
reaches one that is incorporated in the sketch.

\vspace{-0.1in}
\paragraph*{\pri{}  sketches}
With priority ranks~\cite{DLT:sigmetrics04,dlt:pods05}
the rank value of an item with
weight $w$ is selected uniformly at random from $[0,1/w]$.  
This is  equivalent to choosing a rank value $r/w$,  where
 $r\in U[0,1]$, 
 the
uniform distribution on the interval $[0,1]$. 
It is well known that if $r\in U[0,1]$ then
$-\ln(r)/w$ is an exponential random variable
with parameter $w$. Therefore, in contrast with priority
ranks, 
exponential ranks
correspond to using rank values $-\ln r/w$ where  $r\in U[0,1]$.

\pri\ sketches are of interest because one can derive from them
an estimator that (nearly) minimizes the sum of per-item variances
$\sum_{i\in I} \var(\tilde{w}(i))$~\cite{Szegedy:stoc06}.
More precisely, Szegedy showed that
the sum of per-item variances using \pri{} sketches of size $k$
is no larger than the smallest sum of variances attainable
by an estimator that uses sketches with average size $k-1$.\footnote{Szegedy's proof applies
  only to estimators based on adjusted weight assignments.  It also does not apply to estimators on the weight of subpopulations.}

\noindent
Some of our results apply to arbitrary rank functions.  Some basic
properties that hold for both \pri\
and \ws\ ranks are {\em monotonicity}\/ --  if $w_1 \geq w_2$ then
for all $x\geq 0$, ${\bf F}_{w_1}(x) \leq {\bf F}_{w_2}(x)$
(items with larger weight are more likely to have smaller ranks)
and {\em invariability to scaling}\/ -- scaling of all the weights
does not change the distribution of subsets selected to the sketch.

\ignore{
 Observe  that a common property that holds for
both  exponential and priority ranks is that 
for any value $r$, ${\bf F}_w(r)$ is a nondecreasing function of $w$.
Therefore, smaller rank values are more likely for items with larger
weight.
}

\paragraph*{Review of weight estimators for wsr sketches}\label{wsrest:sec}
For a subset $J$, the rank values in the 
$k$-mins sketch $r_1(J),\ldots,r_k(J)$
are $k$ independent samples from an exponential distribution
with parameter $w(J)$.  
The quantity
$\frac{k-1}{\sum_{h=1}^k r_h(J)}$ is an unbiased estimator of $w(J)$.
The standard
deviation of this estimator is equal to $w(J)/\sqrt{k-2}$ and the
average (absolute value of the) relative error is 
approximately $\sqrt{2/(\pi (k-2))}$~\cite{ECohen6f}.
The quantity
$\frac{k}{\sum_{h=1}^k r_h(J)}$ is the maximum likelihood
estimator of $w(J)$.  This estimator is $k/(k-1)$ times the unbiased
estimator.  Hence, it is obviously biased, and the bias is 
equal to $w(J)/(k-1)$.  Since the standard deviation 
is about $(1/\sqrt{k}) w(J)$, the bias is not significant when $k \gg 1$.
The quantity $\frac{\sum_{h=1}^k r_h(J)}{k}$ is an unbiased estimator of 
the {\em inverse weight} $1/w(J)$. 
The standard deviation of this estimate is $1/(\sqrt{k}w(J))$.

\notinproc{
Subpopulation weight estimators for 
\wsr\ sketches when the total weight
is known are the \HT\ estimator, where the adjusted weight is the ratio
of the weight of the item and the probability $1-(1-w(i)/w(I))^k$ that
it is sampled.  This estimator minimizes the sum of per-item variances
but covariances do not cancel out.  Another estimator is the
sum of multiplicities of items in the sketch that are members of
the subpopulation, multiplied by total weight, and divided by $k$.
This estimator has covariances that cancel out, but higher per-item
variances.  With \wsr\ sketches it is not possible to obtain
an estimator with minimum sum of per-item variances and covariances
that cancel out. 
}

\section{Maximum Likelihood estimators for ws  sketches}
\label{ML:sec}
\label{weight-estimates}

\paragraph*{Estimating the total weight} Consider a set $I$ and its bottom-$k$ sketch $s$.  Let
$i_1,i_2,\ldots,i_k$ be the items in $s$ ordered by increasing ranks
(we use the notation $r(i_{k+1})$ for the $(k+1)$st smallest rank).
If  $|I|\leq k$ (and we can
determine this) then $w(I)=\sum w(i_j)$.


Consider the rank differences, $r(i_1), r(i_2)-r(i_1),\ldots,
r(i_{k+1})-r(i_{k})$.  From Lemma~\ref{nextdistk:cor}, they are
independent exponentially distributed random variables.  The joint
probability density function of this set of differences is therefore
the product of the density functions 

\vspace{-0.1in}
{\scriptsize
\begin{displaymath}
 w(I)\exp(-w(I)r(i_1))(w(I)-s_1)\exp(-(w(I)-s_1)(r(i_2)-r(i_1)))\cdots
\end{displaymath}
}
where $s_\ell = \sum_{j=1}^{\ell} w(i_j)$.
Think about this probability density  as a function of $w(I)$.
The maximum likelihood estimate for $w(I)$ is the value that maximizes
this function.
To find the maximum, take the natural logarithm (for simplification) of
the expression and look at the value which makes 
the derivative zero.  
We obtain that the maximum likelihood
estimator $\tilde{w}(I)$ is the solution of the equation
\begin{equation}
\sum_{i=0}^{k} \frac{1}{\tilde{w}(I)-s_i}=r(i_{k+1})\ .\label{mlbottomk}
\end{equation}
The left hand side is a monotone function, and the equation
can be solved by  a binary search on the range
$[s_{k}+1/r(i_{k+1}),s_{k}+(k+1)/r(i_{k+1})]$.  
\notinproc{
We can obtain a tighter estimator (smaller variance)
by redrawing the rank values
of the items $i_1,\ldots,i_{k}$ (see Lemma~\ref{newranks}) and taking
the expectation of the solution of Eq.~(\ref{mlbottomk}) (or average over 
multiple draws).
} 

\paragraph*{Estimating a subpopulation weight}
We  derive maximum likelihood subpopulation weight
estimators that use and do not use the total weight $w(I)$.
 Let $J\subset I$ be a subpopulation.  
Let $j_1,\ldots,j_a$ be the items in $s$
that are in $I\setminus J$.\footnote{We assume
that using meta attributes of items in the sketch 
we can decide which among them
are in $J$.} Let $r'_1,\ldots,r'_a$ be their respective
rank values and let $s'_i=\sum_{h\leq i} w(j_h)$ ($i=1,\ldots,a$).
Define $s'_0\equiv 0$.
Let $i_1,i_2,\ldots,i_c$ be the items in $J\cap s$. 
Let $r_1,\ldots,r_c$ be their respective
rank values and let $s_i=\sum_{h\leq i} w(i_h)$ ($i=1,\ldots,c$).
Define $s_0\equiv 0$.

\smallskip
\noindent
{\bf \ws\ \ml\ subpopulation weight estimator that does not use $w(I)$:}  
Consider rank assignments such that rank values in $I\setminus J$ are fixed
and the order of ranks of the items in $J$ is fixed.  The probability density
of the observed ranks of the first $k$ items in $J$ is that of
seeing the same rank differences (probability density is
$(w(J)-s_i)\exp(-(w(J)-s_i)(r_{i+1}-r_i)$ for the $i$th difference)
and of the rank difference
between the $c+1$ and $c$
 smallest ranks in $J$ being at least
$\tau-r_c$ (where $\tau$ is the $(k+1)$st smallest rank in the sketch),
which is $\exp(-(w(J)-s_c)(\tau-r_{c}))$.  
Rank differences are independent, and therefore, the probability
density as a function of $w(J)$ is the product of the above densities.
The maximum likelihood estimator for $w(J)$ is the value that maximizes
this probability.
If $c=0$, the expression $\exp(-w(J)\tau)$
is maximized for $w(J)=0$.  Otherwise, by
taking the natural logarithm and deriving we find that the value
of $w(J)$ that maximizes the probability density
\ignore{  Let $\tau=\max\{r'_a,r_c\}$ (the largest
rank value in the sketch $s$).}
is the solution of $\sum_{h=0}^{c-1} \frac{1}{\tilde{w}(J)-s_h}=\tau$.
As with the  estimator of the total weight, we can obtain a tighter estimator 
by redrawing the rank values.

\onlyinproc{
{\bf \ws\ \ml\ subpopulation weight estimator that uses $w(I)$:} 
Consider the subspace where  $j_1,\ldots,j_a$
are of smallest rank in $J$ (in this order), and $i_1,\ldots,i_c$
are of smallest rank in $I\setminus J$ (in this order).
We compute the probability density in this subspace,
 as a function of $w(J)$, of obtaining the rank differences that we
see.
We  find the $w(J)$ that maximized this density.
}

\notinproc{
\smallskip
\noindent
{\bf \ws\ \ml\ subpopulation weight estimator that uses $w(I)$:} 
We compute the probability density, as a function of $w(J)$, of
the event that we obtain the sketch $s$ with these ranks given
that the prefix of sampled items from $I\setminus J$ is
$j_1,\ldots,j_a$ and the prefix of sampled items from $J$ is
$i_1,\ldots,i_c$.  
We take the natural logarithms of the
joint probability density and derive with respect to $w(J)$.
If $c=0$, the derivative is positive and the
probability density is maximized for $w(J)=0$. If $a=0$, the derivative is negative 
and the probability density
is maximized for $w(J)=w(I)$.  Otherwise, if $a>0$ and $c>0$, 
the probability density is maximized for $\tilde{w}(J)$
that is the solution of
$$\sum_{h=0}^{c-1} \frac{1}{\tilde{w}(J)-s_h}-\sum_{h=0}^{a-1} \frac{1}{(w(I)-\tilde{w}(J))-s'_h}=0\ .$$
\ignore{
Observe that the solution depends only on
the prefix of $I$ that is in $s$ and not on the actual rank values.
This is expected, since with knowledge of $w(I)$ and conditioned on
the prefix, we have knowledge of the distribution of rank values.
}
The equation is easy to solve numerically, because the left hand
side is a monotone decreasing function of $w(J)$.

\ignore{This estimator is tighter, on any subpopulation,
than the respective maximum likelihood estimator that does not
use $w(I)$.  Observe that this estimator
does not use the $(k+1)$st smallest rank. 
 \edith{Can we prove that it is tighter in terms of mean square error ?}
}
} 

\def\AW{\mbox{\sc AWS}}
\section{Adjusted weights} \label{AW:sec}

\begin{definition} \label{def:aws}
 Adjusted-weight summarization (\AW) of a weighted set $(I,w)$
is a probability distribution $\Omega$ over
weighted sets $b$  of the form $b=(J,a)$ where $J\subset I$ and $a$ is a 
weight function on $J$, such that for all $i\in I$, 
$E(a(i))=w(i)$.
(To compute this expectation we extend the weight function
  from $J$ to $I$ by
assigning $a(i)=0$ for items $i\in I\setminus J$.)
For $i\in J$ we call $a(i)$ the adjusted weight of $i$ in $b$.
\end{definition}

  An {\em \AW\ algorithm\/} is a probabilistic algorithm that inputs a
weighted set $(I,w)$ and returns a weighted set according to some \AW\
of $(I,w)$.  An \AW\ algorithm for $(I,w)$  provides unbiased estimators
for the weight of $I$ and for the weight of 
subsets of $I$: By
linearity of expectation, for any $H\subseteq I$, the sum $\sum_{i\in H}
a(i)$ is an unbiased estimator of $w(H)$.\footnote{A useful
property of adjusted weights is that they provide
unbiased aggregations over 
{\em any other numeric attribute}: For weights $h(i)$, $\sum_{i\in H} h(i)a(i)/w(i)$
is an unbiased estimator of $h(J)$.}

 Let $\Omega$ be a distribution over sketches $s$, where each
sketch consists of a subset of $I$
and some additional information such as the rank values of the items included
in the subset.
  Suppose that given the sampled sketch $s$ 
  we can compute $\Pr\{i\in s |
s\in \Omega\}$ for all $i\in s$ (since $I$ is a finite set, these probabilities
are strictly positive for all $i\in s$). 
Then we can make $\Omega$ into an \AW\ using the Horvitz-Thompson (\HT)
estimator \cite{HT52} which provides for each $i\in s$ the adjusted weight
$$a(i)=\frac{w(i)}{\Pr\{i\in s | s\in \Omega\}}\ .$$ It is well known and
easy to see that
these adjusted weights are unbiased and
have minimal variance {\em for each
item} for the particular distribution $\Omega$ over subsets.

\smallskip
\noindent
{\bf \HT\ on a partitioned sample space (\HTp)} is a method to derive adjusted weights
when we
cannot determine $\Pr\{i\in s | s\in \Omega\}$
from the sketch $s$ alone.
For example if $\Omega$ is a distribution of
bottom-$k$ sketches, then the probability
$\Pr\{i\in s | s\in \Omega\}$ generally depends on all the
weights $w(i)$ for $i\in I$
and therefore, it cannot be determined from the information contained in 
$s$ alone.

  For each item $i$ we
partition $\Omega$ into subsets
$P^i_1,P^i_2\ldots$.  This partition satisfies the following two
requirements
\begin{enumerate}
\item
Given a sketch $s$,
we can determine the set
$P^i_j$ containing $s$.
\item
For every set $P^i_j$ we can compute the 
conditional probability $p^i_j=\Pr\{i\in s | s\in P^i_j\}$.
\end{enumerate}
 For each $i\in s$,
we identify the set $P^i_j$
and  use the adjusted weight $a(i)=w(i)/p^i_j$ (which is
the \HT\ adjusted weight in $P^i_j$).\footnote{In fact all we need is the probability
$p^i_j$. In some cases we can compute it
 from some parameters of $P^i_j$,
without identifying 
 $P^i_j$ precisely.}
Items $i\not\in s$ get an adjusted weight of 0. 
The expected adjusted weight of each
item $i$ within each subspace of the partition is $w(i)$ and therefore 
its expected adjusted weight over $\Omega$ is $w(i)$.

\smallskip
\noindent
{\bf Rank Conditioning (\rc) adjusted weights} for bottom-$k$ sketches
are  an \HTp{} estimator.
The probability
space $\Omega$ includes all rank assignments.  The sketch includes the
$k$ items with smallest rank values and
the $(k+1)$st smallest rank $r_{k+1}$.
The partition $P^i_1,\ldots,P^i_\ell$ which we use
is based on {\em rank conditioning}. For each possible rank value $r$
we have a
set $P^i_r$ containing all rank assignments in which
the $k$th rank assigned to an item other than $i$ is $r$.
(If $i\in s$ then this is the $(k+1)$st smallest rank.)

The probability that $i$ is included in a bottom-$k$ sketch 
given that the rank assignment is from $P^i_r$ is
the probability that its rank value is smaller than $r$.
For \ws\ sketches, this probability is equal to
$1-\exp(-w(i) r)$.
Assume $s$ contains
 $i_1,\ldots,i_k$ and that the
$(k+1)$st smallest rank $r_{k+1}$ is known.
Then  for item $i_j$, the rank assignment 
belongs to $ P^{i_j}_{r_{k+1}}$,
and therefore the adjusted weight of $i_j$ is 
$\frac{w(i_j)}{1-\exp(-w(i_j) r_{k+1})}$. 
The \ws\ \rc\ estimator on the total weight is
$\sum_{j=1}^k \frac{w(i_j)}{1-\exp(-w(i_j) r_{k+1})}\ .$
\ignore{If the $(k+1)$st rank is not available, we can assign adjusted
weight only to the $k-1$ items and use the estimate
$\sum_{i=1}^{k-1} \frac{w_i}{1-\exp(-w_i r_{k})}$.}
 The \pri\ \rc\ adjusted weight for an item $i_j$
(obtained by a tailored derivation
in~\cite{dlt:pods05}), is $\max\{w(i_j),1/r_{k+1}\}.$

\paragraph*{Variance of \rc\ adjusted weights}
 \begin{lemma} \label{lem:covrc}
Consider \rc\ adjusted weights and two items $i$, $j$.
Then, $\cov(a(i),a(j))=0$ (The covariance of the adjusted weight of $i$
and the adjusted weight of $j$ is zero.)
 \end{lemma}
\notinproc{
 \begin{proof}
It suffices to show that $E(a(i) a(j))=w(i) w(j)$.
Consider a partition of the sample space of all
rank assignments according to
the $(k-1)$th smallest 
rank of an item in $I\setminus \{i,j\}$.\footnote{We can use a
finer partitions in which all the ranks in $I\setminus \{i,j\}$
are fixed.}
Consider a subset in the partition and let 
$r_{k-1}$ denote the value of the $(k-1)$th smallest rank
of an item in $I\setminus \{i,j\}$
for rank assignments in this subset.
We show that in this subset $E(a(i) a(j))=w(i) w(j)$.
The product $a(i) a(j)$ is positive in this subset only when $r(i)< r_{k-1}$ and 
$r(j)< r_{k-1}$, which (since rank assignments are independent)
happens with probability $\pr\{r(i)<r_{k-1}\} \pr\{r(j)<r_{k-1}\}$.  
In this case the $k$th smallest rank in $I\setminus\{i\}$
and $I\setminus\{j\}$ is $r_{k-1}$ and therefore,
$a(i)=\frac{w(i)}{\pr\{r(i)<r_{k-1}\}}$, $a(j)=\frac{w(j)}{\pr\{r(j)<r_{k-1}\}}$.
It follows that

\vspace{-0.1in}
{\small
\begin{equation*}
\begin{array}{l}
E(a(i) a(j))=  \\
\pr\{r(i)<r_{k-1}\} \pr\{r(j)<r_{k-1}\}\frac{w(i)}{\pr\{r(i)<r_{k-1}\}}\frac{w(j)}{\pr\{r(j)<r_{k-1}\}} \\
=w(i)w(j)\ . 
\end{array}
\end{equation*}
}
\end{proof}
} 

  This proof also extends to show that for any subset $J\subset I$,
$E(\prod_{i\in I} a(i))=\prod_{i\in I} w(i)$.
 \begin{corollary}
For a subset $J\subset I$,
$$\var(a(J))=\sum_{j\in J} \var(a(j)) \ . $$
  \end{corollary}

 Therefore, with \rc\ adjusted weights, the variance of the
weight estimate of a subpopulation is equal to the sum of the per-item
variances, just like when items are selected independently.  This
Corollary, combined with Szegedy's result~\cite{Szegedy:stoc06}, shows
that when we have a choice of a family of rank functions,
\pri\ weights are the best rank functions to use
when using \rc\ adjusted weights.

\smallskip
\noindent
{\bf Selecting a partition.}
 The variance of the adjusted weight $a(i)$ obtained using \HTp\
depends on the particular partition in the following way.
\begin{lemma} \label{coarser:lemma} \label{lem:coarse}
Consider two partitions of the sample space, such that one partition 
is a refinement of the other, and the \AW s obtained by applying
\HTp\ using these partitions. For each $i\in I$, 
the variance of $a(i)$ using the coarser partition
is at most that of the finer partition.
\end{lemma}
\notinproc{
\begin{proof}
We use the following simple property of the variance.
Consider two random variables $A_1$ and $A_2$ 
over a probability space $\Omega$.  Suppose that
there is a partition $\{B_j\}$ of $\Omega$ such that
for every $B_j$, and for every $s\in B_j$, $A_2(s)=E(A_1(s)|s\in B_j)$.
Then $\var(A_2)\leq \var(A_1)$.

Let  $P^i_j$ be the sets in the fine partition, and
let  $C^i_\ell$ be the sets in the coarse partition such that
 $C^i_\ell = \bigcup_t P^i_{\ell_t}$. 
Let $\overline{P}^i_j$ be the subset
containing all $s\in P^i_j$  such that $i\in s$.
Similarly, let $\overline{C}^i_\ell$ be the subset
containing all $s\in C^i_\ell$ such that $i\in s$.
Let $a(i,s)$ be the adjusted weight of $i$ in a sketch $s$
according to the partition  $P^i_j$, and let
 $\overline{a}(i,s)$ be the adjusted weight of $i$ in a sketch $s$
according to the partition  $C^i_\ell$.
We will show that for $s \in \overline{C}^i_\ell$ such
that $i\in s$, $\overline{a}(i,s) = E_{s'\in \overline{C}^i_\ell}(a(i,s'))$.
From this and the property of the variance stated above the lemma follows.
We   remove the superscript $i$ from the sets 
$P^i_j$, $C^i_\ell$,
$\overline{P}^i_j$, and $\overline{C}^i_\ell$ in the rest of the proof.

Let $p_j= \pr(s\in  \overline{P}_j \mid s\in P_j)$ and 
$\overline{p}_\ell= \pr(s\in  \overline{C}_j \mid s\in C_\ell)$.
Now,
\begin{eqnarray*}
E_{s'\in C^i_\ell}(a(i,s')) & = &
\frac{\sum_{t}\pr(s\in \overline{P}_{\ell_t})\frac{w(i)}{p_{\ell_t}}}
     {\pr(s\in \overline{C}_\ell)} \\
& = &
\frac{\sum_{t}\pr(s\in P_{\ell_t})p_{\ell_t}\frac{w(i)}{p_{\ell_t}}}
     {\pr(s\in C_\ell)\overline{p}_\ell} \\
& = &
\frac{w_i\sum_{t}\pr(s\in P_{\ell_t})}
     {\pr(s\in C_\ell)\overline{p}_\ell} \\
& = &
\frac{w_i}{\overline{p}_\ell} = \overline{a}(i) \ . 
\end{eqnarray*}
\end{proof}
} 

It follows from Lemma \ref{coarser:lemma} that when applying \HTp, it
is desirable to use the coarsest partition for which we can compute
the probability $p^i_j$ from the information in the sketch.  In
particular a partition that includes a single component minimizes the
variance of $a(i)$ (This is the \HT\ estimator).  The \rc\ partition
yields the same adjusted weights as conditioning on the rank values of
all items in $I\setminus i$, so it is in a sense also the {\em finest}
partition we can work with.  It turns out that when the total weight
$w(I)$ is available we can use a coarser partition.

\section{Using the total weight} \label{WI:sec}

When the total weight is available we can use
\HTp{} estimators defined using a coarser partition of
the sample space than the one used by the \rc{} estimator.
The {\em Prefix conditioning estimator\/}  computes the adjusted 
weight of item $i$ by
partitioning the sample space according to the
sequence (prefix) of $k-1$ items with smallest ranks drawn from $I\setminus \{
i\}$. The {\em subset conditioning estimator} (\ssc) 
uses an even coarser partition defined by the
 {\em unordered set} of the first 
$k-1$ items that are different from $i$. By Lemma 
\ref{coarser:lemma}
subset conditioning is the best in terms of per-item variances.
Another advantage of these estimators is that they 
do not need $r_{k+1}$ and thereby require one less sample.
\onlyinproc{In this abstract we discuss only the \ssc{} estimator.}


\notinproc{
\subsection{Prefix conditioning estimator.}
For an item $i\in s$ we partition the sample space according to the
sequence (prefix) of $k-1$ items with smallest ranks drawn from $I\setminus \{
i\}$. That is if $i\not\in s$, then $s$ belongs to the partition
associated with the $k-1$ items in $s$ of smallest ranks.  If $i\in
s$, then $s$ belongs to the partition associated with the sequence of
$k-1$ items in $s\setminus \{i \}$.

We assign adjusted weights as follows. Consider a sketch $s$ and $i\in
s$.  Let $P$ be the set of sketches with the same prefix of $k-1$
items from $I\setminus \{ i \}$ as in $s$.  We compute the probability
$\pr\{i\in s \mid s\in P\}$, that is, the probability that $i$ is in a
sketch from $P$.  We compute the probability of $i$ occurring in each
of the positions $j\in 1,\ldots,k$ and the probability that it does
not occur at all. We use the notation $\pref_J(j_1,\ldots,j_k)$ for
the event that the first $k$ items drawn by weighted sampling without
replacement from a subset $J$ are $j_1,\ldots,j_k$.

 We denote by $i_\ell$ ($1\leq\ell\leq k-1$) 
the $\ell$th item in $s\setminus\{i\}$.
  For each $j=1,\ldots,k$, the probability $e_j$ that $i$
appears in the $j$th position in a sketch from $P$ is
{\small
\begin{eqnarray*}
\lefteqn{p(i\rightarrow j\cap s\in P)=
\pr\{\pref_I(i_1,i_2,i_{j-1},i,i_{j},i_{k-1})\}= } \\ & & 
 \frac{w(i_1)}{w(I)}
 \frac{w(i_2)}{w(I)-w(i_1)} 
 \frac{w(i_{j-1})}{w(I)-\sum_{m=1}^{j-2}w(i_m)} 
 \frac{w(i)}{w(I)-\sum_{m=1}^{j-1}w(i_m)}   \\ & &
 \frac{w(i_{j})}{w(I)-\sum_{m=1}^{j-1}w(i_m)-w(i)} 
 \cdots \frac{w(i_{k-1})}{w(I)-\sum_{m=1}^{k-2}w(i_m)-w(i)}\ .
\end{eqnarray*}
}

\noindent
 The probability that the sketch is from $P$ but $i$
 does not appear in it (technically, appears in
a position $k+1$ or beyond) is
{\scriptsize
\begin{eqnarray*}
\lefteqn{p(i\not\in s\cap s\in P)=
\pr\{\bigcup_{\ell\in\{I\setminus s\}}\pref(i_1,i_2,\ldots,i_{k-1},\ell)\}=}\\
& &  
\frac{w({i_1})}{w(I)} \frac{w(i_2)}{w(I)-w(i_1)} \cdots \frac{w(i_{k-1})}{w(I)-\sum_{m=1}^{k-2}w(i_m)}  \\ & &
\frac{w(I)-w(i)- \sum_{m=1}^{k-1}w(i_m)}
{w(I)-\sum_{m=1}^{k-1}w(i_m)}\ .
\end{eqnarray*}
}

\noindent
  Therefore,
{\small
\begin{equation*}
\pr\{i\in s \mid s\in P\}=
\frac{ \sum_{j=1}^k  p(i\rightarrow j\cap s\in P)}
{ \sum_{j=1}^k  p(i\rightarrow j\cap s\in P) + p(i\not\in s\cap s\in P)} \ .
\end{equation*}
}

  The computation of the prefix conditioning adjusted weights
is quadratic in
 $k$ for each item $i$.
 \rc\ adjusted weights, on the other hand, can be computed in
 constant number of operations per item.

\subsection{Subset conditioning estimator.} 
}
\label{subscond}


The \ssc{} estimator has 
the following two additional important properties. In contrast with
\rc{}, the
adjusted weights of different items
  have negative covariances, and the covariances
cancel out:  
the sum of the adjusted weights equals the total weight of the set.
This implies that the variance of the estimator of a large subset
is smaller than the sum of the variances of the individual
items in the subset, and in particular, the variance of the estimator
for the entire set is zero. We now define this estimator precisely.

  For a set $s=\{i_1,i_2,\ldots,i_k\}$ and $\ell \geq 0$, we define
\begin{equation}\label{intsubset}
f(s,\ell)=\int_{x=0}^\infty \ell\exp(-\ell x)\prod_{j\in s}(1-\exp(-w(i_j)x))dx\ .
\end{equation}
This 
 is the probability that a random rank assignment with exponential
ranks for items in $s$ and an  additional set of items $X$
such that $w(X) = \ell$, assigns the $|s|$ smallest ranks to
 the items in 
$s$ and the $(|s|+1)$st smallest rank to an item from $X$.
 For exponential ranks, this probability  depends only on $w(X)$
(the total weight of items in $X$), and does not depend on how the weight
of $X$ is divided between items.  This is a critical property that allows
us to compute  adjusted weights
with  subset conditioning.

 Recall that for an item $i$, we use the 
subspace
with all rank assignments in which among the items in
$I\setminus\{i\}$, the items in $s\setminus \{ i \}$
have the $(k-1)$st smallest ranks. 
 The probability, conditioned on this subspace, that 
 item $i$ is contained in the sketch is
 $\frac{f(s,w(I\setminus s))}{f(s\setminus\{i\},w(I\setminus s))}$,
 and so the adjusted weight assigned to $i$ is
$$a(i)=w(i)\frac{f(s\setminus\{i\},w(I\setminus s))}{f(s,w(I\setminus s))}\ .$$
The following lemma shows that \ssc{} estimate the
entire set with zero variance.

 \begin{lemma}
Let $s$ be a \ws\ sketch of $I$ and
let $a(i)$ be \ssc\ adjusted weights.
Then, $\sum_{i\in s} a(i) = w(I)$.
 \end{lemma}
\notinproc{
 \begin{proof}
  Observe that for any sketch $s$, $i\in s$, and $\ell \geq 0$
\begin{equation}\label{manip}
f(s,\ell)=f(s\setminus\{i\},\ell)-f(s\setminus\{i\},\ell+w(i)) \frac{\ell}{\ell+w(i)}\ .
\end{equation}
  This relation follows by manipulating Eq.~(\ref{intsubset}), or by
the following argument:  Let $X=I\setminus s$ and $w(X)=\ell$.
The probability that the items with
smallest ranks in $s\cup X$ are the items in $s$ is equal to the
probability that the $|s|-1$
items of  smallest ranks in $(s\setminus\{i\})\cup X$
are $s\setminus\{i\}$ minus the probability that the
$|s|-1$ items of 
smallest ranks in $s\cup X$ are $s\setminus\{i\}$ and the $|s|$th
smallest rank is from $X\setminus\{i\}$.  This latter probability
is equal to $$f(s\setminus\{i\},w(X\cup \{i\})) \frac{\ell}{\ell+w(i)}\ .$$

\noindent
Using Equation (\ref{manip}) we obtain that
{\scriptsize

\begin{eqnarray*}
\lefteqn{\sum_{i\in s} a(i)=} \\
 & = & \frac{\sum_{i\in s}w(i)f(s\setminus\{i\},w(I\setminus s))}{f(s,w(I\setminus s))} \\
&=& \frac{\sum_{i\in s}w(i)(f(s,w(I\setminus s))+f(s\setminus\{i\},w(I\setminus \{s\setminus \{i\}\})) \frac{w(I\setminus s)}{w(i)+w(I\setminus s)})}{f(s,w(I\setminus s))} \\
&=&\sum_{i\in s}w(i)+w(I\setminus s)\sum_{i\in s}\frac{\frac{w(i)}{w(i)+w(I\setminus s)}f(s\setminus\{i\},w(I\setminus \{s\setminus \{i\}\}))}{f(s,w(I\setminus s))}\\
&=& w(I) \ .
\end{eqnarray*}


}

 To verify the last equality, observe that
$$\frac{w(i)}{w(i)+w(I\setminus s)}f(s\setminus\{i\},w(I\setminus \{s\setminus \{i\}\})) $$
is the probability that the first $|s|-1$ items drawn from $I$
are $s\setminus{i}$ and the $|s|$th item is $i$.  These are disjoint
events and their union is the event that the first $|s|$ items 
drawn from $I$ are $s$.  The probability of this union is 
$f(s,w(I\setminus s))$.
  \end{proof}
} 

 \begin{lemma}  \label{lem:cov}
Consider \ssc\ adjusted weights of two items $i \not= j$.
Then, $\cov(a(i),a(j))<0$.
 \end{lemma}
\notinproc{
\begin{proof}
Consider a partition of rank assignments according to
the items in $I\setminus\{i,j\}$ that have the $k-2$ smallest ranks.
Consider a part in this partition and denote this set of $k-2$ items
by $c$.
We compute the expectation of $a(i)a(j)$ conditioned on this part.
Let $\ell=w(I)-w(c)-w(i)-w(j)$.
The probability of this part is $f(c,\ell)$, the probability
that $a(i)a(j)>0$ in $c$ is equal to $f(c\cup\{i,j\},\ell)$.
Therefore, the conditional probability is
$\frac{f(c\cup\{i,j\},\ell)}{f(c,\ell)}$.
In this case, the adjusted weight assigned to $i$ is
set according to items $c\cup\{j\}$ having the $(k-1)$ smallest ranks
in $I\setminus i$.  Therefore, this weight is
\onlyinproc{$}\notinproc{$$}a(i)=w(i) \frac{f(c\cup\{j\},\ell)}{f(c\cup\{i,j\},\ell)}\ .\onlyinproc{$}\notinproc{$$}
Symmetrically for $j$,
\onlyinproc{$}\notinproc{$$}a(j)=w(j) \frac{f(c\cup\{i\},\ell)}{f(c\cup\{i,j\},\ell)}\ .\onlyinproc{$}\notinproc{$$}
We therefore obtain that $E(a(i)a(j))$ conditioned on this part is
$$w(i)w(j)\frac{f(c\cup\{j\},\ell)f(c\cup\{i\},\ell)}{f(c\cup\{i,j\},\ell)f(c,\ell)}\ .$$
It suffices to show that
$$\frac{f(c\cup\{j\},\ell)f(c\cup\{i\},\ell)}{f(c\cup\{i,j\},\ell)f(c,\ell)}\leq 1\ .$$
To show that, we apply Eq.~\ref{manip} and substitute in the numerator
\onlyinproc{$}\notinproc{$$}f(c\cup\{j\},\ell)=f(c,\ell)-f(c,\ell+w(j))\frac{\ell}{\ell+w(j)}\onlyinproc{$}\notinproc{$$}
and in the denominator
\onlyinproc{$}\notinproc{$$}f(c\cup\{i,j\},\ell)=f(c\cup\{i\},\ell)-f(c\cup\{i\},\ell+w(j))\frac{\ell}{\ell+w(j)}\onlyinproc{$}\notinproc{$$}
The numerator being at most the denominator therefore follows from the
immediate inequality
$$f(c,\ell)f(c\cup\{i\},\ell+w(j))\leq f(c,\ell+w(j))f(c\cup\{i\},\ell)\ .$$
\end{proof}
}

 \begin{lemma}
Consider \ws{} sketches of a weighted set $(I,w)$ and 
subpopulation $J\subset I$.  The \ssc\ estimator for the
weight of $J$
has smaller variance than 
 the \rc\ estimator for the
weight of $J$.
 \end{lemma}
\begin{proof}
By Lemma \ref{lem:covrc} 
the variance of the \rc{}
estimator for  $J$ is
$\sum_{j\in J}\var_{{\rm RC}}(a(j))$. So using Lemma
 \ref{lem:coarse} 
we obtain that $\sum_{j\in J}\var_{{\rm SC}}(a(j))$
is no larger than the variance of the \rc{}
estimator for $J$.
Finally since
{\small
 $$\var_{{\mbox{\ssc}}}(\sum_{j\in J} a(j))=\sum_{j\in J}\var_{{\mbox{\ssc}}}(a(j)) + \sum_{i\not= j, i,j\in J}\cov_{{\mbox{\ssc}}}(a(i),a(j))\ ,$$ 
}
and Lemma \ref{lem:cov} that implies that the second term is negative the lemma follows.
\end{proof}

\subsection{Computing \ssc\ adjusted weights.} \label{sec:compsc}
The adjusted weights can be computed by numerical integration.
We propose (and implement) an alternative method based on a Markov chain
that is faster and easier to implement.
The method converges to the subset conditioning adjusted weights as
the number of steps grows.  It can be used with a fixed
number of steps and provides unbiased adjusted weights.

 As an intermediate step we define a new estimator as
follows.
We partition the rank assignments into subspaces, each
consisting of all
rank assignments with the same ordered set 
 of  $k$ items of smallest ranks.
Let $P$ be a subspace in the partition. 
For each rank assignment in $P$ and item $i$ the
adjusted weight of $i$ is the expectation of the 
\rc\ adjusted weight of $i$ over all rank assignments in $P$.\footnote{Note that
this is not
an instance of \HTp, we simply average 
another estimator in each part.}


  These adjusted weights are
unbiased because the underlying \rc\ adjusted weights
are unbiased.  By the convexity of the variance, they have smaller
per-item variance than \rc.

\notinproc{
It is also
easy to see that the variance of this estimator is higher than the
variance of the prefix conditioning estimator: Rank assignments with
the same prefix of items from $I\setminus i$, but where the item $i$
appears in different positions in the $k$-prefix, can have different
adjusted weights with this assignment, whereas they have the same
adjusted weight with prefix conditioning.
}

The distribution of $r_{k+1}$ in each subspace $P$ is the sum
of $k$ independent exponential random variables with parameters
$w(I)$, $w(I)-w(i_1)$,\ldots,$w(I)-\sum_{h=1}^{k}w(i_h)$ where
$i_1,\ldots,i_k$ are the items of $k$ smallest ranks in 
rank assignments of $P$ (see
Lemma~\ref{nextdistk:cor}).  So the adjusted weight of $i_j$
($j=1,\ldots,k$) is $a(i_j)=E(w(i_j)/(1-\exp(-w(i_j)r_{k+1})))$ where
the expectation is over this distribution of $r_{k+1}$.


 Instead
of computing the expectation, we 
average the \rc\ adjusted weights
$w(i_j)/(1-\exp(-w(i_j)r_{k+1}))$ over multiple draws of $r_{k+1}$.
This average is clearly an unbiased
estimator of $w(i_j)$  and its variance decreases 
with the number
of draws.  Each repetition can be implemented in $O(k)$ time (drawing
and summing $k$ random variables.).
\ignore{
 Observe, however, that even if we could compute the expectation, the
resulting estimator is worse than the prefix conditioning estimator.
\footnote{This is a consequence of convexity of the variance and
the following: The adjusted weight of an item $i$
is the expectation of the rank conditioning estimator
over rank assignments where the position of the item in the
order is fixed, and 
the permutation of the other items is fixed.  With prefix conditioning,
the adjusted weight is the expectation of the \rc\
estimator over all rank assignments such that 
the permutation of other items is fixed, but our item can be anyone of
the first $k$ items.}
}

We define a Markov chain over permutations of the $k$ items
$\{i_1,\ldots,i_k\}$. Starting with a permutation
$\pi$ we continue to a permutation $\pi'$ by applying the following
process.
We  draw $r_{k+1}$ as described above from the distribution
of $r_{k+1}$ in the subspace corresponding to $\pi$.  We then 
redraw rank values for
the items $\{i_1,\ldots,i_k\}$ as
described in 
Section~\ref{sketches:sec} following Definition \ref{def:botk}.
 The permutation $\pi'$ is
obtained by reordering $\{i_1,\ldots,i_k\}$ 
 according to the new rank values. This Markov
chain has the following property.

\begin{lemma} \label{lem:stationary}
Let $P$ be a (unordered) set of  $k$ items.
Let $p_\pi$ be the conditional probability that
 in a random rank assignment whose prefix consists of items of 
$P$, the order of these items in the prefix is as in
$\pi$. Then $p_\pi$ is 
the stationary distribution
of the Markov chain described above.
\end{lemma}
\notinproc{
\begin{proof}
 Suppose we draw a permutation 
$\pi$ of the items in  $P$ with probability $p_\pi$ and
then draw $r_{k+1}$ as described above. 
Then this is equivalent to drawing a random rank assignment
whose prefix consists of items in $P$ and
taking $r_{k+1}$ of this assignment.

Similarly assume we draw $r_{k+1}$ as
we just described,  draw ranks
for items in ${P}$, and order ${P}$ by these ranks.
Then this is equivalent to drawing a permutation $\pi$ with
probability $p_\pi$.
\end{proof}
} 

  Our implementation
is controlled by two parameters $\inperm$ and  $\permnum$.
$\inperm$ is the number of times the rank value $r_{k+1}$ is
redrawn for a permutation $\pi$ (at each step of the Markov
chain).  $\permnum$ is the number of steps of the Markov chain
(number of permutations in the sequence).

We start with the permutation $(i_1,\ldots,i_k)$ obtained in 
the \ws{} sketch.  We apply this Markov chain to obtain a sequence
of $\permnum$ permutations 
 of 
$\{i_1,\ldots,i_k\}$.  For each permutation $\pi_j$, $1\le j\le \permnum$,
we draw $r_{k+1}$ from $P_{\pi_j}$ $\inperm$ times as described above.
  For each such draw we compute the \rc\
adjusted weights for all items.  The final adjusted weight is
the average of the \rc{} adjusted weights assigned to the item in
the $\permnum *\inperm$ applications of the \rc\ method.
\onlyinproc{The total running time is 
$O(\permnum \cdot k\log k  + \inperm \cdot k)$.}

\notinproc{
We  redraw a permutation 
in this Markov chain in $O(k\log k)$ time ($O(k)$ time to
redraw $k$ rank values and $O(k\log k)$ to sort).  Redrawing
$r_{k+1}$ given a permutation takes $O(k)$ time.  Therefore,
the total running time is $O(\permnum \cdot k\log k  + \inperm \cdot k)$.
}

 The expectation of the \rc\ adjusted weights
over the stationary distribution
is the subset conditioning adjusted weight.
  An important property of this process is that if we apply it
for a {\em fixed}\/ number of steps, and average over a fixed number
of draws of $r_{k+1}$ within each step, we still obtain unbiased estimators. 
Our experimental section shows that these estimators perform very well.

\ignore{
 Explicit computation of the adjusted weight can be exponential.
 We define the rational expression in $y$
{\scriptsize
$$g(w_1,w_2,\ldots,w_k,y)=y^{-1}\int_{x=0}^\infty y\exp(-y x)\prod_{j=1}^k(1-\exp(-w_jx))dx\ .$$
}
We have that $g(y)=1/y$,
{\scriptsize
$$g(w_1,w_2,\ldots,w_k,y)=g(w_1,w_2,\ldots,w_{k-1},y)-g(w_1,w_2,\ldots,w_{k-1},y+w_k)\ .$$
}
 We obtain that
$$g(w_1,\ldots,w_k,y)=\sum_{s| s\subset \{w_1,\ldots,w_k\}}(-1)^{|s|}\frac{1}{y+\sum_{i\in s} w_i}\ .$$

}

  The subset conditioning estimator has powerful properties.  Unfortunately,
it seems specific to \ws\ sketches.   Use of subset conditioning 
requires that given a weighted set $(H,w)$ of $k-1$ weighted items, an item
$i$ with weight $w(i)$, and a weight $\ell>0$, we can compute the
probability that the bottom-$k$ sketch of a set $I$ that
includes $H$, $\{i\}$ and has total weight $\ell+w(H)+w(i)$ contains the
items $H\cup\{i\}$.  
This probability is determined from the distribution of 
the smallest rank of items with
total weight $\ell$.  
In general, however, this probability depends on
the weight distribution of the items in $I\setminus\{H\cup\{i\}\}$.
The exponential distribution has the property that the distribution
of the smallest rank depends only on $\ell$ and not on the weight
distribution.

\section{Confidence bounds} \label{conf:sec}

Let $r$ be a rank assignment of a weighted set $Z=(H,w)$.  Recall that
for $H'\subseteq H$, $r(H')$ is the minimum rank of an item in $H'$.
In this section it will be useful to denote by
$\overline{r}(H')$ the maximum rank of an item in $H'$.
We define 
$r(\emptyset)=+\infty$ and $\overline{r}(\emptyset)=0$.
For a distribution $D$ over a totally ordered set (by $\prec$) and $0<\alpha<1$, we
denote by $Q_{\alpha}(D)$ the $\alpha$-quantile of $D$.  That is,
$\pr_{y\in D}\{y\prec Q_{\alpha}(D)\} \leq  \alpha$ and
$\pr_{y\in D}\{y\succeq Q_{\alpha}(D)\} \geq 1-\alpha$.


\subsection{Total weight}
For two weighted sets $Z_1=(H_1,w_1)$ and 
$Z_2=(H_2,w_2)$, let
$\Omega(Z_1,Z_2)$ be the probability
subspace that contains all rank assignments $r$ over 
$Z_1\cup Z_2$ such that
$\orr(H_1) < r(H_2)$.

Let $(I,w)$ be a weighted set, let $r$ be a rank assignment for
$(I,w)$, $s$ be the bottom-$k$ sketch  that corresponds to $r$
(we also use $s$ as the set of $k$ items with smallest ranks).
Let
$\oWW((s,w),r_{k+1},\delta)$ be the set containing all 
weighted sets $Z' = (H,w')$ such that
$\pr\{r'(H)\geq r_{k+1} \mid r'\in \Omega((s,w),Z')\} \geq \delta$.
Define  $\oww((s,w),r_{k+1},\delta)$ as follows.
If $\oWW((s,w),r_{k+1},\delta)=\emptyset$, then $\oww((s,w),r_{k+1},\delta)=0$.
Otherwise,
let $\oww((s,w),r_{k+1},\delta)=\sup\{w'(H)\mid (H,w')\in \oWW((s,w),r_{k+1},\delta)\}\ .$
(This supremum is well defined for ``reasonable'' families of rank
functions, otherwise, we allow it to be $+\infty$)

Let 
$\uWW((s,w),r_{k+1},\delta)$ be the set of all 
weighted sets $Z' = (H,w')$ such that
$\pr\{r'(H)\leq r_{k+1} \mid r'\in \Omega((s,w),Z')\} \geq \delta$.
Let $\uww((s,w),r_{k+1},\delta)$  be as follows:
We have
$\uWW((s,w),r_{k+1},\delta)\not=\emptyset$ for ``reasonable'' families of
rank functions, but if it is empty, we define $\uww((s,w),r_{k+1},\delta)=+\infty$.  Otherwise,
let $\uww((s,w),r_{k+1},\delta)=\inf\{w'(H)|(H,w')\in \uWW((s,w),r_{k+1},\delta)\}\ .$
(This infimum is well defined since
weighted sets have nonnegative weights.)

\begin{lemma} \label{conflemmatot}
Let $r$ be a rank assignment for the weighted set
$(I,w)$, and let $s$ be the bottom-$k$ sketch  that corresponds to $r$
Then $w(s) +  \oww((s,w),r_{k+1},\delta)$ is a 
$(1-\delta)$-confidence upper bound on $w(I)$,
and
 $w(s) +  \uww((s,w),r_{k+1},\delta)$ is a 
$(1-\delta)$-confidence lower bound on $w(I)$.
\end{lemma}
\begin{proof}
We prove (1). The proof of (2) is analogous.

  We show that in each subspace $\Omega((s,w),(I\setminus s,w))$ of
  rank assignments our bound is correct with probability $1-\delta$.
  Since these subspaces,  specified by $s\subset
  I$ of size $|s|=k$, form a partition of the rank assignments over
  $(I,w)$, the lemma follows.

Let $D_{k+1}$ be the distribution of the $(k+1)$st smallest rank 
over rank assignments in
$\Omega((s,w),(I\setminus s,w))$ (the smallest rank in $I\setminus s$).
Assume that $r$ is a rank assignment in $\Omega((s,w),(I\setminus s,w))$. We
show that if
 $r_{k+1}\leq Q_{1-\delta}(D_{k+1})$ then our upper bound is correct.  Since
by the definition of a quantile $r_{k+1}\leq Q_{1-\delta}(D_{k+1})$ with probability $\ge
(1-\delta)$ in $\Omega((s,w),(I\setminus s,w))$, it follows that our bound is correct with probability
$\ge (1-\delta)$ in $\Omega((s,w),(I\setminus s,w))$.

If $r_{k+1}\leq Q_{1-\delta}(D_{k+1})$ then
{\small
\begin{eqnarray*}
\pr\{r'(I\setminus s)\geq r_{k+1} \mid r'\in \Omega((s,w),(I\setminus s,w))\} 
& \ge & \\
\pr\{r'(I\setminus s)\geq  Q_{1-\delta}(D_{k+1}) \mid r'\in \Omega((s,w),(I\setminus s,w))\}
& \ge & \delta \ .
\end{eqnarray*}
}
So we obtain that $(I\setminus s,w) \in  \oWW((s,w),r_{k+1},\delta)$ 
and therefore $w(I\setminus s) \le \oww((s,w),r_{k+1},\delta)$.
\end{proof} 

This lemma also holds for a variant, where we consider
rank assignments $r$ (and corresponding subspaces) where the items in
$s$ appear in the same order as in $r'$.

\subsection{Subpopulation weight}
We derive confidence bounds for 
the weight of a subpopulation $J\subset I$.  
The arguments are more delicate, as
the number of items from $J$ that we see in the sketch can
vary between $0$ and $k$ and we do not know if the $(k+1)$th
smallest rank belongs to an item in $J$ or in $I\setminus J$.  
We will work with weighted lists instead of weighted sets.

A {\em weighted list} $(H,w,\pi)$ consists of a weighted set $(H,w)$
and a linear order (permutation) $\pi$ on the elements of $H$. 
We will find it convenient to
sometimes specify the permutation $\pi$ as the order induced
by a rank assignment $r$ on $H$.

The {\em concatenation} $(H^{(1)},w^{(1)},\pi^{(1)})\oplus (H^{(2)},w^{(2)},\pi^{(2)})$
of two weighted lists, is a weighted list with items
$H^{(1)}\cup H^{(2)}$, corresponding weights as defined by $w^{(i)}:H^{(i)}$
and order such that each $H^{(i)}$ is ordered according to $\pi^{(i)}$
and the elements of $H^{(1)}$ precede those of $H^{(2)}$.
Let $\Omega((H,w,\pi))$ be 
the probability subspace of rank assignments over $(H,w)$
such that the rank order is according to $\pi$.

Let $r$ be a rank assignment, $s$ be the corresponding sketch, and
$\ell$ be the weighted list $\ell=(J\cap s,w,r)$.
Let $\oWW(\ell,r_{k+1},\delta)$ be the set of all weighted lists
$h=(H,w',\pi)$
such that
\begin{equation*}
\pr\{r'(H)\geq r_{k+1}|r'\in \Omega(\ell\oplus h)\}\geq \delta\ .
\end{equation*}
Let $\oww(\ell,r_{k+1},\delta)=\sup\{w'(H)|(H,w',\pi)\in \oWW(\ell,r_{k+1},\delta)\}$.
(If $\oWW(\ell,r_{k+1},\delta)=\emptyset$, then $\oww(\ell,r_{k+1},\delta)=0$.  If unbounded, then $\oww(\ell,r_{k+1},\delta)=+\infty$.)
Let $\uWW(\ell,r_{k},\delta)$ be the set of all weighted lists
$h=(H,w',\pi)$ such that
\begin{equation*}\label{lb2}
\pr\{\overline{r'}(J\cap s)\leq r_{k} |r'\in \Omega(\ell\oplus h)\}\geq \delta\ .
\end{equation*}
Let $\uww(\ell,r_{k},\delta)=\inf\{w'(H)|(H,w',\pi)\in \uWW(\ell,r_{k},\delta)$.
(If $\uWW(\ell,r_{k},\delta)=\emptyset$, then $\uww(\ell,r_{k},\delta)=+\infty$).
We prove the following.
 
\begin{lemma} \label{conflemma}
Let $r$ be a rank assignment, $s$ be the corresponding sketch, and
$\ell$ be the weighted list $\ell=(J\cap s,w,r)$.
Then
$w(J\cap s)+\oww(\ell,r_{k+1},\delta)$ is a $(1-\delta)$-confidence
upper bound on $w(J)$ and
$w(J\cap s)+ \uww(\ell,r_k,\delta)$
is a $(1-\delta)$-confidence
lower bound on $w(J)$.
\end{lemma}
\notinproc{
\begin{proof}
 The bounds are conditioned on the subspace of rank assignments over $(I,w)$
where the ranks of items in $I\setminus J$ are fixed and the 
order of the ranks of the items in $J$ is fixed.  These subspaces are a
partition of the sample space of rank assignments over $(I,w)$.  We show that
the confidence bounds hold within each subspace. 

Consider such a subspace $\Phi\equiv \Phi(J,\pi:J,a:(I\setminus J))$, where $\pi:J$  is a permutation over $J$, representing the order of the ranks
of the items in $J$.
 and
$a:(I\setminus J)$ are the rank values of the elements in $I\setminus J$.



Let $D_{k+1}$ be the distribution of
$r_{k+1}$ for $r\in\Phi$ and let $D_k$ be the distribution of
$r_k$ for $r\in \Phi$. Over rank assignments in $\Phi$ we have
$\pr\{r_{k+1}\leq Q_{1-\delta}(D_{k+1})\}\geq 1-\delta$
and
$\pr\{r_{k}\geq Q_{\delta}(D_k)\}\geq 1-\delta$.\footnote{
Note that
these distributions have some discrete values with positive probabilities,
therefore, it does not necessarily holds that
  $\pr\{r_{k}\leq Q_{\delta}(D_k)\}\leq\delta$ and
  $\pr\{r_{k+1} \geq  Q_{1-\delta}(D_{k+1})\}\leq \delta$.}
 We  show that 
\begin{itemize}
\item
The upper bound is correct for rank assignments $r\in\Phi$
such that $r_{k+1}\leq Q_{1-\delta}(D_{k+1})$.
 Therefore, it is correct with probability at least
$(1-\delta)$.
\item
The lower bound is correct for rank assignments $r\in\Phi$ 
such that
$r_{k}\geq Q_{\delta}(D_k)$.  Therefore, it
 is correct with probability at least $(1-\delta)$.
\end{itemize}

  Consider a rank assignment $r\in\Phi$. Let $s$ be the
items in the sketch. Let
$\ell=(J\cap s,w,r)$ and
$\ell^{(c)}=(J\setminus s,w,r)$
be the weighted lists of 
the items in $J\cap s$ or $J\setminus s$, respectively,
as ordered by $r$.
There is bijection between rank assignments in 
$\Omega(\ell\oplus \ell^{(c)})$
and rank assignments in $\Phi$ by augmenting the rank assignment in $\Omega(\ell\oplus \ell^{(c)})$ with the
ranks $a(j)$ for items  $j\in I\setminus J$.  
For a rank assignment $r\in \Phi$ let $\hat{r}\in \Omega(\ell\oplus \ell^{(c)})$  be its restriction to $J$.

  A rank assignment $r'\in \Phi$ has $r'_{k+1}\geq r_{k+1}$ if and only if
$\widehat{r'}(J\setminus s)\geq r_{k+1}$.\footnote{Note that the statement with strict inequalities does not necessarily hold.}
So if $r\in \Phi$ such that $r_{k+1}\leq Q_{1-\delta}(D_{k+1})$ then
\begin{eqnarray*}
\pr_{r'\in \Omega(\ell\oplus \ell^{(c)})} \{r'(J\setminus s)\geq r_{k+1}\} & = &
\pr_{r'\in \Phi} \{r'_{k+1}\geq r_{k+1}\} \\
\ge \pr_{r'\in \Phi} \{r'_{k+1} \geq  Q_{1-\delta}(D_{k+1})\}
&\geq &\delta \ .
\end{eqnarray*}
Therefore, $\ell^{(c)}\in \oWW(\ell,r_{k+1},\delta)$, and hence
$w(J\setminus s)\leq \oww(\ell,r_{k+1},\delta)$ and the upper 
bounds holds.

  A rank assignment $r'\in \Phi$ has $r'_{k}\leq r_{k}$ if and only if
the maximum rank that  $\widehat{r'}$ gives to an item in $J\cap s$
is $\leq r_k$. So if $r\in \Phi$ such that  $r_{k}\geq Q_{\delta}(D_k)$ 
\begin{eqnarray*}
\pr_{r'\in \Omega(\ell\oplus \ell^{(c)})} \{ \overline{r'}(J\cap s)  \leq r_{k}\} & = &
\pr_{r'\in \Phi} \{r'_{k} \leq r_{k}\} \\ 
\geq
\pr_{r'\in \Phi} \{r'_{k}\leq Q_\delta(D_k)\} & \geq &
\delta
\end{eqnarray*}
Therefore, $\ell^{(c)}\in \uWW(\ell,r_{k},\delta)$, and hence
$w(J\setminus s)\geq \uww(\ell,r_{k},\delta)$ and the lower bound holds.
\end{proof}
}

\subsection{Subpopulation weight using w(I)} \label{ciwswt:sec}

  We derive tighter confidence intervals that use the total weight $w(I)$.
For weighted lists $h^{(i)}=(H^{(i)},w^{(i)},\pi^{(i)})$ ($i=1,2$), 
the probability
space $\Omega(h^{(1)},h^{(2)})$ contains all
rank assignments $r$ over the weighted set
$(H^{(1)},w^{(1)})\cup (H^{(2)},w^{(2)})$
such that
for each $i=1,2$,
the  order of $H^{(i)}$ induced by the
rank values $r:H^{(i)}$ is $\pi^{(i)}$.
We define the functions $c_{h^{(1)},h^{(2)}}(r)$ and
$d_{h^{(1)},h^{(2)}}(r)$ for $r\in \Omega(h^{(1)},h^{(2)})$
as follows:
$c_{h^{(1)},h^{(2)}}(r)$ is the number of items amongst those
with $k$ smallest ranks that are in $H^{(1)}$ (equivalently, it is
$i$ such that $r_i(H^{(1)})<r_{k-i+1}(H^{(2)})$ and $r_{k-i}(H^{(2)})<r_{i+1}(H^{(1)})$);
$$d_{h^{(1)},h^{(2)}}(r)=r_{k-c_{h^{(1)},h^{(2)}}(r)}(H^{(2)})-r_{c_{h^{(1)},h^{(2)}}(r)}(H^{(1)})$$
is the difference between the largest rank values of items in $H^{(2)}$
and $H^{(1)}$ that are amongst the $k$ least ranked items.

We use the notation $(c_1,d_1)\preceq (c_2,d_2)$ for the 
lexicographic order over pairs.

Let $r$ be a rank assignment,
and let $s$ be the sketch corresponding to $r$. Let
$\Delta=\overline{r}((I\setminus J)\cap s)-\overline{r}(J\cap s)$, and let
 $\ell_1=(J\cap s,w,r:J\cap s)$ and
$\ell_2=( (I\setminus J) \cap s,w,r:(I\setminus J) \cap s)$.  

Let $\oWW(\ell_1,\ell_2,\Delta,\delta)$ be the set of all pairs $(h_1,h_2)$
of weighted lists $h_1=(H_1,w_1,\pi_1)$ and $h_2=(H_2,w_2,\pi_2)$
such that $w_1(H_1)+w_2(H_2)=w(I)-w(s)$ and

\vspace{-0.15in}
{\small
\notinproc{\begin{equation} \label{upperlex} }
\onlyinproc{ $$ }
\pr\{(c_{\ell_1\oplus h_1,\ell_2\oplus h_2}(r'),d_{\ell_1\oplus h_1,\ell_2\oplus h_2}(r'))\succeq (|J\cap s|,\Delta)\}\geq  \delta\ ,
\notinproc{\end{equation}}
\onlyinproc{ $$ }
}
over  the
probability  space of all  $r'\in \Omega(\ell_1\oplus h_1, \ell_2\oplus h_2)$.

If $\oWW(\ell_1,\ell_2,\Delta,\delta)=\emptyset$, then
$\oww(\ell_1,\ell_2,\Delta,\delta)=0$.  Otherwise,
$\oww(\ell_1,\ell_2,\Delta,\delta)=\sup\{w_1(H_1) \mid (h_1,h_2)\in\oWW(\ell_1,\ell_2,\Delta,\delta)\}$.

Let $\uWW(\ell_1,\ell_2,\Delta,\delta)$ be the set of all pairs $(h_1,h_2)$
of weighted lists $h_1=(H_1,w_1,\pi_1)$ and $h_2=(H_2,w_2,\pi_2)$
such that $w(H_1)+w(H_2)=w(I)-w(s)$ and

\vspace{-0.15in}
{\small
\notinproc{ \begin{equation} \label{lowerlex} }
\onlyinproc{ $$ }
\pr\{(c_{\ell_1\oplus h_1,\ell_2\oplus h_2}(r'),d_{\ell_1\oplus h_1,\ell_2\oplus h_2}(r'))\preceq (|J\cap s|,\Delta)\}\geq \delta\ ,
\notinproc{ \end{equation} }
\onlyinproc{ $$ }
}
over  the probability space of all $r'\in\Omega(\ell_1\oplus h_1, \ell_2\oplus h_2)$.

If $\uWW(\ell_1,\ell_2,\Delta,\delta)=\emptyset$, then 
$\uww(\ell_1,\ell_2,\Delta,\delta)=w(I)-w(s)$.  Otherwise,\\
$\uww(\ell_1,\ell_2,\Delta,\delta)= \inf\{w_1(H_1)\mid (h_1,h_2)\in\uWW(\ell_1,\ell_2,\Delta,\delta)\}$.

\begin{lemma} \label{lexconflemmaww}
Let $r$ be a rank assignment, $s$ be the corresponding sketch, let
$\Delta=\overline{r}((I\setminus J)\cap s)-\overline{r}(J\cap s)$, and let
 $\ell_1=(J\cap s,w,r:J\cap s)$ and
$\ell_2=((I\setminus J)\cap s,w,r:(I\setminus J)\cap s )$.  
Then $w(J\cap s)+\oww(\ell_1,\ell_2,\Delta,\delta)$ is a $(1-\delta)$-confidence upper bound on $w(J)$, and
$w(J\cap s)+\uww(\ell_1,\ell_2,\Delta,\delta)$ is a $(1-\delta)$-confidence lower bound on $w(J)$.
\end{lemma}
\notinproc{
\begin{proof}
 The lower bound on $w(J)$ 
is equal to $w(I)$ minus a $(1-\delta)$-confidence
 upper bound,
$w((I\setminus J)\cap s)+ \oww(\ell_2,\ell_1,-\Delta,\delta)$
 on $w(I\setminus J)$.
Therefore it suffices to prove the upper bound.

  We show that the bound holds with probability at least $(1-\delta)$
in  the subspace of rank assignments over $(I,w)$ 
where the rank order of the items in $J$ and the rank order of the items
in $I\setminus J$ are fixed.  
These subspaces are a partition of the 
space of rank assignments.
Consider such a subspace $\Phi=\Omega(\ell'_1,\ell'_2)$.  
Let $\ell'_1=(J,w,\pi_1)$ and
$\ell'_2=(I\setminus J,w,\pi_2)$ be the weighted lists that corresponds
to the rank order of the items in $J$ and in $I\setminus J$, respectively,
for $r\in\Phi$.

 Let $D$ be the distribution over the pairs 
$(c_{\ell'_1,\ell'_2}(r),d_{\ell'_1,\ell'_2}(r))$ for $r\in\Phi$.
We define the quantile
$Q_{1-\delta}(D)$  with respect to the
lexicographic order over the pairs.

  We show that the upper bound is correct for all $r\in\Phi$ such that
$(c_{\ell'_1,\ell'_2}(r),d_{\ell'_1,\ell'_2}(r))\preceq Q_{1-\delta}(D)$.  
Therefore, it holds with
probability at least $1-\delta$.

Let $r\in\Phi$ such that
$(c_{\ell'_1,\ell'_2}(r),d_{\ell'_1,\ell'_2}(r))\preceq Q_{1-\delta}(D)$.
 Let $s$ be the corresponding sketch,
$\ell_1=(J\cap s,w,r)$, $\ell_2=((I\setminus J)\cap s,w,r)$, 
$\ell_1^{(c)}=(J\setminus s,w,r)$,
$\ell_2^{(c)}=((I\setminus J)\setminus s,w,r)$.  By definition,
$c_{\ell'_1,\ell'_2}(r)=|J\cap s|$,
$\Delta=d_{\ell'_1,\ell'_2}(r)=  
\overline{r}((I\setminus J)\cap s)
-\overline{r}(J\cap s)$,
 $\ell'_1=\ell_1\oplus \ell_1^{(c)}$, and 
$\ell'_2=\ell_2\oplus \ell_2^{(c)}$. It follows that
\begin{eqnarray*}
\pr\{(c_{\ell'_1,\ell'_2}(r),d_{\ell'_1,\ell'_2}(r)) & \succeq & (|J\cap s|,\Delta)\mid r\in\Phi\}\geq \\
\pr\{(c_{\ell'_1,\ell'_2}(r),d_{\ell'_1,\ell'_2}(r))& \succeq & Q_{1-\delta}(D)\mid r\in\Phi\}\geq \delta\ .
\end{eqnarray*}

Therefore, 
$(\ell_1^{(c)},\ell_2^{(c)})\in \oWW(\ell_1,\ell_2,\Delta,\delta)$,
and hence, $$w(J\setminus s) \leq \oww(\ell_1,\ell_2,\Delta,\delta)\ .$$
\end{proof}
} 

\notinproc{
}

\notinproc{
We formulate the conditions in the statement of 
Lemma~\ref{lexconflemmaww} in
terms of predicates on the rank assignment.
Inequality~(\ref{upperlex}) is equivalent to
$\pr\{U_{h_1,h_2}(r)\mid r\in \Omega(\ell_1\oplus h_1,\ell_2\oplus h_2)\}
 \geq \delta\ ,$  where $U_{h_1,h_2}(r)$
is the predicate 
(that depends on $\ell_1,\ell_2,\Delta$):

\vspace{-0.15in}
{\small
\begin{equation}\label{prcond:upper}
\begin{array}{l}
U_{h_1,h_2}(r)=(r(H_2)>\overline{r}(J\cap s))\wedge\\
 \left( \begin{array}{l}
\left(r(H_1)<\overline{r}(s\cap (I\setminus J)) \right) \vee \\
\left(r(H_1)>\overline{r}(s\cap (I\setminus J)) \wedge (\overline{r}((I\setminus J)\cap s)-\overline{r}(J\cap s)>\Delta)\right) \end{array} \right)\ .
\end{array}
\end{equation}
}
The first line guarantees that we have at least 
$|J \cap s|$ items of $J$ among the $k$ items of
smallest ranks. If the second line holds then we have strictly more
than $|J \cap s|$ items of $J$ among the $k$ items of
smallest ranks. If the third line holds then we have 
exactly $|J \cap s|$ items of $J$ among the $k$ items of
smallest ranks and $(\overline{r}((I\setminus J)\cap s)-\overline{r}(J\cap s)>\Delta)$

Similarly, the condition in Inequality~(\ref{lowerlex}) is equivalent to
$\pr\{L_{h_1,h_2}(r)\mid r\in \Omega(\ell_1\oplus h_1,\ell_2\oplus h_2)\}
 \geq \delta\ ,$  where $L_{h_1,h_2}(r)$
is the predicate :

\vspace{-0.15in}
{\small
\begin{equation}\label{prcond:lower}
\begin{array}{l}
L_{h_1,h_2}(r)=(r(H_1)>\overline{r}(s\cap (I\setminus J)))\wedge \\
\left( \begin{array}{l} \left(r(H_2)<\overline{r}(J\cap s) \right) \vee \\
 \left( r(H_2)>\overline{r}(J\cap s) \wedge (\overline{r}((I\setminus J)\cap s)-\overline{r}(J\cap s)<\Delta ) \right)\end{array} \right)\ .
\end{array}
\end{equation}
}
(Either the $k$ items with smallest ranks include strictly less
than $|J\cap s|$ items from $J$ or they  include exactly 
$|J\cap s|$ items from $J$ and 
$\overline{r}((I\setminus J)\cap s)-\overline{r}(J\cap s)< \Delta$.)

}

 \subsection{Confidence bounds for wsr sketches} \label{confwsr}
\onlyinproc{
The \wsr\ estimator is the average 
the $k$ minimum ranks which are 
independent exponential random variables with (the same)
parameter $w(I)$. We used the normal approximation to
this distribution in order to compute  \wsr\ confidence bounds.
}

\notinproc{
 In our simulations, we apply the normal approximation to obtain
confidence bounds on total weight using \wsr\ sketches:
The average of the $k$ minimum ranks $\overline{r}=\sum_{i=1}^k r_i/k$ is an
average of $k$ independent exponential random variables with (the same)
parameter $w(I)$ (This is a Gamma distribution). 
 The expectation of the sum is $k/w(I)$ and the variance
is $k/w(I)^2$.  The confidence bounds are
the $\delta$ and $1-\delta$ quantiles of $\overline{r}$.
Let $\alpha$ be the Z-value that corresponds to 
confidence level $1-\delta$ in the standard normal distribution.
By applying the normal approximation, the approximate upper bound
is the solution of $k/w(I) + \alpha \sqrt{k/w(I)^2}=k\overline{r}$, and
the approximate
lower bound is the solution of $k/w(I) - \alpha
\sqrt{k/w(I)^2}=k\overline{r}$.  Therefore, the approximate bounds
are $(1\pm \alpha/\sqrt{k})/\overline{r}$.
}

 \subsection{Confidence bounds for ws sketches} \label{confws}

  The confidence bounds make ``worst case'' assumptions on the
weight distribution of ``unseen'' items.  
 \ws\ sketches have the nice property that
the distribution of the 
$i$th largest rank in a weighted set, conditioned on either
the set or the list of the $i-1$ items of smallest rank values, depends
only on the total weight of the set (and not on the 
particular
partition of the ``unseen'' weight into items). Therefore,
the confidence  bounds are {\em tight} in the respective probability
subspaces:  for 
any distribution and any subset, the probability that the bound is
violated is {\em exactly} $\delta$.

\vspace*{-0.1in}
\paragraph*{Bounds on the total weight (w(I))}
We apply Lemma~\ref{conflemmatot}.
 For a weighted set $(s,w)$, $|s|=k$,  and $\ell \geq 0$, consider 
a weighted set $U$ of weight $w(s)+\ell$ containing $(s,w)$.
Let $y$ be the
$(k+1)$th smallest
rank value, over rank assignments over $U$ such that
the $k$ items with smallest rank values are the elements of $s$.
The probability density function of $y$ is
(see Section~\ref{subscond} and Eq. (\ref{intsubset}))
\begin{equation} \label{subconfws}
D(\ell,y)=\frac{\exp(-\ell y)\prod_{j\in s}(1-\exp(-w(i_j)y))}{\int_{x=0}^\infty \exp(-\ell x)\prod_{j\in s}(1-\exp(-w(i_j)x))dx}
\end{equation}

Let $r_{k+1}$ be the observed $k+1$ smallest rank.
 The $(1-\delta)$-confidence upper bound 
is $w(s)$ plus the value of $\ell$ 
that solves the equation $\int_0^{r_{k+1}} D(\ell,y) dy=1-\delta$.
The function $\int_0^{r_{k+1}} D(\ell,y)$ is an increasing 
function of $\ell$ (the probability of the $(k+1)$st smallest
rank being at most $r_{k+1}$ is increasing with $\ell$.)
If $\int_0^{r_{k+1}} D(0,y) dy> 1-\delta$, then there is no
solution and the upper bound is $w(s)$.

 The lower bound is $w(s)$ plus the value of $\ell$ 
that solves the equation $\int_0^{r_{k+1}} D(\ell,y) dy=\delta$.
If there is no solution ($\int_0^{r_{k+1}} D(0,y) dy > \delta$), then
the lower bound is $w(s)$.

\vspace*{-0.1in}
\paragraph*{ Conditioning on the order of items}
We consider bounds that use the stronger conditioning, where we fix
the rank order of the items.
 For $0\leq s_0\leq\cdots\leq s_h< t$, 
we use the notation $v(t,s_0,\ldots,s_h)$  for the
random variable that is the sum of $h+1$ independent
exponential random variables with parameters 
$t-s_j$ ($j=0,\ldots,h$).
\notinproc{
From linearity of expectation, 
$$E(v(t,s_0,\ldots,s_h))=\sum_{j=0}^{h} 1/(t-s_j)\ .$$
From independence, the variance is the sum of variances
of the exponential random variables and is
$$\var(v(t,s_0,\ldots,s_h))=\sum_{j=0}^{h} 1/(t-s_j)^2\ .$$
}

 Consider a weighted set $(I,w)$ and a subspace of rank assignments where
the set and the order of the $h$ items of smallest rank 
is fixed to be $i_1,i_2,\ldots,i_{h}$.
Let $s_{j}=\sum_{\ell=1}^{j} w(i_\ell)$.  For convenience
we define $s_0\equiv 0$ and  $r_0=0$.
By Lemma \ref{nextdistk:cor}, for $j=0,\ldots,h$,
the rank difference $r(i_{j+1}) - r(i_j)$ is an exponential r.v.\
with parameter $w(I)- s_{j}$.
These rank differences are
independent, and for $i\in \{0,\ldots,h\}$, 
the distribution of the $i$th smallest rank,
$r_{i}$ (also the sum of the first
$i$ rank differences) is
$v(w(I),s_0,\ldots,s_{i-1})$
 in the subspace that
we conditioned on.
\onlyinproc{We show how to specialize Lemma \ref{conflemma} 
to \ws\ sketches to obtain confidence bounds for
subpopulation weight. We similarly specialize 
Lemma \ref{conflemmatot} and obtain confidence bounds for
the total weight.}

\notinproc{
We obtain confidence bounds for the total weight and for 
subpopulation weight when the total weight is not provided,
by solving
 an equation of the form:
\begin{equation}\label{sumexpoeq}
\pr\{v(x,s_0,\ldots,s_h)\leq \tau\}=\delta
\end{equation}
 for $x> s_h$ (where
$0\leq s_0<\cdots< s_h$, $\tau>0$, and $0<\delta<1$ are provided.)

Since for $x>y> s_h$ and any $\tau$,
$\pr\{v(x,s_0,\ldots,s_h)\leq \tau\}\geq \pr\{v(x,s_0,\ldots,s_h)\leq \tau\}$,
it is easy 
to approximately 
solve equations like this numerically.
Observe that 
the probability $\pr\{v(x,s_0,\ldots,s_h\leq \tau)$ is
minimized as $x$ approaches $s_h$ from above.  If the limit is at least
$\delta$, then the equation has no solution.

\vspace*{-0.1in}
\paragraph*{ The weight w(I)}
Let $i_1,i_2,\ldots,i_{k}$ be the items in the current sketch, ordered
by increasing rank values, and
let $s_{j}=\sum_{\ell=1}^{j} w(i_\ell)$.
The distribution of
$(k+1)$ smallest rank
 (for any fixed possible order
of the remaining items) is 
the random variable
$v(w(I),s_0,\ldots,s_{k})$.
Using an ordered variant of Lemma~\ref{conflemmatot} we obtain that
the $(1-\delta)$-confidence lower bound 
is the solution of the equation
$$\pr\{v(x,s_0,\ldots,s_k)\leq r_{k+1} \} = \delta\ $$
and is $s_k$ if there is no solution $x> s_k$.
The $(1-\delta)$-confidence upper bound 
is the solution of the equation
$$\pr\{v(x,s_0,\ldots,s_k)\leq  r_{k+1} \}= 1-\delta\ $$
(and is $s_k$ if there is no solution $x> s_k$.)
} 

\vspace*{-0.1in}
\paragraph*{ Subpopulation weight (with unknown w(I))}

Let $J$ be a subpopulation.  For a rank assignment, let $s$ be
the corresponding  sketch and let
 $s_h$ ($1\leq h\leq |J\cap s|$) be the sum of the weights of the 
$h$ items of smallest rank values from $J$ (we define $s_0\equiv 0$).
Specializing Lemma~\ref{conflemma} to \ws\ sketches we obtain
that the $(1-\delta)$-confidence upper bound on $w(J)$ is the solution of
the equation
$$\pr\{v(x,s_0,\ldots,s_{|J\cap b|})\leq r_{k+1}\} = 1-\delta\ $$
(and is $s_{|J\cap b|}$ if there is no solution $x>s_{|J\cap b|}$.)
The $(1-\delta)$-confidence lower bound is $0$ if
$|J\cap b|=0$.  Otherwise,
let $x> s_{|J\cap b|-1}$ be
the solution of
$$\pr\{v(x,s_0,\ldots,s_{|J\cap b|-1})\leq r_k\}= \delta\ .$$
The lower bound is $\max\{s_{|J\cap b|},x\}$ if there is a solution
and is $s_{|J\cap b|}$ otherwise.

To solve these equations, 
we either used the {\em normal approximation} 
to the respective sum of exponentials distribution \onlyinproc{(not described in
this abstract)} or used
the {\em quantile method} which we developed.

\notinproc{
\vspace*{-0.1in}
\paragraph*{ Normal approximation}
We apply the normal approximation to the
quantiles of a sum of exponentials distribution.
For $\delta \ll 0.5$, let $\alpha$ be the Z-value that corresponds to 
confidence level $1-\delta$. 
The approximate $\delta$-quantile of $v(x,s_0,\ldots,s_h)$ is
$E(v(x,s_0,\ldots,s_h))-\alpha\sqrt{\var(v(x,s_0,\ldots,s_h))}$
and the approximate $(1-\delta)$-quantile is
$E(v(x,s_0,\ldots,s_h))+\alpha\sqrt{\var(v(x,s_0,\ldots,s_h))}$.

To approximately solve $\pr\{v(x,s_0,\ldots,s_h)\leq \tau\}=\delta$ 
($x$ such that $\tau$ is the $\delta$-quantile of
$v(x,s_0,\ldots,s_h)$), 
we solve the equation
$$E(v(x,s_0,\ldots,s_h))-\alpha\sqrt{\var(v(x,s_0,\ldots,s_h))}=\tau\ .$$
To approximately solving
$\pr\{v(x,s_0,\ldots,s_h)\leq \tau\}=1-\delta$,
we solve
$$E(v(x,s_0,\ldots,s_h))+\alpha\sqrt{\var(v(x,s_0,\ldots,s_h))}=\tau\ .$$

We solve these equations (to the desired approximation level) by
searching over values of $x> s_h$ using standard numerical methods. The
function $E(v(x))+\alpha\sqrt{\var(v(x))}$ is
 monotonic decreasing in the range $x>s_h$.
The function $E(v(x))-\alpha\sqrt{\var(v(x))}$
 is decreasing or bitonic (first increasing then
decreasing) depending on the value of $\alpha$.
} 

\vspace*{-0.1in}
\paragraph*{ The quantile method}
Let $D^{(x)}$ be a parametric family of probability spaces
such that there is a total order $\prec$  over the union of the domains
of $\{D^{(x)}\}$.
Let $\tau$ be a value in the union of the domains of $\{D^{(x)}\}$ such that
the probability $\pr\{y\preceq \tau \mid y\in D^{(x)}\}$ is increasing with $x$.
So the solution ($x$) to the equation $\pr\{y\preceq \tau \mid y\in D^{(x)}\}=\delta$ ($Q_\delta(D^{(x)})=\tau$) is unique.
(We refer to this property as {\em monotonicity of $\{D^{(x)}\}$ with respect
to $\tau$.})

  We assume the following two ``black box'' ingredients.
The first ingredient is drawing {\em independent  monotone
 parametric samples\/} $s(x)\in D^{(x)}$.
That is, for any $x$, $s(x)$ is a sample from $D^{(x)}$ and
if $x\geq y$ then $s(x)\preceq s(y)$. Two different
parametric samples are independent: That is for every $x$,
$s^1(x)$ and $s^2(x)$ are independent draws from $D^{(x)}$.
  The second ingredient is a solver of equations of the form
$s(x)=\tau$ for a parametric sample $s(x)$.

We define a distribution $\overline{D}^{(\tau)}$ such that
a sample from from $\overline{D}^{(\tau)}$ is obtained by drawing 
a parametric sample $s(x)$ and returning the solution of
$s(x)=\tau$.  The two black box ingredients allow us
to draw samples from $\overline{D}^{(\tau)}$.
Our interest in $\overline{D}^{(\tau)}$ is due to the following property:
\begin{lemma} \label{quantmethod:lemma}
For any $\delta$, the solution of
$Q_\delta(D^{(x)})=\tau$ is the $\delta$-quantile of $\overline{D}^{(\tau)}$.
\end{lemma}

 The quantile method for approximately solving equations of the form
$\pr\{y\preceq \tau \mid y\in D^{(x)}\}=\delta$ draws multiple samples
from $\overline{D}^{(\tau)}$ and returns the $\delta$-quantile of the
set of samples.

We apply the quantile method to approximately solve Equations of the form
\onlyinproc{
\begin{equation}
\pr\{v(x,s_0,\ldots,s_h)\leq \tau\}=\delta
\end{equation}
}

\notinproc{Eq.~(\ref{sumexpoeq})} (as an alternative to 
the normal approximation).  The family of distributions that 
we consider is $D^{(x)}=v(x,s_0,\ldots,s_h)$.
This family  has the monotonicity property with respect to any $\tau>0$.
  A parametric sample $s(x)$ from $v(x,s_0,\ldots,s_h)$ is obtained
by drawing $h+1$
independent random variables $v_0,\ldots,v_h$ from $U[0,1]$.
The parametric sample is $s(x)=\sum_{j=0}^h -\ln v_h/(x-s_j)$
and is a monotone decreasing function of $x$.
A sample from $\overline{D}^{(\tau)}$ is then the solution of the
equation
$\sum_{j=0}^h -\ln v_h/(x-s_j)=\tau\ .$
  Since $s(x)$ is monotone, 
the solution can be found using standard search.

\vspace*{-0.1in}
\paragraph*{ Subpopulation weight using w(I)}
\onlyinproc{
We get confidence intervals in this case by 
specializing the conditions in Lemma~\ref{lexconflemmaww} to \ws\
sketches. The confidence bounds then become quantiles 
of an appropriately defined parameterized family of distributions that
we find using the quantile method.
}

\notinproc{
We specialize the conditions in Lemma~\ref{lexconflemmaww} to \ws\
sketches.  Consider the distribution of $(c_{\ell_1\oplus h_1,\ell_2\oplus
  h_2}(r),d_{\ell_1\oplus h_1,\ell_2\oplus h_2}(r))$ for
$r\in\Omega(\ell_1\oplus h_1,\ell_2\oplus h_2)$.
We shall refer to items of $h_1$ as items of $J$
and to items of $h_2$ as items of $I\setminus J$.
This distribution in general depends on the
decomposition of the weighted lists
$h_1$ and $h_2$ into items. However from
Equation (\ref{prcond:lower}) we learn that  
$$\pr\{(c_{\ell_1\oplus h_1,\ell_2\oplus h_2}(r),d_{\ell_1\oplus h_1,\ell_2\oplus h_2}(r))\preceq (|J\cap s|,\Delta)\}$$ 
 where 
$\Delta=\overline{r}((I\setminus J)\cap s)-\overline{r}(J\cap s)$,
 depends only on 
 $x=w(H_1)$, and $w(H_2)=w(I)-x$.
Indeed, 
let $\tau = (|J\cap s|,\Delta)$,
$\pr\{(c_{\ell_1\oplus h_1,\ell_2\oplus h_2}(r),d_{\ell_1\oplus h_1,\ell_2\oplus h_2}(r))\leq \tau\}$ 
is the probability of the predicate $L_{h_1,h_2}$ stated in
Eq.~(\ref{prcond:lower}).  This predicate depends on the rank values of
the $|J\cap s|$ and $|J\cap s|+1$ smallest ranks in $J$ and of 
the $|(I\setminus J)\cap s|$ and $|(I\setminus J)\cap s|+1$
smallest ranks in $I\setminus J$.  For $\ws$ sketches,
the distribution of these ranks is determined by the weighted
lists $\ell_1,\ell_2$ and $x$.

So we pick a weight list $h_1$ with a single item of weight
$x$, and 
a weighted list $h_2$ with a single item of
weight $w(I)-x$,
and let $D^{(x)}$ be the
 distribution of $$(c_{\ell_1\oplus h_1,\ell_2\oplus
  h_2}(r),d_{\ell_1\oplus h_1,\ell_2\oplus h_2}(r))$$ for
$r\in\Omega(\ell_1\oplus h_1,\ell_2\oplus h_2)$.
To emphasis the dependency of $r$ on $x$ 
we shall denote
by $r^{(x)}$ a rank assignment drawn from 
$\Omega(\ell_1\oplus h_1,\ell_2\oplus h_2)$ where
$w(H_1) = x$.

Since the largest rank in $J\cap s$ and the smallest rank
of an item in $H_1$ decrease with $x$, and the largest rank
in $(I\setminus J) \cap s$ and the smallest rank in $H_2$
increase with $x$ (decrease with $w(I) - x$) it follows that
the  family $D^{(x)}$ has the monotonicity property with respect to
$\tau = (|J\cap s|,\Delta)$.\footnote{The precise statements here is
that the probability that $r(J\cap s)$ is smaller than some threshold $t$,
increases with $x$ etc.}

Obviously, 
$w(J\setminus s)\in [0,w(I)-w(s)]$.  Therefore, 
we can truncate the bounds to be in this range.
So the upper bound on $w(J\setminus s)$ is the minimum of
$w(I)-w(s)$ and 
$x$ such that
$Q_{1-\delta}(D^{(x)})=(|J\cap s|,\Delta)$.
 If there is no
solution then the upper bound is $0$.
The lower bound on $w(J\setminus s)$ is the 
value of $x$ such that
$Q_{\delta}(D^{(x)})=(|J\cap s|,\Delta)$.
If there is no solution, then the lower bound is $0$.
The respective (upper or lower)
bounds on $w(J)$ are $w(J\cap s)$ plus the bound on $w(J\setminus s)$.

  We apply the quantile method to solve the equations
$$Q_{1-\delta}(D^{(x)})=(|J\cap s|,\Delta) \ , $$ and 
$$Q_{\delta}(D^{(x)})=(|J\cap s|,\Delta) \ .$$

The first black box ingredient
that we need for the quantile method is
drawing a monotone parametric sample $s(x)$ from 
$D^{(x)}$.
Let $s_i$ ($i\in (0,1,\ldots,|J\cap s|)$)
 be the sum of the weights of the first $i$ items from $J$
in $\ell_1$. Let $s'_i$ ($i\in (0,1,\ldots,k-|J\cap s|)$) be the
respective sums for $I\setminus J$.
We draw a rank assignment 
$r^{(x)}\in\Omega(\ell_1\oplus h_1,\ell_2\oplus h_2)$ as follows.
We draw
$k+2$ independent random variables
 $v_0,\ldots,v_{|J\cap s|},v'_0,\ldots,v'_{k-|J\cap s|}$
from $U[0,1]$.  We let the $j$th rank difference 
between items from $J$ be
$-\ln(v_j)/(x-s_j)$,  and the $j$th rank difference 
between items from $(I\setminus J)$ be
$-\ln(v'_j)/(x-s'_j)$.  These rank differences determine
$\overline{r}(J\cap s)$ and $r(H_1)$ (sums of $|J\cap s|$ and
$|J\cap s|+1$ first rank differences from $J$, respectively), and
$\overline{r}((I\setminus J)\cap s)$ and  $r(H_2)$
(sums of $|(I\setminus J)\cap s|$ and
$|(I\setminus J)\cap s|+1$ first rank differences from $I\setminus J$, respectively). Then $s(x)$ is the pair $(c(r^{(x)}),d(r^{(x)}))$.

  The second black box ingredient is solving the equation $s(x)=\tau$.
 Let $i=|J\cap s|$ and let $i'=k-i=|(I\setminus J)\cap s|$
as before.
The solver has three phases:
 We first compute the range $(L,U)$ of values of $x$ such that
the first coordinate of the pair $s(x)$ is equal to $|J\cap s|$.
That is, the rank assignment $r$ has exactly $|J\cap s|$ items from $J$ among the first $k$ items.   Let $d(r^{(x)})=r_{i'}(I\setminus J)-r_{i}(J)$ denote
the second coordinate in the pair $s(x)$.
In the second phase 
we look for a value $x\in(L,U)$ (if there
is one) such that $d(r^{(x)})=\Delta$
(the second coordinate of $s(x)$ is equal to $\Delta$).
The function $d(r^{(x)})$ is monotone increasing in this range, which simplifies
numeric solution.
The third phase is truncating the solution to be in $[0,w(I)-w(s)]$.
Details are provided in Figure~\ref{solver:fig}.

\begin{figure}[htnp]
{\scriptsize
\noindent
{\bf Computing the range $(L,U)$.}
\begin{itemize}
\item
If $i'=0$, let $U=w(I)-w(s)$.  Otherwise ($i'>0$), $U$ is
the solution of
$\sum_{h=0}^i \frac{-\ln v_h}{x-s_h}-\sum_{h=0}^{i'-1} \frac{-\ln v'_h}{w(I)-x-s'_h} =0\ .$
(There is always a solution $U\in (s_i,w(I)-s'_{i'-1})$.)
\item
If $i=0$, let $L=0$. Otherwise ($i>0$), $L$ is the solution of
$\sum_{h=0}^{i-1} \frac{-\ln v_h}{x-s_h}-\sum_{h=0}^{i'} \frac{-\ln v'_h}{w(I)-x-s'_h} =0\ .$
(There is always a solution $L\in (s_{i-1},w(I)-s'_{i'})$.)
\end{itemize}
\noindent
{\bf Search for $x\in (L,U)$ such that $d(x)=\Delta$.}
\begin{itemize}
\item
If $i=0$ (we must have $\Delta>0$) we set $M$ to be the solution of
$ \sum_{h=1}^{i'-1}\frac{-\ln v'_h}{w(I)-x-s_h} =\Delta\ $
in the range $(L,U)$.  If there is no solution, we set
$M\leftarrow L$.
\item
If $i'=0$ (we must have $\Delta<0$), we set $M$ to be the solution of 
$\sum_{h=0}^{i-1} \frac{-\ln v_h}{x-s_h}=-\Delta\ $ in the range $(L,U)$.
If there is no solution, we set $M\leftarrow U$.
\item
Otherwise, if $i>0$ and $i'>0$, we set $M$ to be the solution of
$\sum_{h=0}^{i-1} \frac{-\ln v_h}{x-s_h}-\sum_{h=0}^{i'-1}\frac{-\ln v'_h}{w(I)-x-s_h} =\Delta\ .$
There must be a solution in the range $(L,U)$.
\end{itemize}
\noindent
{\bf Truncating the solution.}
\begin{itemize}
\item
  We can have $L\in (s_{i-1},s_{i})$ and hence possibly  $M<s_{i}$.
In this case we set $M=s_i$.
Similarly, we can have $U\in (w(I)-s_{i'},w(I)-s_{i'-1})$ and hence possibly
$M>w(I)-s_{i'}$.  In this case we set $M=w(I)-s_{i'}$.
\item 
 We return $M$.  
\end{itemize}
\caption{Solver for $s(x)=\tau$ for subpopulation weight with known $w(I)$.}
\label{solver:fig}
}
\end{figure}
} 


\ignore{\edith{ check:
 We can extend the second method, to reduce the
variance of the bounds $\underline{w}_{\delta}(J)$ and $\overline{w}_{\delta}(J)$
by re-drawing the permutation $\pi$ of the items
$\{i_1,\ldots,i_k\}$ conditioned on the 
$(k+1)$st rank being $\tau$~\cite{bottomk07:ds}.
}}

\subsection{Confidence bounds for priority sketches}

 We review the
confidence bounds for \pri\ sketches obtained by 
Thorup~\cite{Thorup:sigmetrics06}.
 We denote
$p_{\tau}(i)=\pr\{r(i)<\tau\}$. The number of items in $J\cap s$
with $p_{\tau}(i)<1$ 
is used to bound $\sum_{i\in J|p_{\tau}(i)<1} p_{\tau}(i)$ (the expectation
of the sum of independent Poisson trials).
These bounds are then used to obtain bounds on the weight
$\sum_{i\in J|p_{\tau}(i)<1} w(i)$, exploiting
the correspondence (specific for \pri\ sketches)
between $\sum_{i\in J|p_{\tau}(i)<1} p_{\tau}(i)$ and
$\sum_{i\in J|p_{\tau}(i)<1} w(i)$:
For \pri\ sketches, $p_{\tau}(i)=\min\{1,w(i)\tau\}$.
If $w(i)\tau\geq 1$ then $p_{\tau}(i)=1$ (item is included in the sketch) and
if $w(i)\tau< 1$ then $p_{\tau}(i)=w(i)\tau$.   
Therefore, $p_{\tau}(i)<1$ if and only if $p_{\tau}(i)=w(i)\tau$ and 
$$\sum_{i\in J|p_{\tau}(i)<1} w(i) = \tau^{-1} \sum_{i\in J|p_{\tau}(i)<1} p_{\tau}(i)\ .$$

For $n'\geq 0$, define $\overline{n}_{\delta}(n')$ (respectively,
$\underline{n}_{\delta}(n')$) to be the infimum (respectively, supremum)
over all $\mu$, such that for all sets of 
independent Poisson trials with
sum of expectations $\mu$, the sum is less than $\delta$
likely to be at most $n'$ (respectively, at least $n'$).
If $n'=|\{i\in J\cap s|w(i)\tau < 1\}|$,
then $\underline{n}_{\delta}(n')$ and $\overline{n}_{\delta}(n')$ 
are $(1-\delta)$-confidence bounds on 
$\sum_{i\in J\cap s |w(i)\tau < 1} p_{\tau}(i)$.
Since 
$$w(J)=\sum_{i\in J\cap s |w(i)\tau \geq 1} w(i) +
\tau^{-1}\sum_{i\in J\cap s |w(i)\tau < 1} p_{\tau}(i)\ ,$$
we obtain $(1-\delta)$-confidence upper and lower bounds on
$w(J)$ by substituting $\overline{n}_{\delta}(J)$
and $\underline{n}_{\delta}(J)$ for 
$\sum_{i\in J\cap s |w(i)\tau < 1} p_{\tau}(i)$ in this formula, respectively.

Chernoff bounds provide an upper bound on
$\overline{n}_{\delta}(n')$ of $-\ln\delta$ if $n'=0$ and 
the solution of $\exp(n'-x)(x/n')^{n'}=\delta$ otherwise; and a lower bound on
$\underline{n}_{\delta}(n')\leq n'$ that is 
the solution of $\exp(n'-x)(x/n')^{n'}=\delta$ and  $0$ if there is
no solution.

\notinproc{
With other families of rank functions, this approach provides bounds on
the sum $\sum_{i\in J} p_{\tau}(i)$.  We then need to consider
the distribution of the $p_{\tau}(i)$'s, given the sum,
that maximizes or minimizes the respective
sum of the weights of items.  
For \ws\ sketches, $w(i)$ can be arbitrarily large when $p(i)$
approaches $1$, which precludes good upper bounds using this approach.
}

We point on three sources of slack in the bounds used 
in~\cite{Thorup:sigmetrics06}. 
\notinproc{
As a result,
the bounds are not ``tight'' since they are correct with probability
strictly higher than $(1-\delta)$.  
}
The first is the use of
Chernoff bounds rather than exactly computing 
$\overline{n}_\delta(n')$ and $\underline{n}_\delta(n')$.
The other two sources of slack are due to the fact that
the actual distribution of the sum of independent Poisson trials
depends not only on the sum of their expectations but also on how 
they are distributed
(variance is higher when there are more items with smaller $p_i$'s).
The second slack is that
these bounds make ``worst case'' assumptions on the
distribution of the items.
\notinproc{
(This is
present even if we compute $\underline{n}_\delta(n')$
and $\overline{n}_\delta(n')$ exactly).
}    
\ignore{(the bounds for \ws\ sketches are tight as there are no ``worst case''
assumptions on the distribution.)}
The third  slack is that the derivation of the
bounds does
not 
use the weights of the items in $J\cap s$ 
with $w(i)\tau < 1$ that we see in the sketch.
\notinproc{
Thus the ``worst case'' assumptions are extended to
the distribution of the sampling probabilities of these items.
}

\notinproc{

The first and third sources of slack can be addressed by assuming
Poisson distribution on the ``unseen'' part of the distribution (the
``worst case'' is having many tiny items) and using simulations for the items 
in $J\cap s$. 
 Alternatively, instead of bounding the weight through
the sum of probabilities, 
we can apply Lemma~\ref{conflemmatot} to 
bound the weight of $I\setminus s$. 
Since we use the weights of the items in $s$, we
address the third source of slack in the bounds
of~\cite{Thorup:sigmetrics06}.

The maximum weight of an item in
$I\setminus s$ is $\tau^{-1}$.  For any $\ell \geq 0$, 
we consider the distribution
of item weights with total weight equal to $\ell$
that maximizes the probability that the
minimum rank of these items is at least $\tau$ (for the lower bound)
or is at most $\tau$ (for the upper bound.)

\vspace*{-0.1in}
\paragraph*{Lower bound on W(I)}

For a fixed $\ell$ (which is the tentative bound on
the weight of $I\setminus s$), consider the maximum probability that
the minimum rank of an item in  a set $Z$ ($=I\setminus s$)  with
total weight $\ell$ and maximum weight $1/y$, is at most $y$.
This probability is maximized if we make the items of
$Z$ as large as possible: It is $1$ if $\ell \geq 1/y$ (we put in
$Z$ at least one item of weight $1/y$), and it is 
 $y\ell$ if  $\ell < 1/y$ ($Z$ consists of one item of weight
$\ell$).

The respective 
probability density
of the minimum rank $y$ as a function of $\ell$ 
is $0$ for $y>1/\ell$ and $\ell$ otherwise.
 Applying a similar derivation to that of Eq.(\ref{subconfws}), we obtain
that the probability density of the event that
the items in $s$ have smaller ranks
than items in $I\setminus s$ and the smallest rank among items
in $I\setminus s$ is equal to $y$ is $0$ for $y>1/\ell$ and otherwise
it is $\ell \prod_{j\in s}\min\{1,w(i_j)y\}$.  This probability density
conditioned on subspace where the items in $s$ have smaller ranks
than the items in $I\setminus s$ is

\vspace{-0.1in}
{\small
\begin{eqnarray} 
D^{(\pri,low)}(\ell,y)& = &\frac{\ell \prod_{j\in s}\min\{1,w(i_j)y\} }
{\int_{x=0}^{1/\ell} \ell \prod_{j\in s}\min\{1,w(i_j)x\}dx} \nonumber \\
&= &\frac{\prod_{j\in s}\min\{1,w(i_j)y\} }{\int_{x=0}^{1/\ell} \prod_{j\in s}\min\{1,w(i_j)x\}dx} \label{subconfpril}
\end{eqnarray}
}
The lower bound on $w(I\setminus s)$
is the value of $\ell<\tau^{-1}$ 
that solves the equation 
$\int_0^{\tau} D^{(\pri,low)}(\ell,y) dy=\delta$.\footnote{Lower bound obtained using this method is at most $\tau^{-1}$.}

\vspace*{-0.1in}
\paragraph*{Upper bound on W(I)}
 For  total weight $\ell$, the
probability that the minimum rank is at least $\tau$ is maximized
at the limit when there are many small items and is equal to
$\exp(-\ell \tau)$.  The probability
density function of the minimum rank value 
being equal to $\tau$ is $\ell \exp(-\ell \tau)$.

 Applying a similar consideration to that of Eq.~(\ref{subconfpril})
using a similar derivation to that of Eq.(\ref{subconfws}), we obtain
that the probability density of the event that
the items in $s$ have smaller ranks
than items in $I\setminus s$ and the smallest rank among items
in $I\setminus s$ is equal to $y$ is
\begin{eqnarray*}
D^{(\pri,u)}(\ell,y)&=&\frac{\ell\exp(-\ell y)\prod_{j\in s}\min\{1,w(i_j)y\} }{\int_{x=0}^\infty \ell\exp(-\ell x)\prod_{j\in s}\min\{1,w(i_j)x\}dx}\\
&= &\frac{\exp(-\ell y)\prod_{j\in s}\min\{1,w(i_j)y\} }{\int_{x=0}^\infty \exp(-\ell x)\prod_{j\in s}\min\{1,w(i_j)x\}dx}  \label{subconfpriu}
\end{eqnarray*}

 The upper bound on $w(I\setminus s)$ is 
the value of $\ell$ 
that solves the equation $\int_0^{\tau} D^{(\pri,u)}(\ell,y) dy=1-\delta$.

For the lower bound,
the integrand is a piecewise polynomials with breakpoints at
$w(i)^{-1}$ ($i\in s$). For the upper bound, the integrand is a
piecewise function of the form of a polynomial multiplied by an
exponential.  Both forms are simple to integrate.
} 

%

\section{Simulations} \label{simul:sec}


\vspace{-0.1in}
\paragraph*{Total weight}
 We compare estimators and confidence bounds on 
the total weight $w(I)$ using
three distributions of 1000 items each with weights
independently 
drawn from Pareto distributions with parameters $\alpha\in\{1,1.2,2\}$,
and also on a uniform distribution.

\smallskip
\noindent
{\bf Estimators. }
 We evaluate
the maximum likelihood \ws\ estimator (\ws\ \ml), 
the rank conditioning \ws\ estimator (\ws\ \rc), the rank conditioning
\pri\ estimator (\pri\ \rc)~\cite{dlt:pods05}, and 
the \wsr\ estimator~\cite{ECohen6f} (Section~\ref{wsrest:sec}).

\begin{figure*}[ht]
\centerline{\begin{tabular}{ccc}
 \epsfig{figure=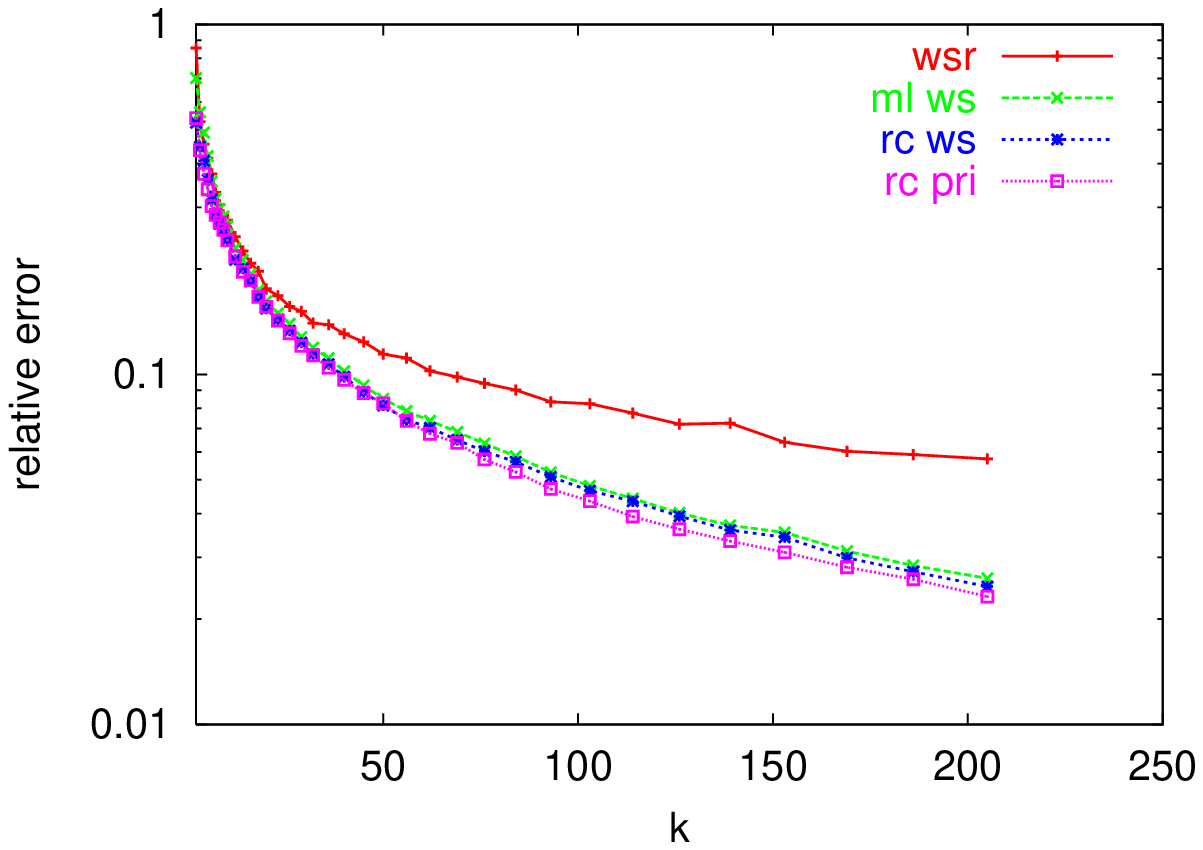,width=0.32\textwidth} &
 \epsfig{figure=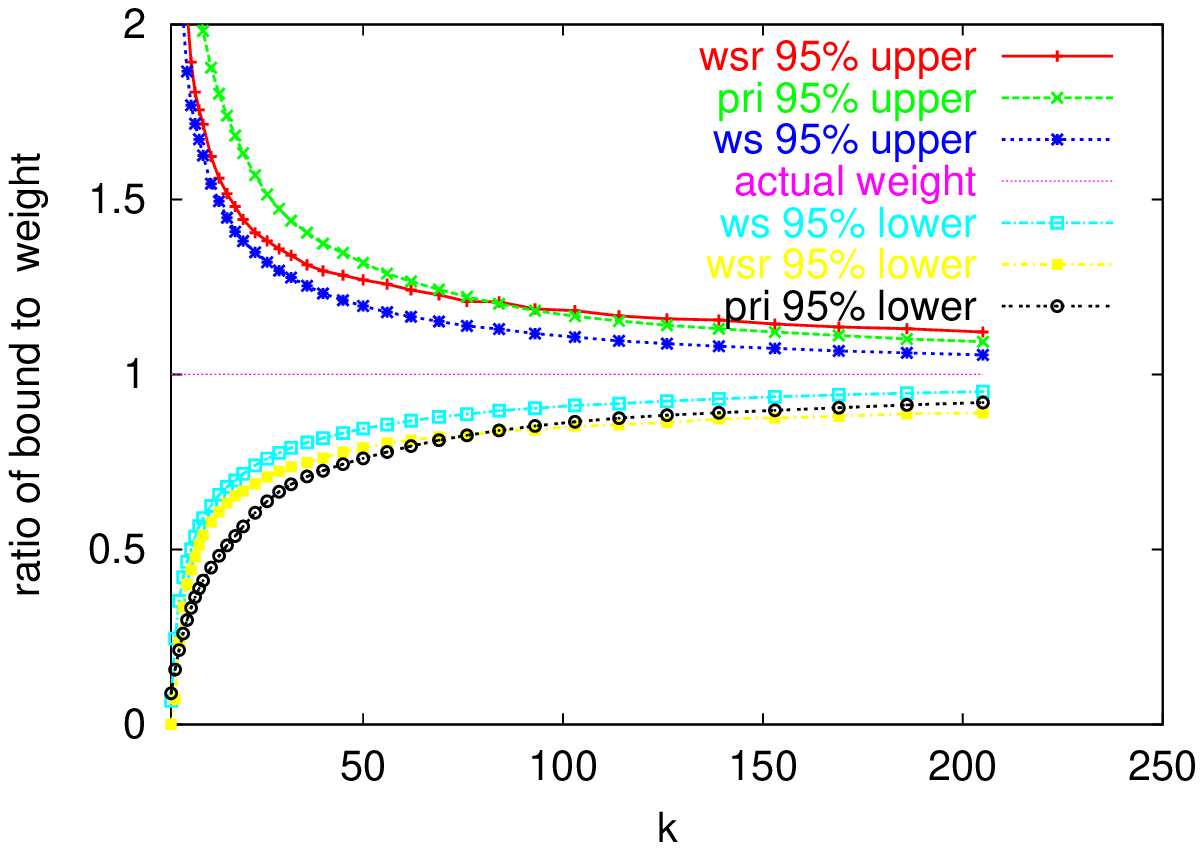,width=0.32\textwidth} &
\epsfig{figure=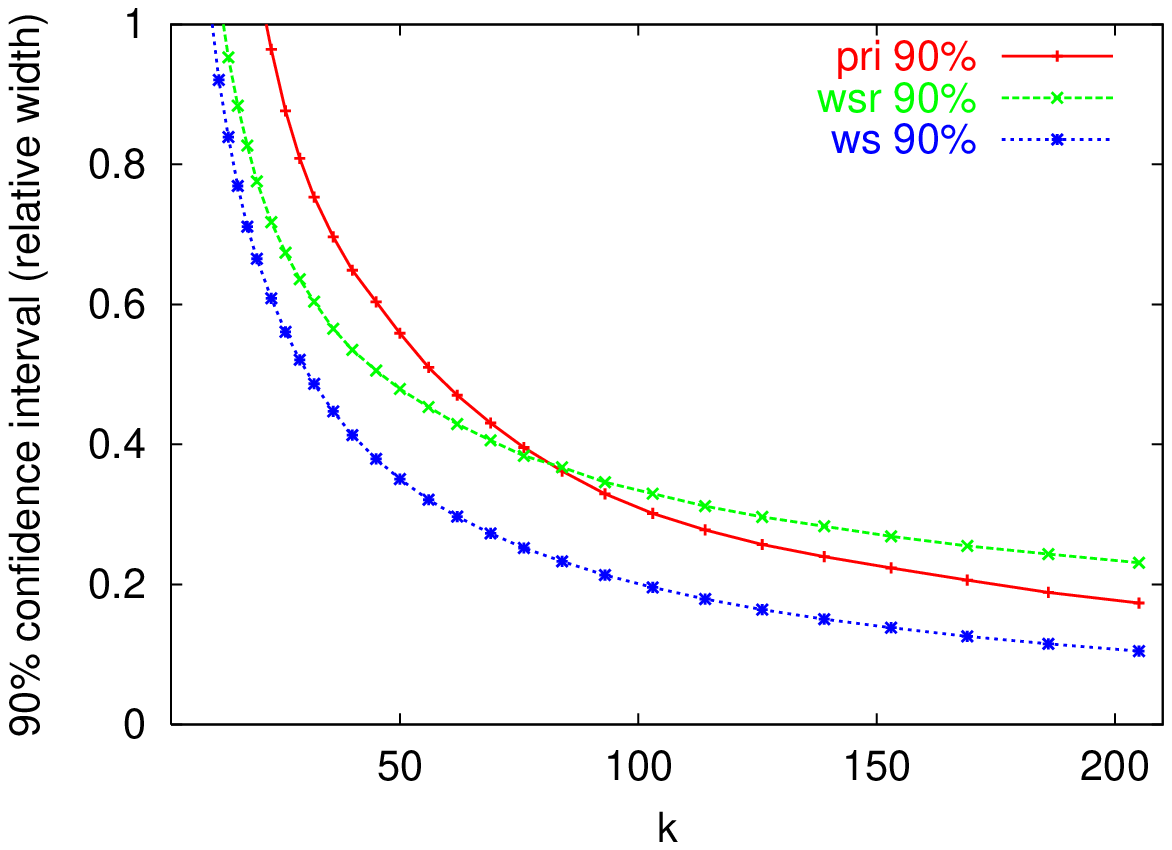,width=0.32\textwidth} \\
\notinproc{
 \epsfig{figure=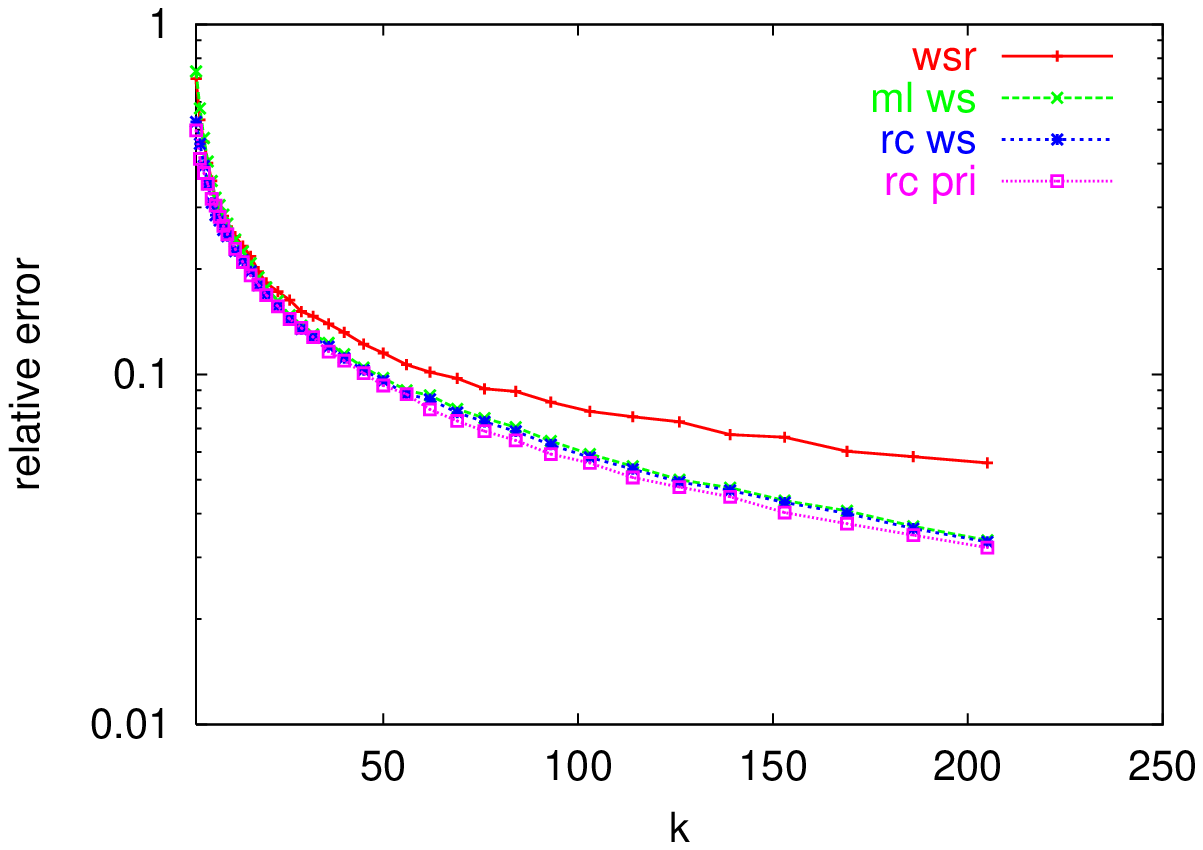,width=0.32\textwidth} &
 \epsfig{figure=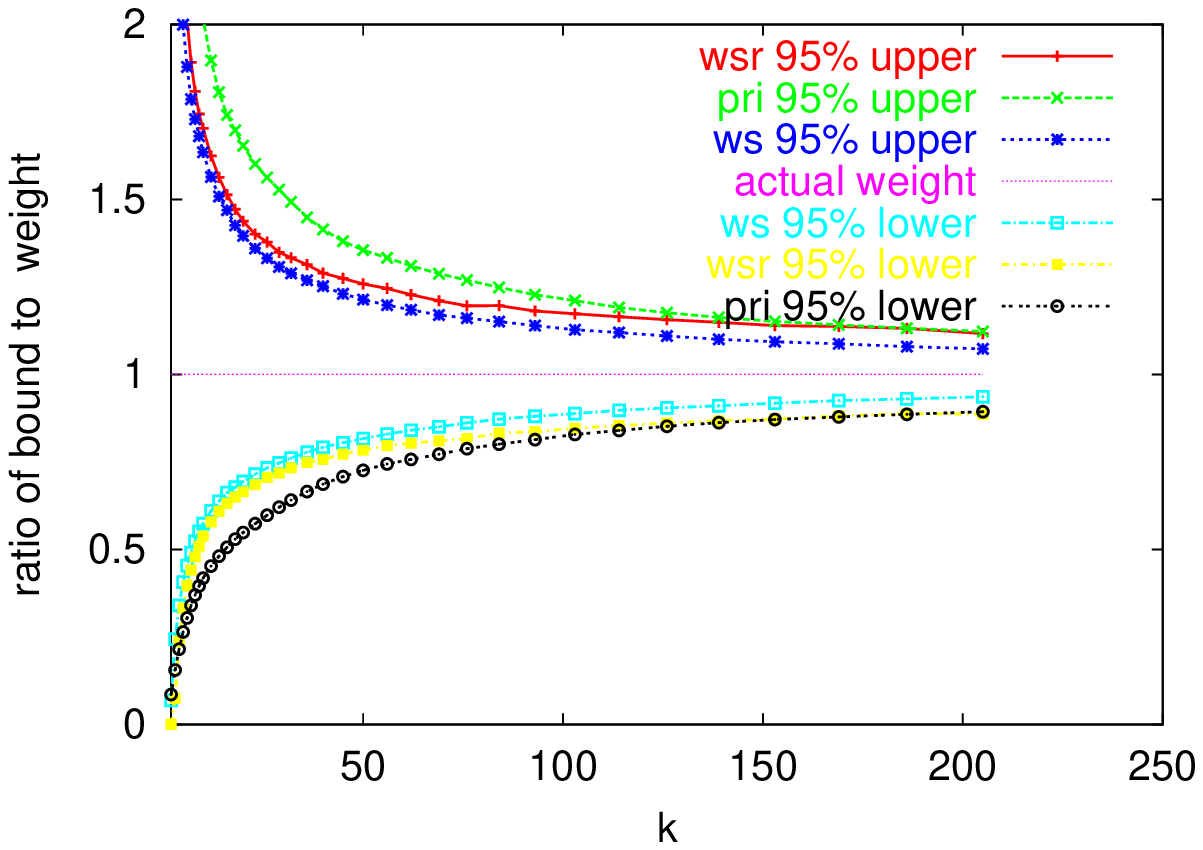,width=0.32\textwidth} &
 \epsfig{figure=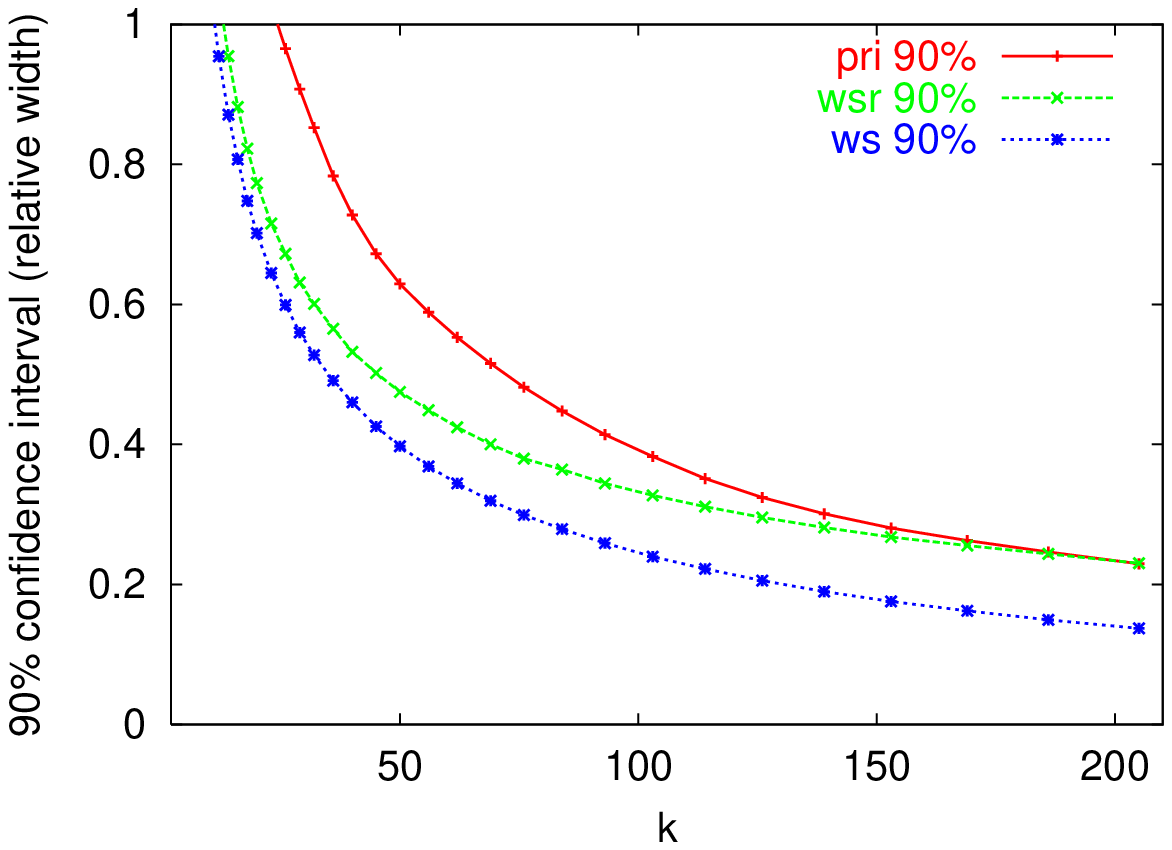,width=0.32\textwidth }\\}
 \epsfig{figure=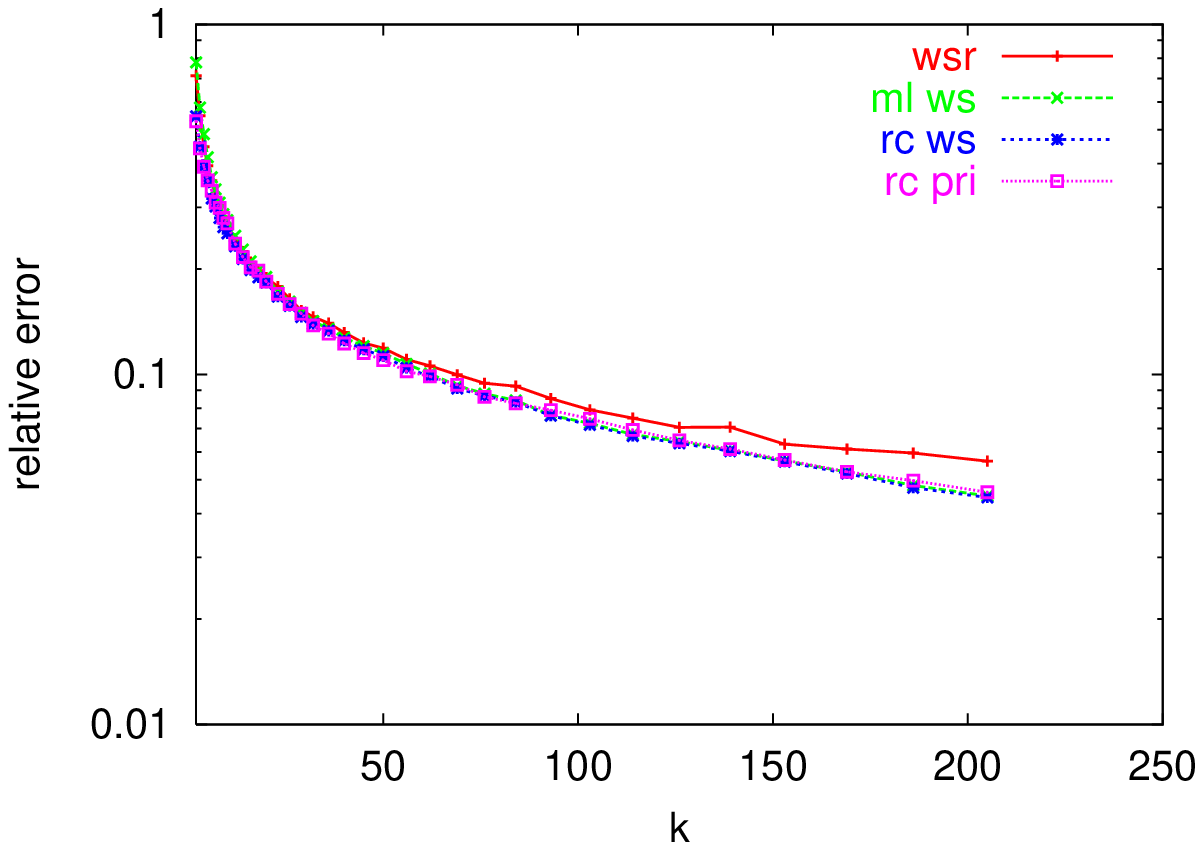,width=0.32\textwidth} &
 \epsfig{figure=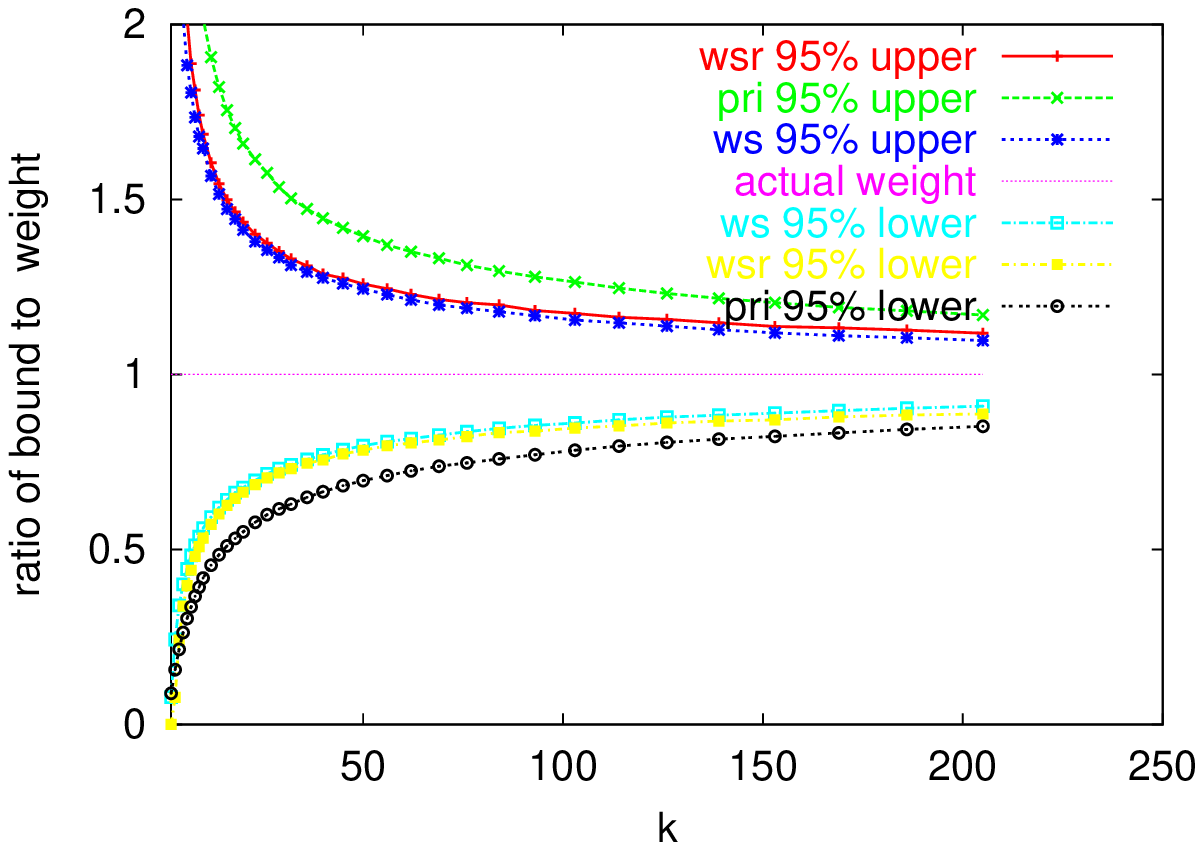,width=0.32\textwidth} &
 \epsfig{figure=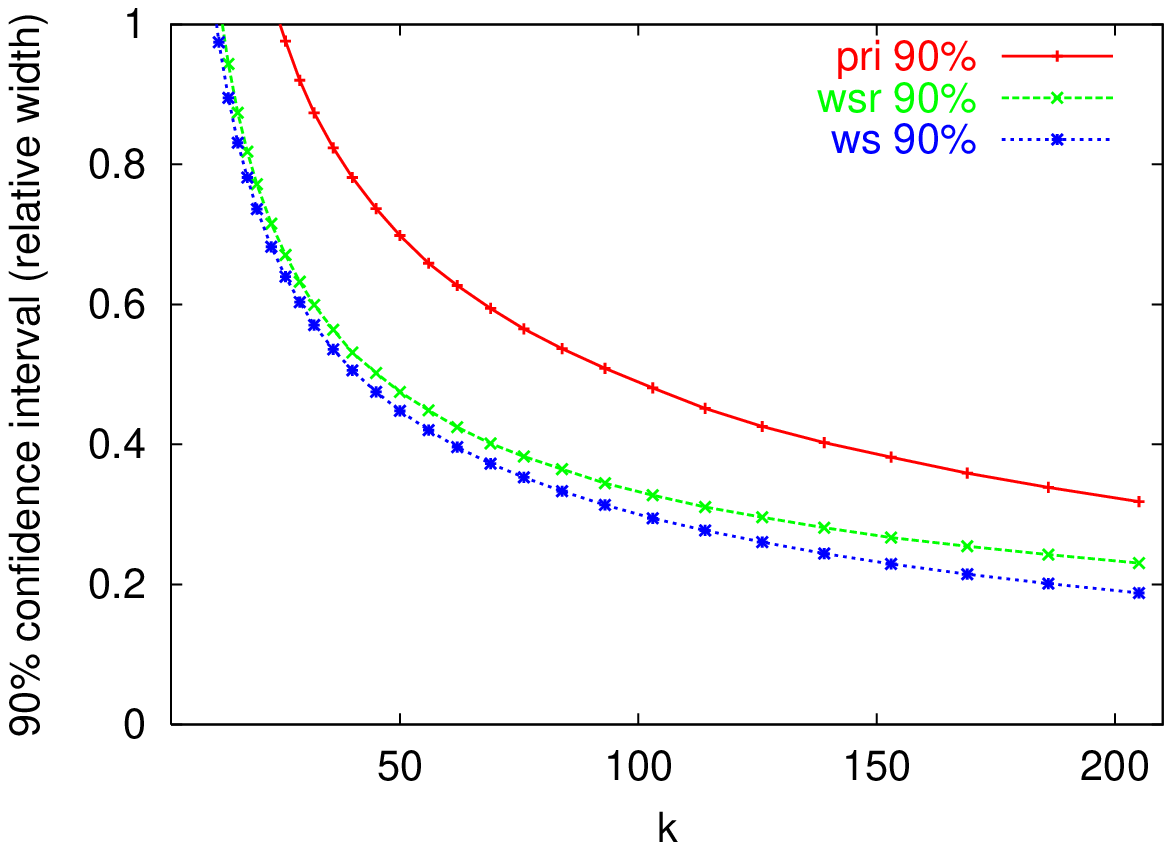,width=0.32\textwidth} 
\notinproc{\\
 \epsfig{figure=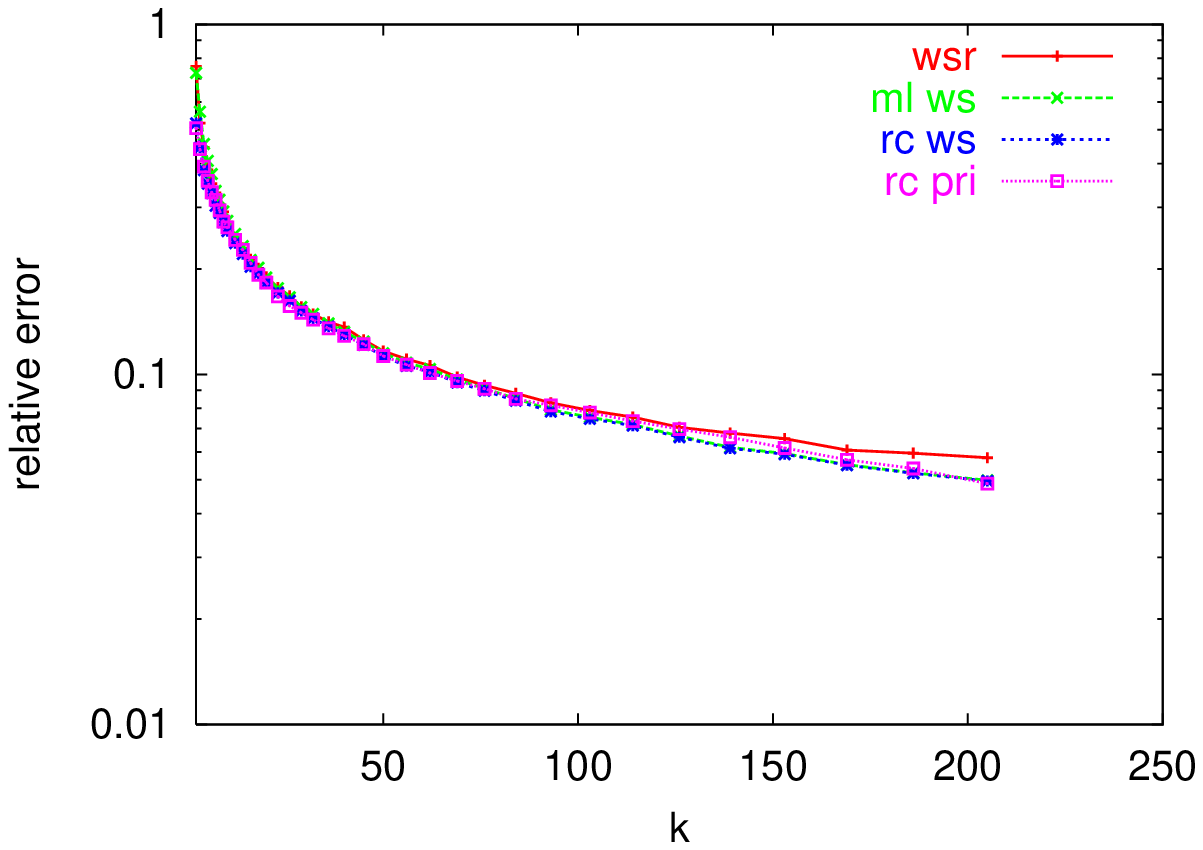,width=0.32\textwidth} &
 \epsfig{figure=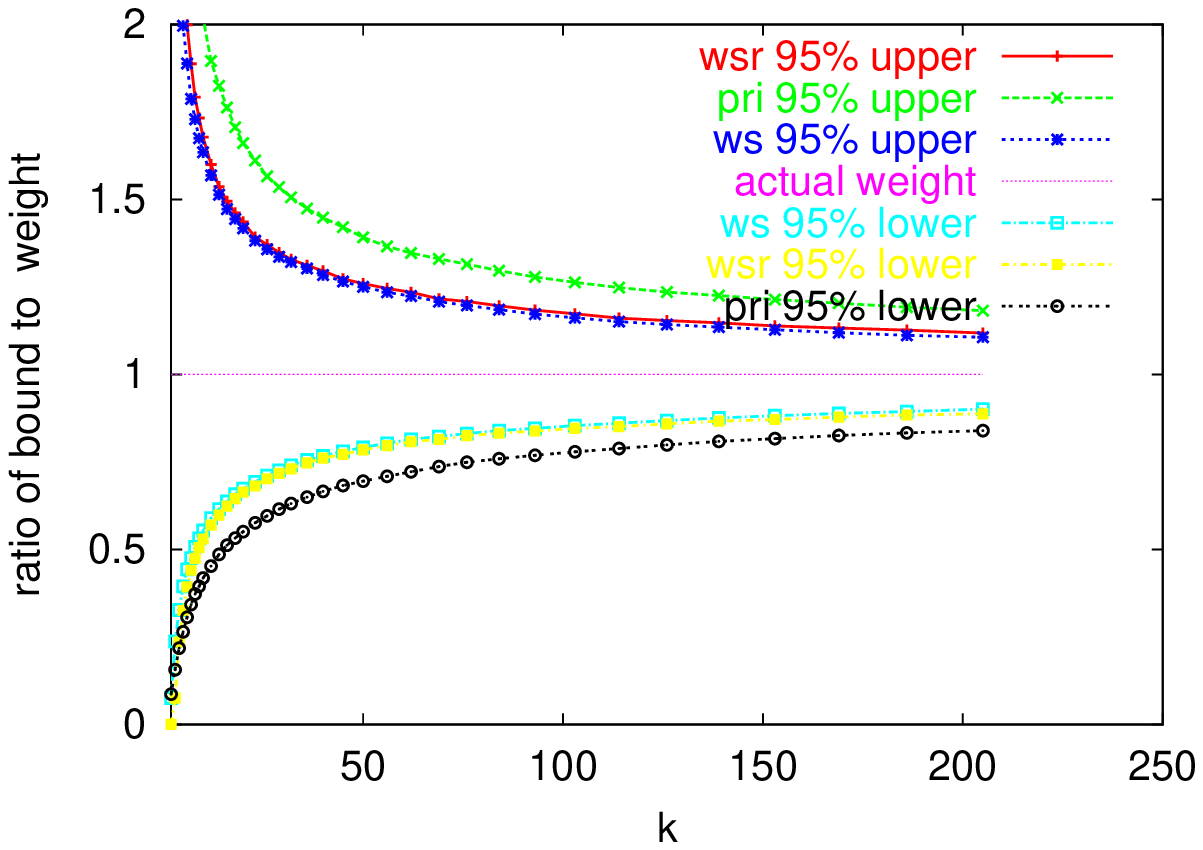,width=0.32\textwidth} &
\epsfig{figure=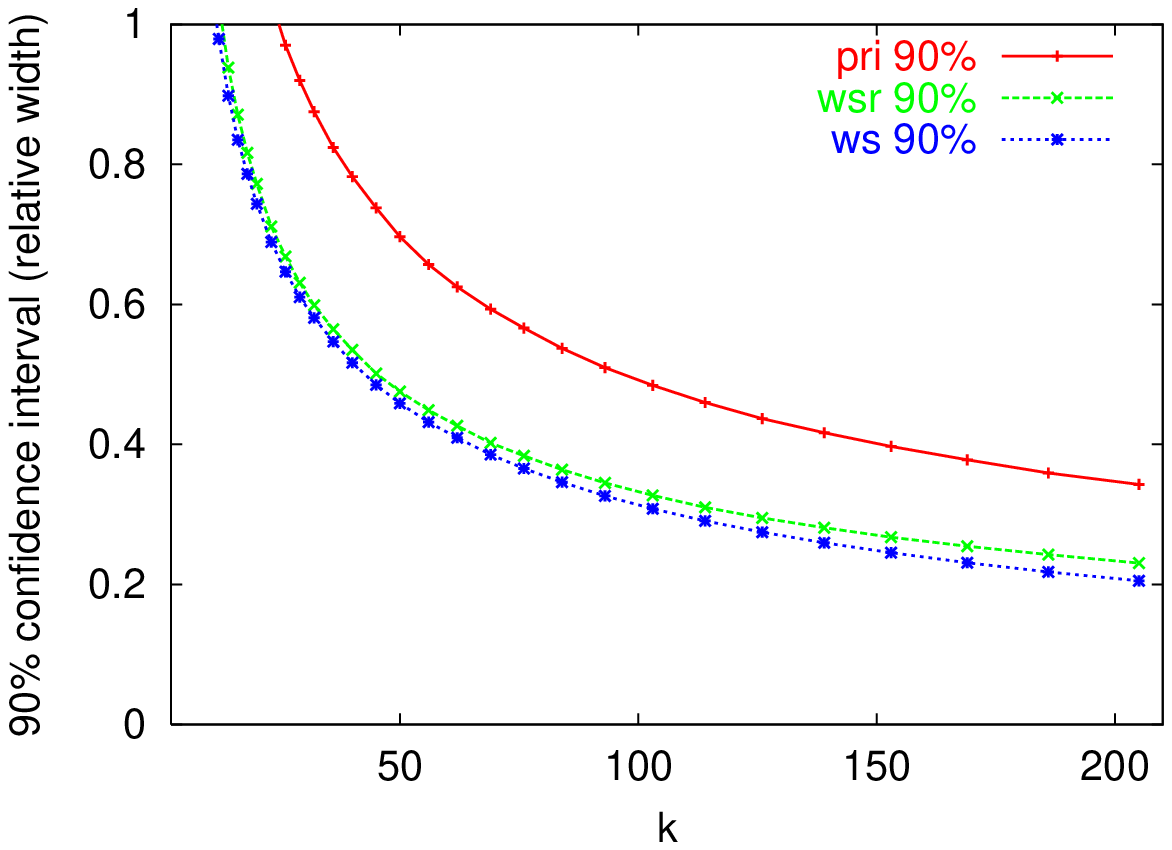,width=0.32\textwidth} 
}
\end{tabular}
}
\caption{Left: Absolute value of the relative error of the estimator
of $w(I)$ averaged
over 1000 repetitions. Middle: 95\% confidence upper and lower bounds for estimating $w(I)$.  Right: width of 90\% confidence interval for estimating $w(I)$.
\onlyinproc{ 
We show results for $\alpha=1$ (top row) and $\alpha=2$ (bottom row).
}
\notinproc{
We show results for $\alpha=1$ (top row), $\alpha=1.2$ (second row), $\alpha=2$ (third row), and uniform weights (bottom row).}
}
\label{estqual3:fig} \label{estqualuni:fig}
\end{figure*}

Figure~\ref{estqual3:fig} (left) shows the
absolute value of the relative error, averaged over
1000 runs, as a function of $k$.
 We can see that all three bottom-$k$ based estimators outperform the
\wsr\ estimator,  demonstrating the advantage of the added information
when sampling ``without replacement'' over sampling ``with
replacement'' (see also~\cite{bottomk07:ds}).  The advantage
of these estimators grows with the skew.  The quality 
of the estimate is
 similar among the bottom-$k$ estimators (\ws\ \ml, \ws\ \rc, and \pri\ \rc).  The maximum likelihood estimator (\ws\ \ml), which is biased,
has worse performance for very small values of $k$ where the bias is
more significant.  \pri\ \rc\ has a slight
advantage especially if the distribution is more skewed.  This is because,
in this setting, with unknown $w(I)$, \pri\ \rc\ is a nearly optimal
adjusted-weight based estimator.

\smallskip
\noindent
{\bf Confidence bounds. }
We compare the Chernoff based \pri\
 confidence bounds from~\cite{Thorup:sigmetrics06} and the \ws\ and
\wsr\ confidence bounds we derived.  We apply the normal
approximation with the stricter (but easier to compute) conditioning
on the order for the \ws\ confidence bounds and the normal
approximation for the \wsr\ confidence bounds (see
Sections~\ref{confwsr} and \ref{confws}).  The 
 95\%-confidence upper and lower bounds and the 90\% confidence
interval ({\em the width}, which is the difference between the upper
and lower bounds), averaged over 1000 runs, are shown in
Figure~\ref{estqual3:fig} (middle and right).  We can see
that the \ws\ confidence bounds are tighter, and often significantly so,
than the \pri\ confidence
bounds.  In fact \pri\ confidence bounds 
were worse than
the \wsr-based bounds on less-skewed distributions (including
the uniform distribution on 1000 items). This perhaps surprising behavior is explained
by the large ``slack'' between the bounds in~\cite{Thorup:sigmetrics06}
 and the actual variance of
the (nearly-optimal) \pri\ \rc\ estimator.  

The \ws\ bounds in Eq.~(\ref{subconfws}) (that do not use conditioning on the order) should be tighter than the bounds that use this conditioning.
\notinproc{
The \pri\
bounds in Eq.~(\ref{subconfpriu}) and Eq.~(\ref{subconfpril})
(that address some of the ``slack'' factors) may be tighter. }
We have not implemented these alternative bounds
 and  leave this comparisons for
future work.

The normal approximation provided fairly accurate confidence bounds
for the total weight.  The \ws\ and \wsr\ bounds were evidently more
efficient, with real error rate that closely corresponded to the
desired confidence level.  For the 90\% confidence interval, across
the three distributions with $\alpha=1,1.2,2$, and value of $k$, the
highest error rate was 12\%.  
The true weight was within the  \ws\ confidence bounds on average in 
90.5\%, 90.2\%, 90\% of the time for the
different values of $\alpha$.
The corresponding in-bounds rates for
 \wsr\ were 90.6\%, 90.3\%, and 90.0\%, and for 
 \pri\ 99.2\%, 99.1\%, and 98.9\%.  (The high in-bounds rate for the \pri\
bounds reflects the slack in these bounds).

\vspace{-0.1in}
\paragraph*{Subpopulation weight} 
{\bf Estimators. }
 We implemented an approximate version of \ws\ \ssc\ using
the Markov chain and averaging method. We showed that
this approximation
provides unbiased estimators that are better than the plain
\ws\ \rc\ estimator (better per-item variances and negative covariances for
different items), but attains zero sum of covariances only at the limit.
We
quantified this improvement
of \ws\ \ssc\ over \ws\ \rc\ and its
dependence on the size of the subpopulation.
We  evaluated the quality of approximate \ws\ \ssc\ 
as a function of the parameters \inperm, and \permnum\ (see
Section \ref{sec:compsc}), and we
compared \ws\ \ssc\ to the \pri\ \rc\ estimator.

To 
evaluate how
the quality of the estimator depends on 
the size of the subpopulation 
we introduce a {\em group size\/} parameter $g$.
We order the items by their weights 
 and 
partition then sequentially into 
$|I|/g$  groups each consisting of
 $g$ items.
 For each group size, we compute the sum,
over subsets in this partition,
 of the
square error of the estimator 
(averaged over multiple runs).  This sum corresponds to the
sum of the variances of the estimator over the subsets of
the partition.
For $g=1$, this sum corresponds to the sum of the variances
of the  items.  

The \rc\ estimators have zero covariances, and therefore, the
sum of square errors should remain constant when sweeping $g$.
The \ws\ \ssc\ estimator has negative covariances and therefore
we expect the sum to decrease as a function of $g$.
For $g=n$, we obtain the variance of the sum of the adjusted weights,
which should be $0$ for the \ws\ \ssc\ estimator (but not for the
approximate versions).

We used two distributions generated by drawing $n=20000$ items from a
Pareto distribution with parameter $\alpha\in\{1.2,2\}$.
\onlyinproc{
Figure~\ref{varovergrouping:fig} shows some representative plots: 
}
The sum of square errors, as a function of $g$, is constant for the 
\rc\ estimators, but decreases with the \ws\ \ssc\ estimator.
For $g=1$, the \pri\ \rc\ estimator (that obtains the
minimum sum of per-item variances by a sketch of size $k+1$)
performs slightly better than the
\ws\ \rc\ estimator when the data is more skewed 
(smaller $\alpha$).  The \ws\ \ssc\ estimator, however, 
performs very closely and better for small values of $k$
(it uses one fewer sample).
 For $g>1$, the \ws\ \ssc\ estimator outperforms
both \rc\ estimators and has significantly smaller
variance for larger subpopulations.
\notinproc{
Figure~\ref{varovergrouping:fig} shows the results for
$k\in\{4,40,500\}$.  For each value of $k$, we show the sum of square
errors over subsets in the partition, averaged over 1000 repetitions,
as a function of the partition parameter $g$.
Figure~\ref{varoverkfixedgroup:fig} shows the sum of square errors
(again, averaged over 1000 repetitions) as a function of $k$ for
partitions with $g\in\{1,5000\}$.
}

 We conclude that in applications when $w(I)$ is
provided, the \ws\ \ssc\ estimator emerges as a considerably better choice than
the \rc\ estimators.
It also shows that the metric of the sum of per-item variances,
that  \pri\ \rc\ is nearly optimal~\cite{Szegedy:stoc06}
with respect to it,
is not a sufficient notion of optimality.

\onlyinproc{
\begin{figure*}[ht]
\centerline{\begin{tabular}{ccc}
 \epsfig{figure=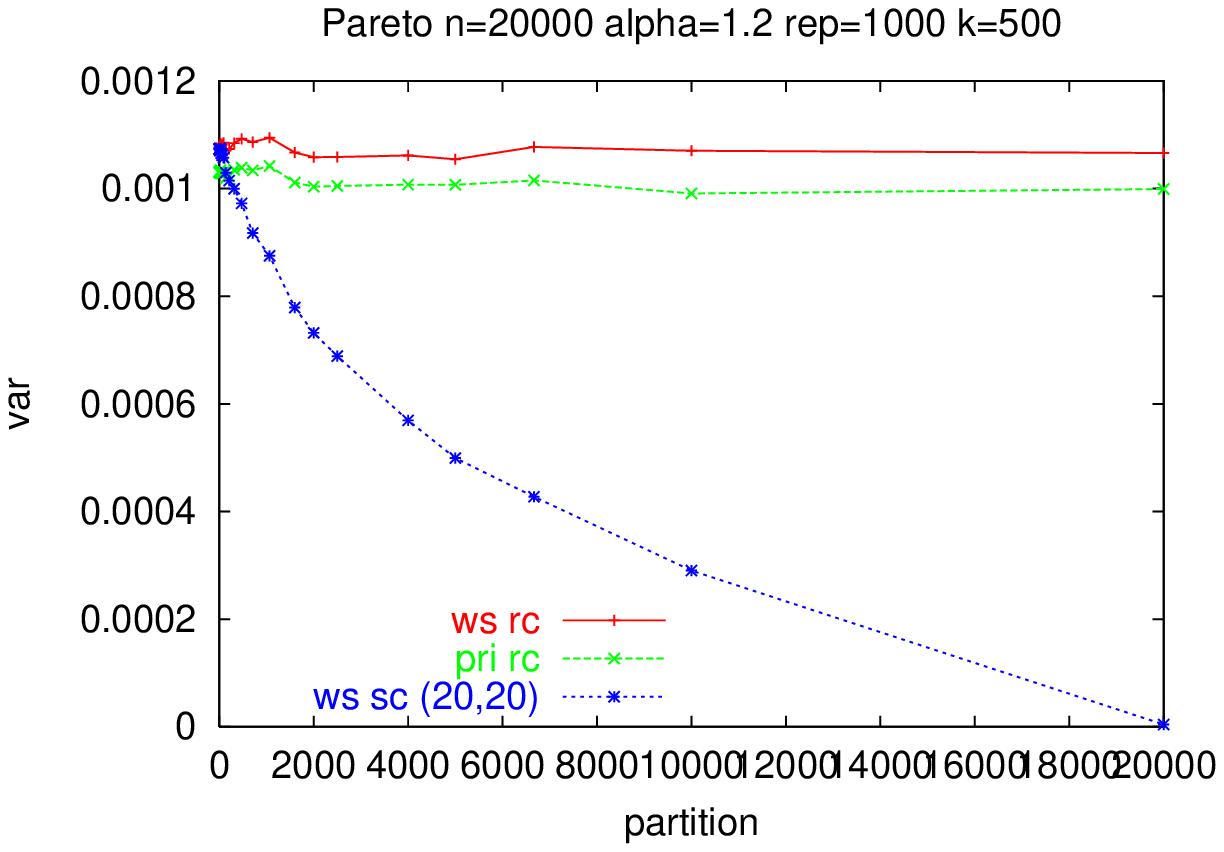,width=0.32\textwidth} &
 \epsfig{figure=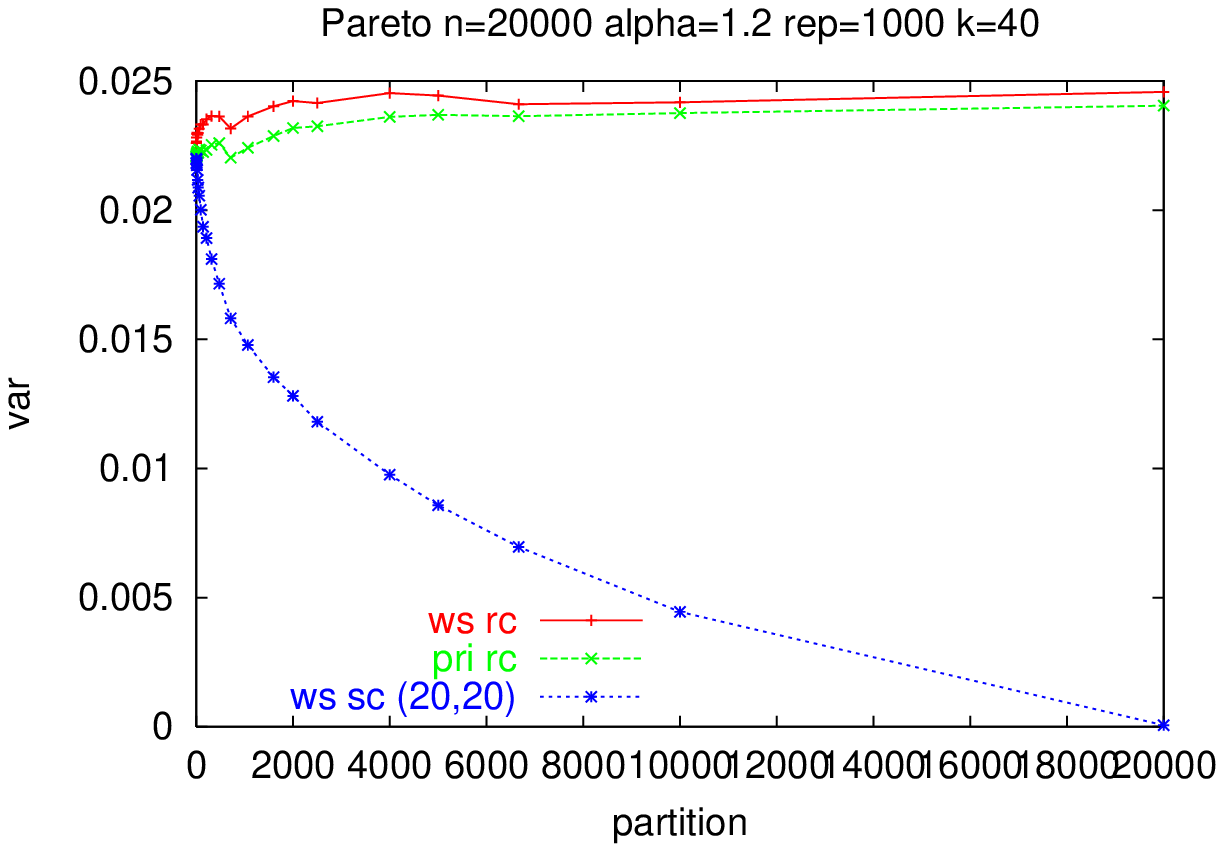,width=0.32\textwidth} &
 \epsfig{figure=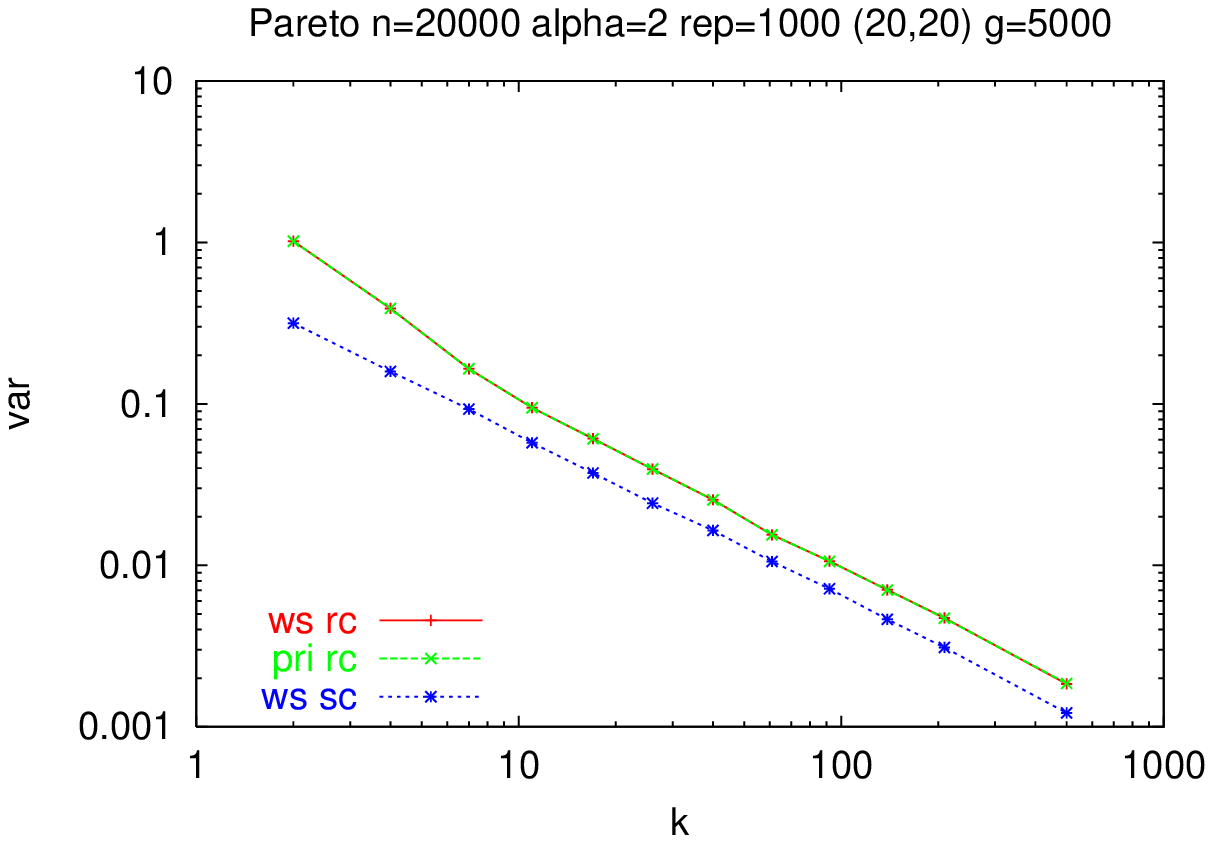,width=0.32\textwidth} 
\end{tabular}
}
\caption{Sum of square errors (averaged over 1000 repetitions) over a partition. Left and middle:  as
a function of group size for $k=500$ and $k=40$ and $\alpha = 1.2$. Right:
as a function of $k$ for
$g=5000$ and $\alpha=2$.
\label{varovergrouping:fig}}
\end{figure*}

}

\notinproc{
\begin{figure*}[htbp]
\centerline{\begin{tabular}{ccc}
 \epsfig{figure=code2/results/k_500_alpha_1.2_n_20000_rep_1000_20_20_var.eps,width=0.32\textwidth} &
 \epsfig{figure=code2/results/k_40_alpha_1.2_n_20000_rep_1000_20_20_var.eps,width=0.32\textwidth} &
 \epsfig{figure=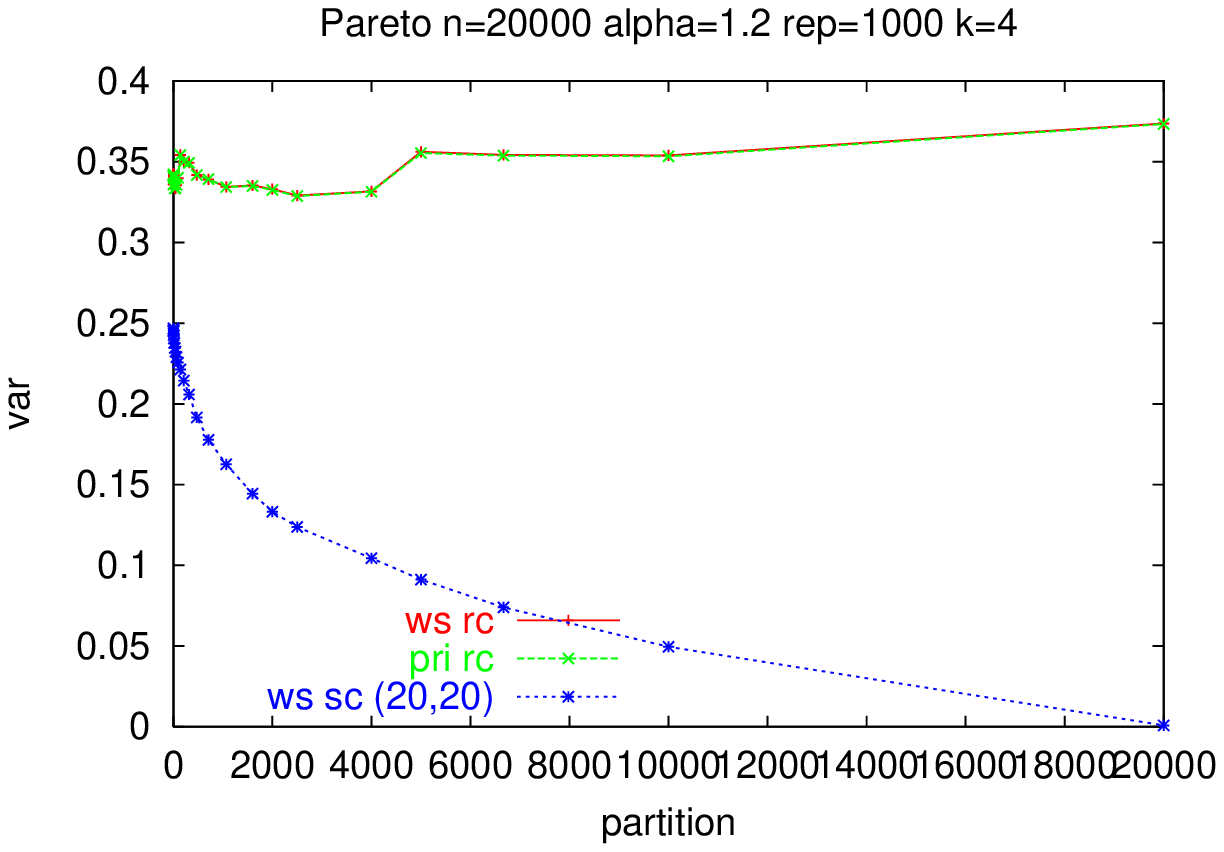,width=0.32\textwidth} \\
 \epsfig{figure=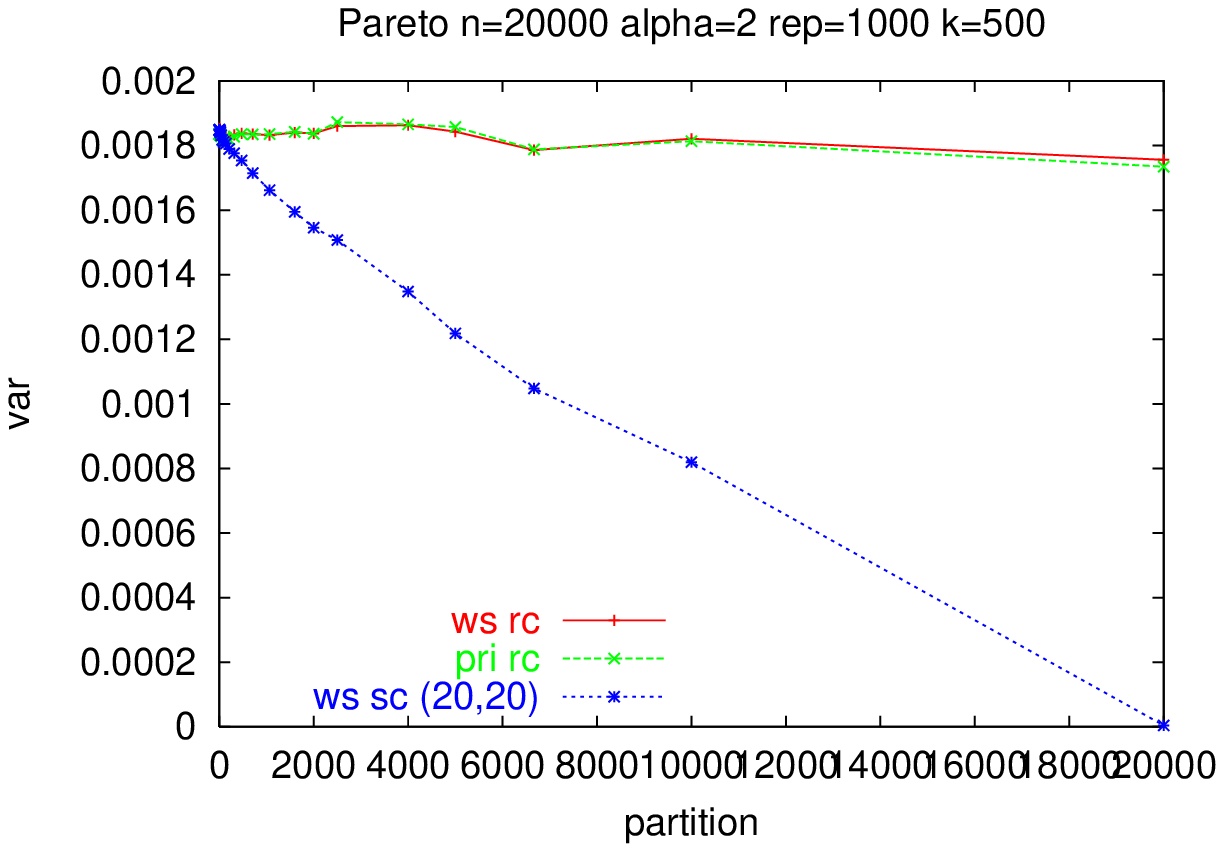,width=0.32\textwidth} &
 \epsfig{figure=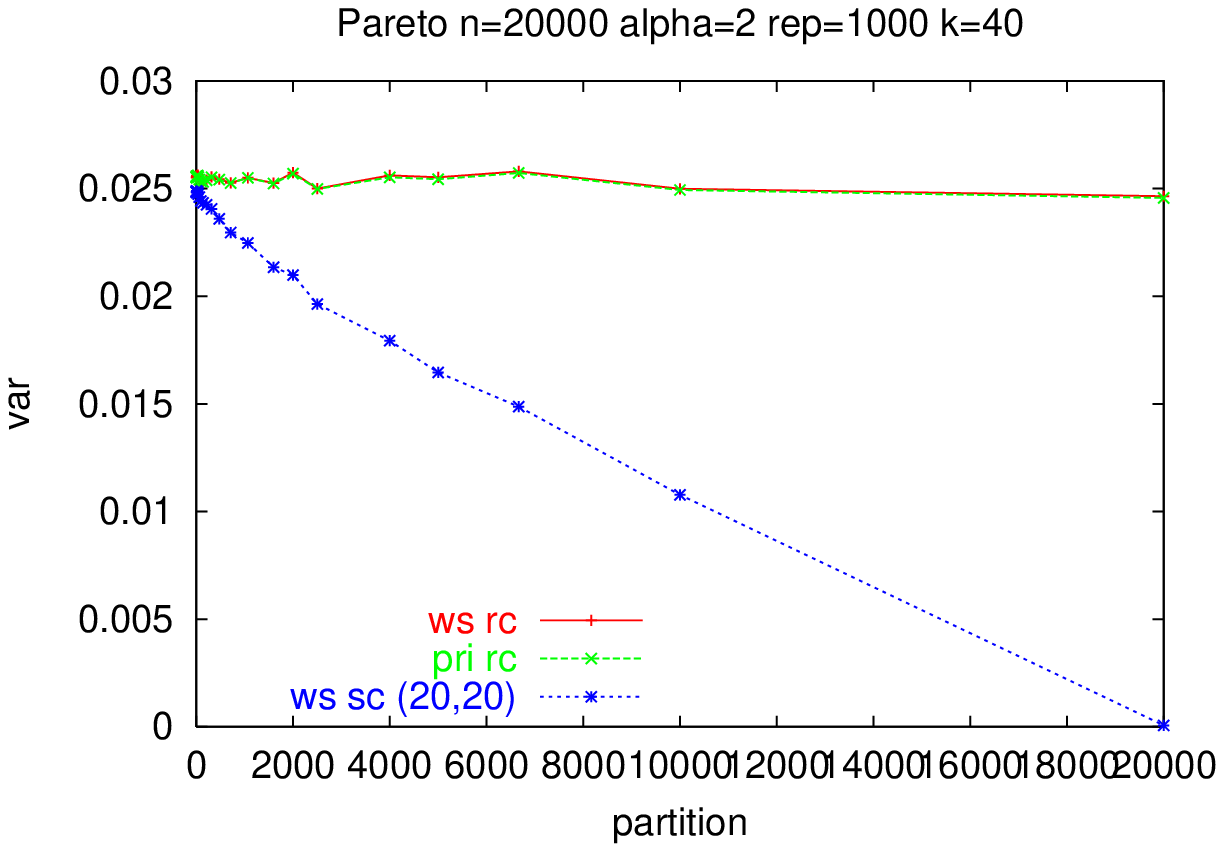,width=0.32\textwidth} &
 \epsfig{figure=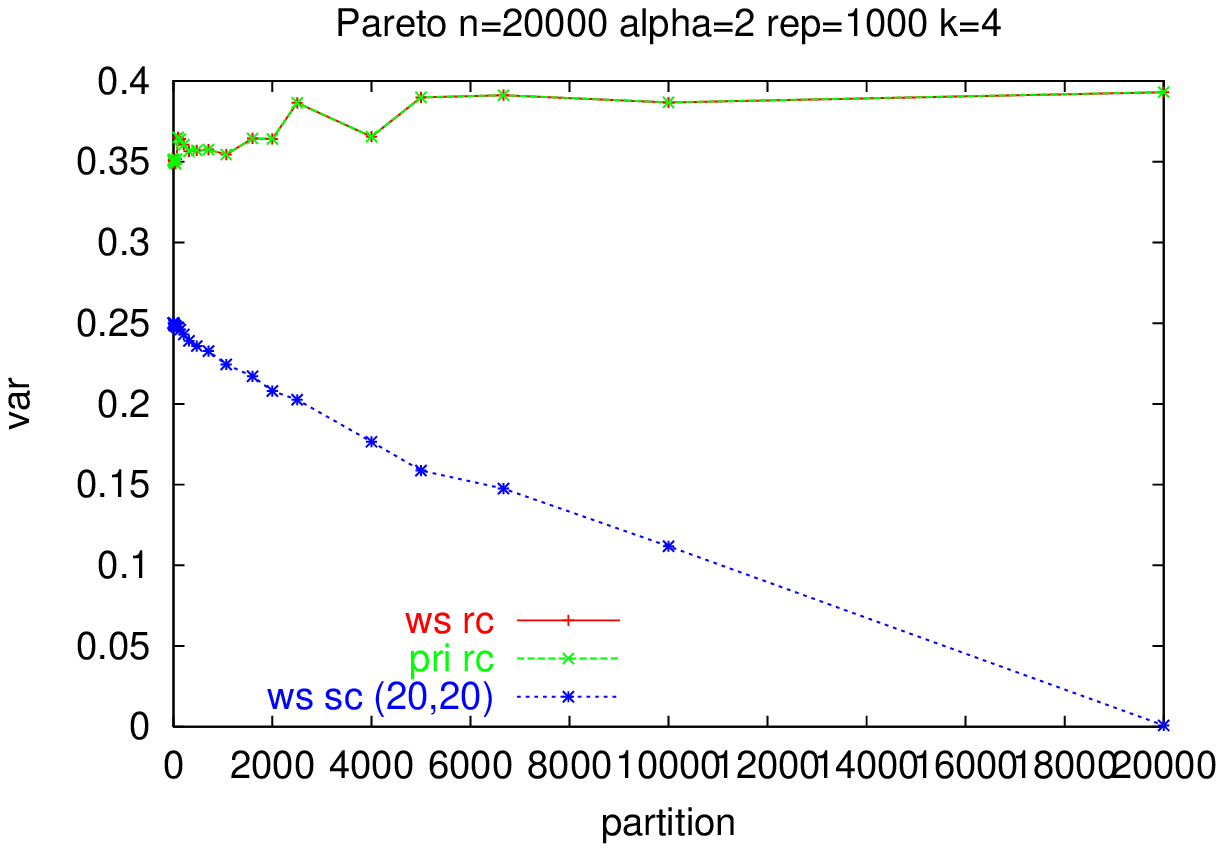,width=0.32\textwidth} \\
 $k=500$ & $k=40$ & $k=4$
\end{tabular}
}
\caption{Sum of variances over a partition as
a function of group size for fixed values of $k$.
We used  20000 items drawn
from Pareto distributions with $\alpha=1.2$ (top) and $\alpha=2$ (bottom).
To compute the variance in a group
we averaged over 1000 repetitions.
We used the approximation of \ws{} \ssc{} with  $\inperm=20$, $\permnum=20$.}
\label{varovergrouping:fig}
\end{figure*}

{\small
\begin{figure}[htbp]
\centerline{\begin{tabular}{cc}
 \epsfig{figure=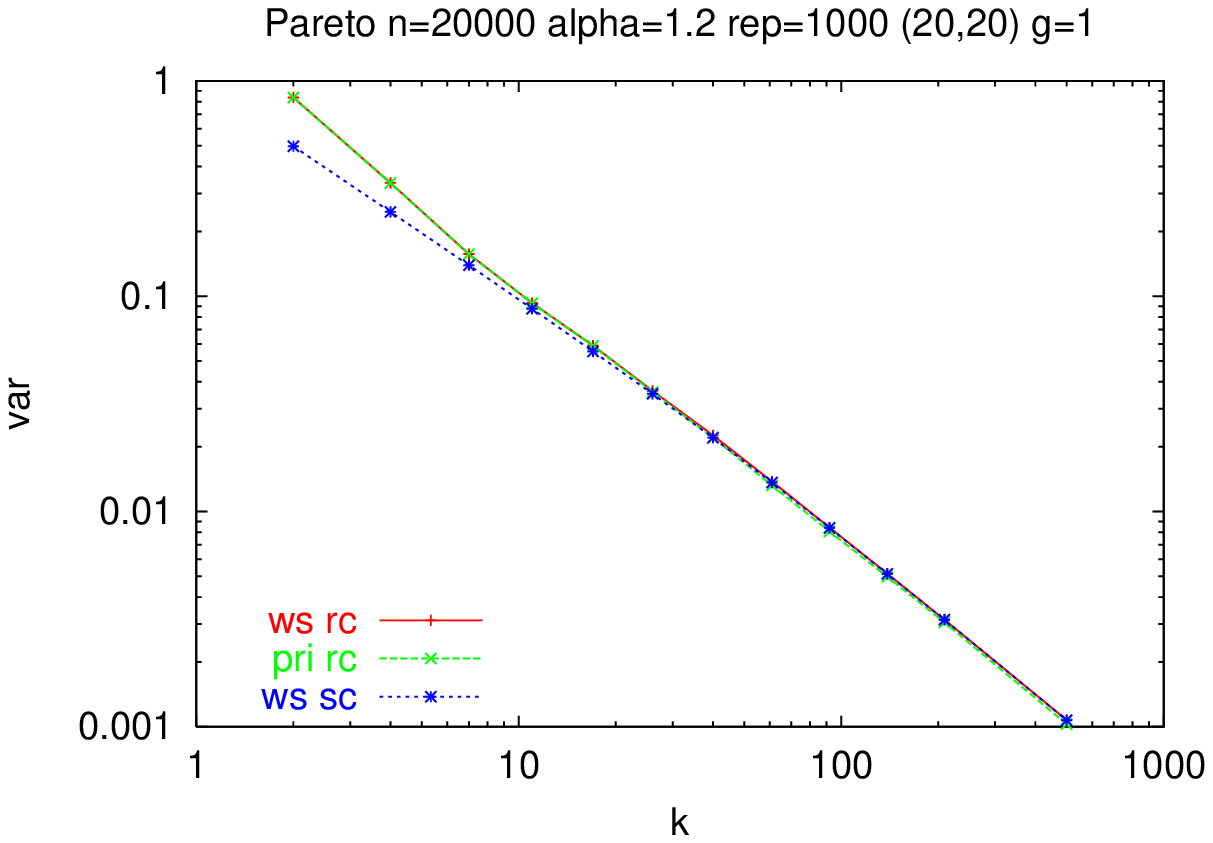,width=0.22\textwidth} &
 \epsfig{figure=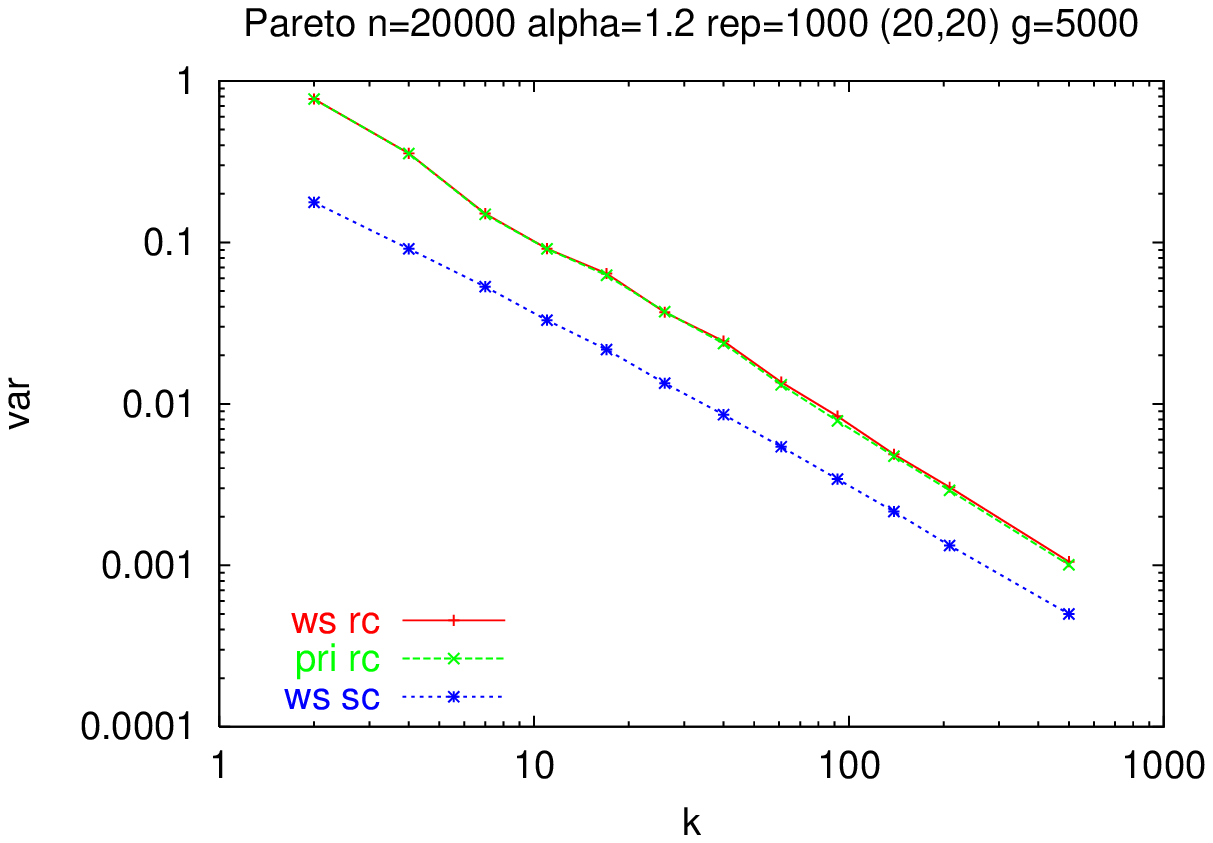,width=0.22\textwidth} \\
 \epsfig{figure=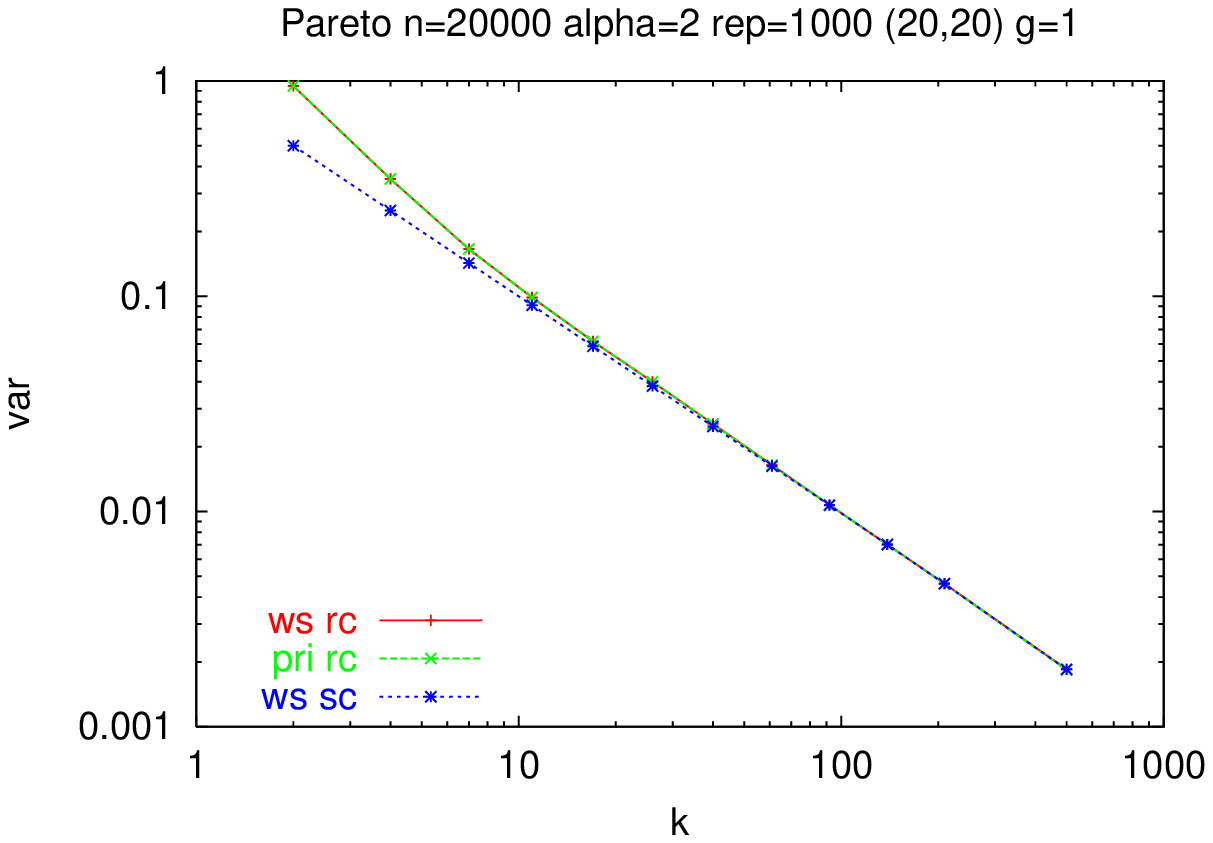,width=0.22\textwidth} &
 \epsfig{figure=code2/results/g_5000_alpha_2_n_20000_rep_1000_20_20_var.eps,width=0.22\textwidth} \\
 $g=1$ & $g=5000$ 
\end{tabular}
}
\caption{Estimator quality as sum of variances over partition, as
a function of $k$ for a fixed grouping. We use Pareto distributions with 20000 items $\alpha=1.2$ (top) and $\alpha=2$ (bottom).  Averaging is over 1000 repetitions, and  $\inperm=20$, $\permnum=20$.}
\label{varoverkfixedgroup:fig}
\end{figure}
}
}

Figure~\ref{param:fig}  
compares different  choices
of the parameters $\inperm$, and $\permnum$
for the approximate 
(Markov chain based) \ws\ \ssc\ estimator.  
We denote each such choice as  a pair $(\inperm, \permnum)$.
 We compare estimators with parameters $(400,1)$, $(20,20)$,
$(1,400)$, and $(5,2)$.  We conclude the following: 
(i)~A lot of the benefit of \ws\ \ssc\ on moderate-size subsets is
obtained for small values: $(5,2)$ performs nearly as well as the variants
that use more steps and iterations.
(ii)~There is a considerable benefit of redrawing
within a permutation: $(400,1)$ that iterates within a single permutation
performs well.  
(iii)~Larger subsets, however, benefit from larger \permnum: $(1,400)$ performs
better than $(20,20)$ which in turn is better than $(400,1)$.

{\small
\begin{figure}[ht]
\centerline{\begin{tabular}{cc}
 \epsfig{figure=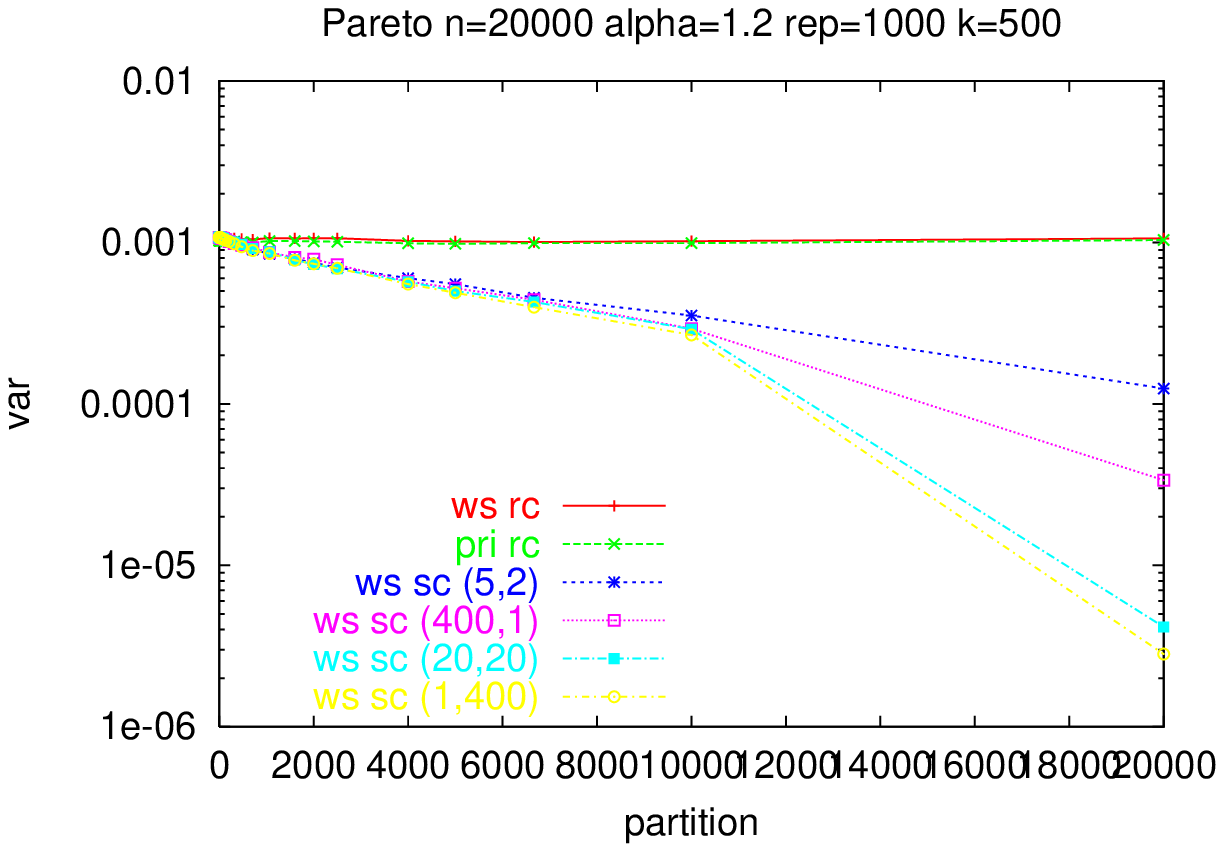,width=0.22\textwidth} &
 \epsfig{figure=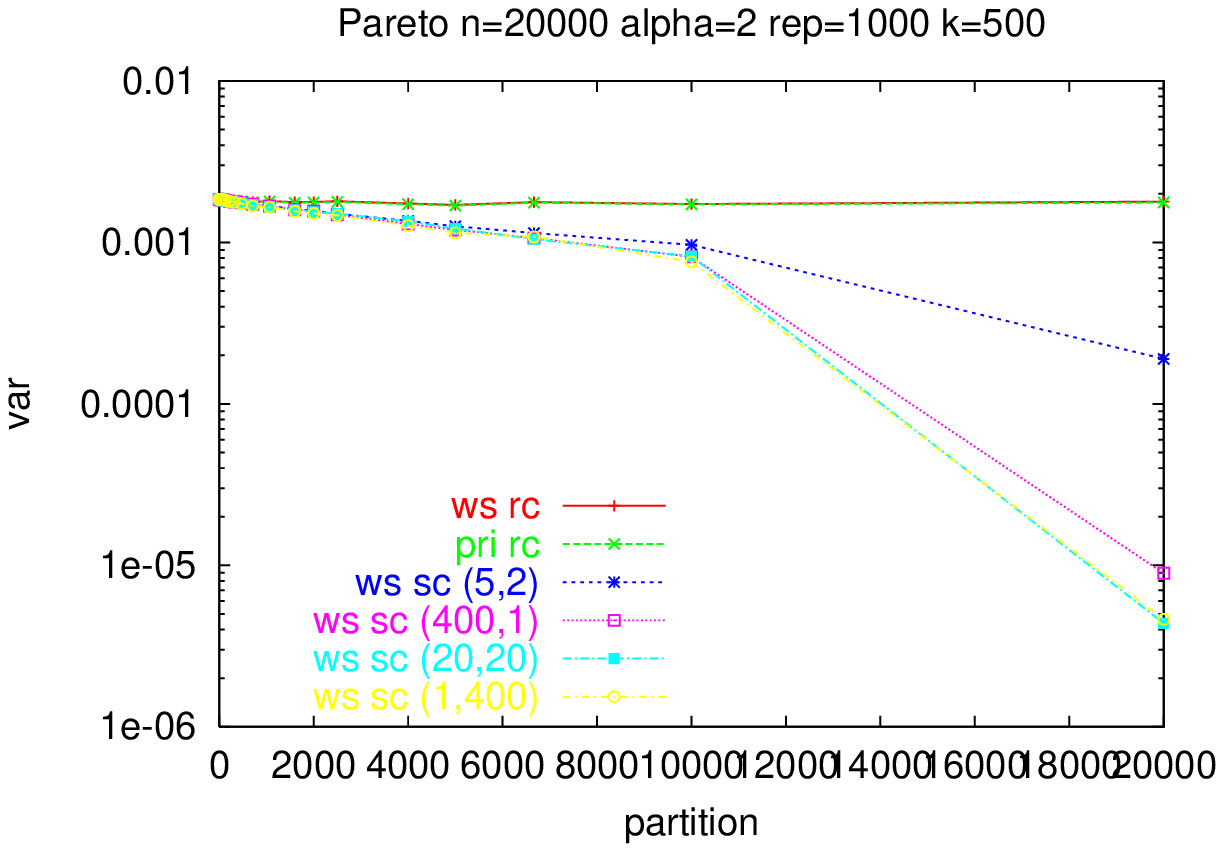,width=0.22\textwidth} \\
 $\alpha=1.2$ & $\alpha=2$ 
\end{tabular}
}
\caption{Sum of variances in  a partition for 
$k=500$ as a function of group size for different combinations of
$\inperm$ and $\permnum$.}
\label{param:fig}
\end{figure}
}

\smallskip
\noindent
{\bf Confidence bounds.}
We evaluate  confidence bounds on subpopulation weight
using the \pri\ 
Chernoff-based bounds~\cite{Thorup:sigmetrics06} (\pri), and
the \ws\ bounds that use $w(I)$ (\ws\ $+w(I)$)
or do not use $w(I)$ (\ws\ $-w(I)$), that are derived in
 Section~\ref{confws}.
The \ws\ bounds are computed using the quantile method with 
200 draws from the appropriate distribution.
\ignore{This method is accurate for estimations where 
 the normal approximation
may not be sufficiently accurate such as 
small supopulations (when the sketch includes few or no
members of the subpopulation) and for the distribution of
the difference of two sums of independent exponentials
(which is needed for the bounds when $w(I)$ is provided.).}

  We used three distributions of 1000 items drawn from a
 Pareto distributions with  parameters
$\alpha\in\{1,1.2,2\}$ and
group sizes
of $g=200$ and $g=500$ (5 groups and 2 groups).
We also use two distributions of $20000$ items drawn from a
 Pareto distributions with  parameters
$\alpha\in\{1.2,2\}$ and $g=4000$.

  We consider the relative error of the bounds, the width of the
confidence interval (difference between the upper and lower bounds), 
and the square error of the bounds (square of the difference between
the bound and the actual value).
  The confidence bounds, intervals, and square errors,
 were normalized using the 
weight of the corresponding subpopulation.
For each distribution and values of $k$ and $g$, the normalized bounds
were then averaged across 500 repetitions and across 
all subpopulations of size $g$.
Across these distributions, 
the \ws\ $+w(I)$ confidence bounds are tighter (more so for larger $g$)
than \ws\ $-w(I)$
and both are significantly tighter 
than the \pri\ confidence bounds.
Representative results  are shown in 
Figure~\ref{conf_subpop200:fig}).

\notinproc{
\begin{figure*}[ht]
\centerline{\begin{tabular}{ccc}
 \epsfig{figure=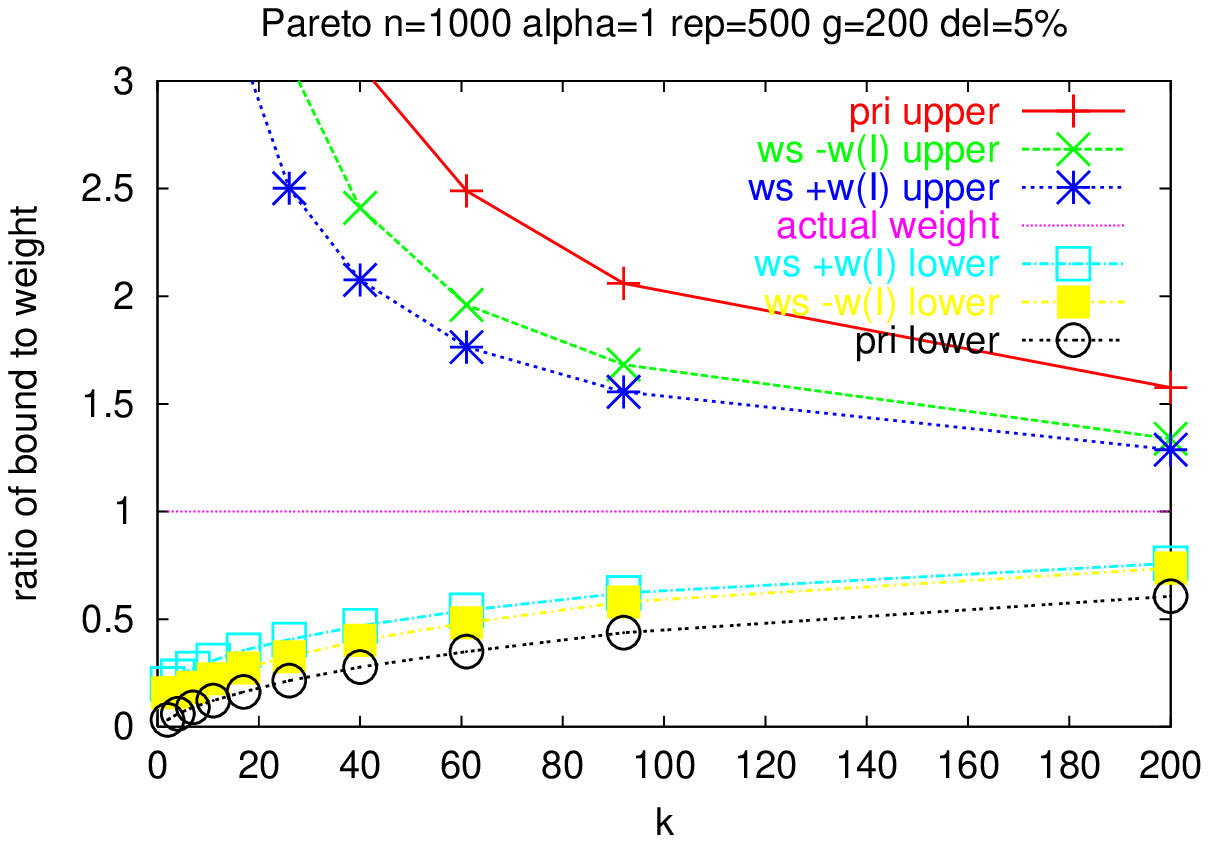,width=0.32\textwidth} &
 \epsfig{figure=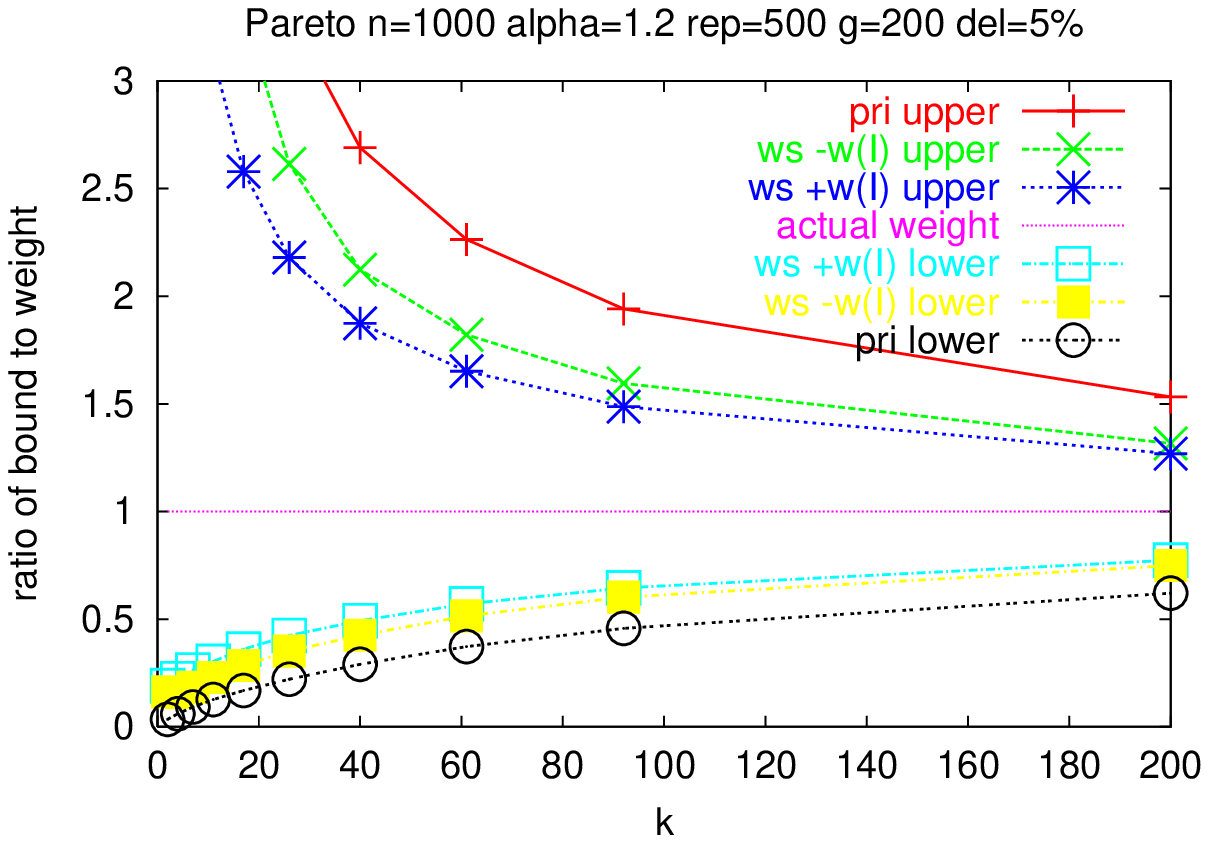,width=0.32\textwidth} &
 \epsfig{figure=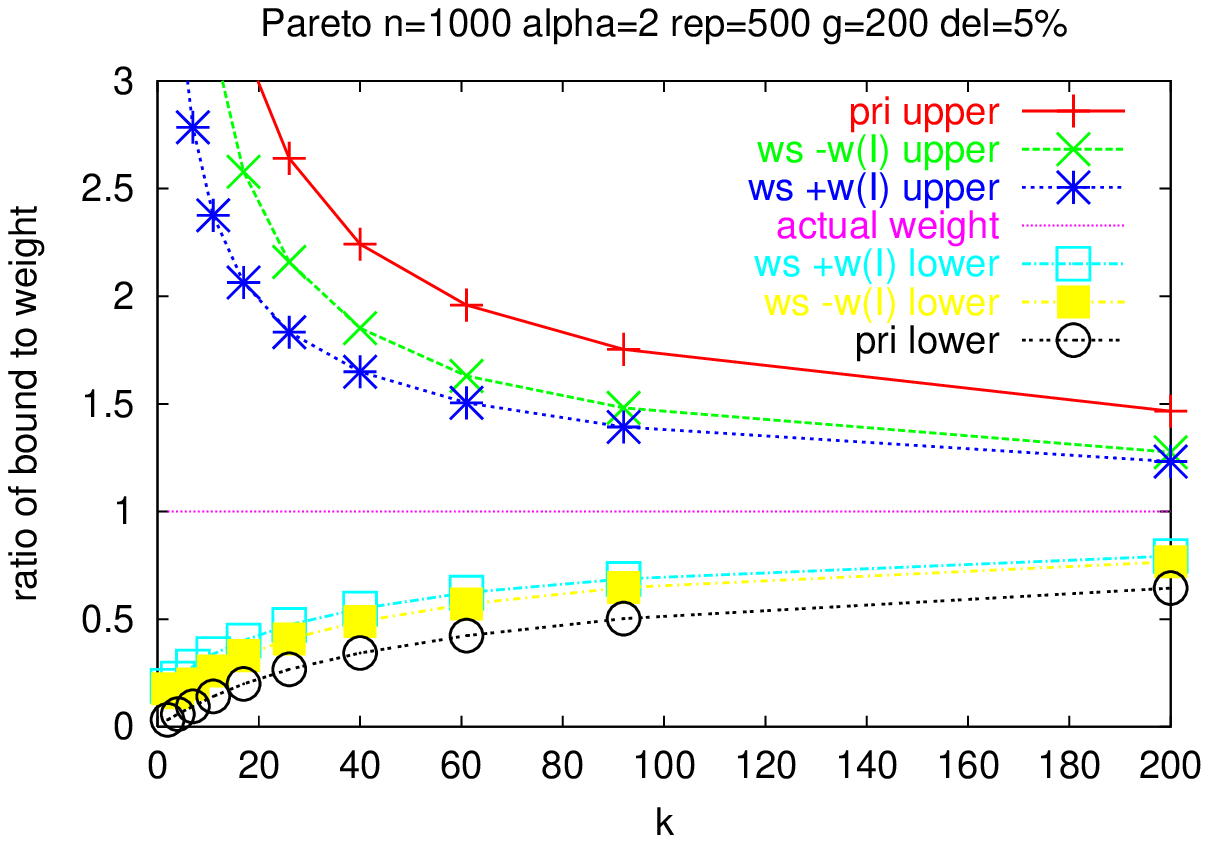,width=0.32\textwidth} \\
 \epsfig{figure=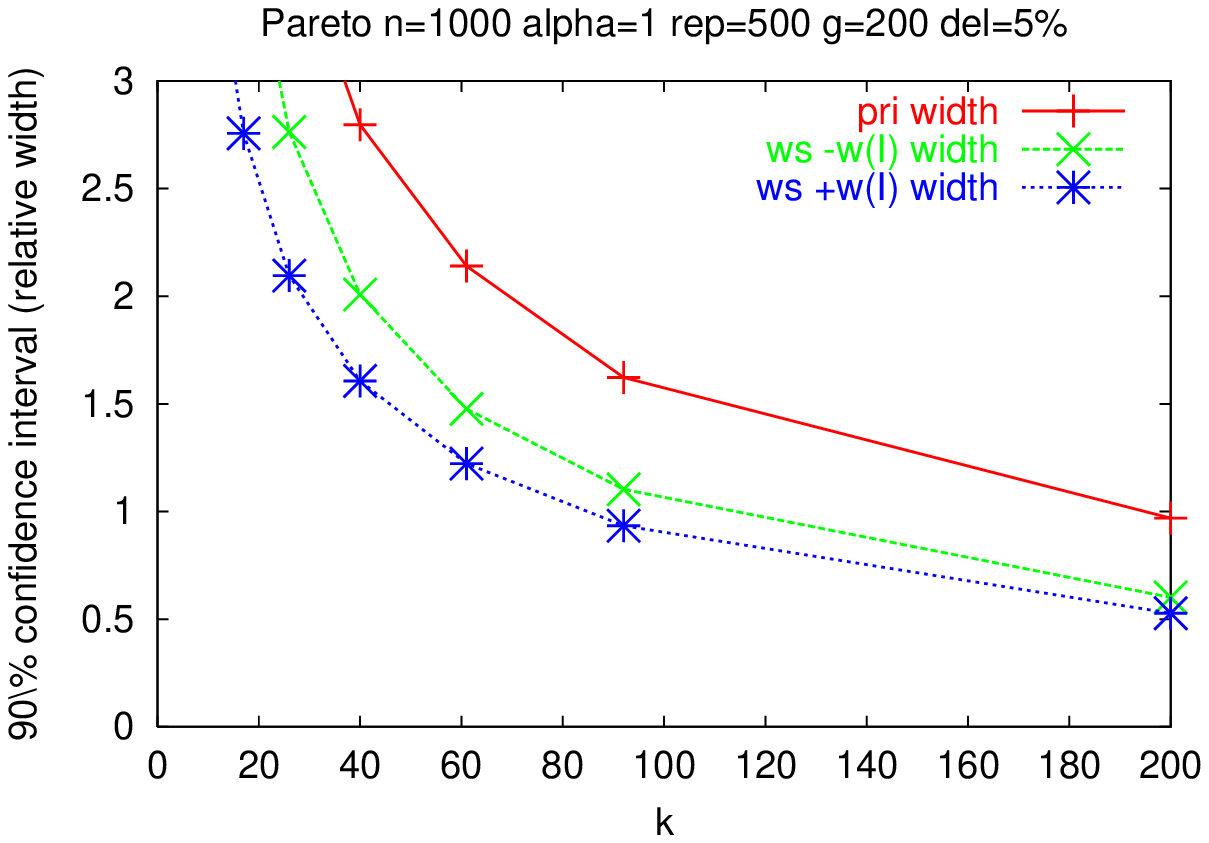,width=0.32\textwidth} &
 \epsfig{figure=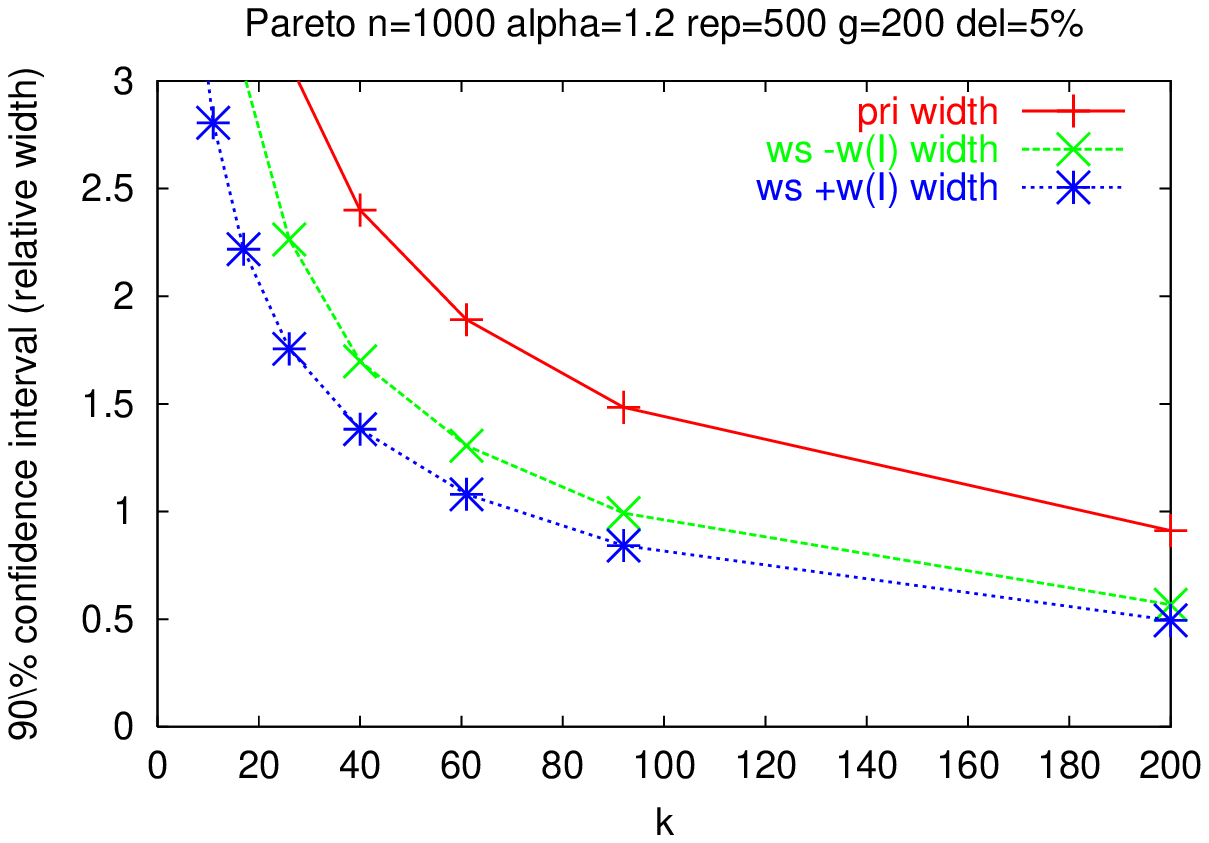,width=0.32\textwidth} &
 \epsfig{figure=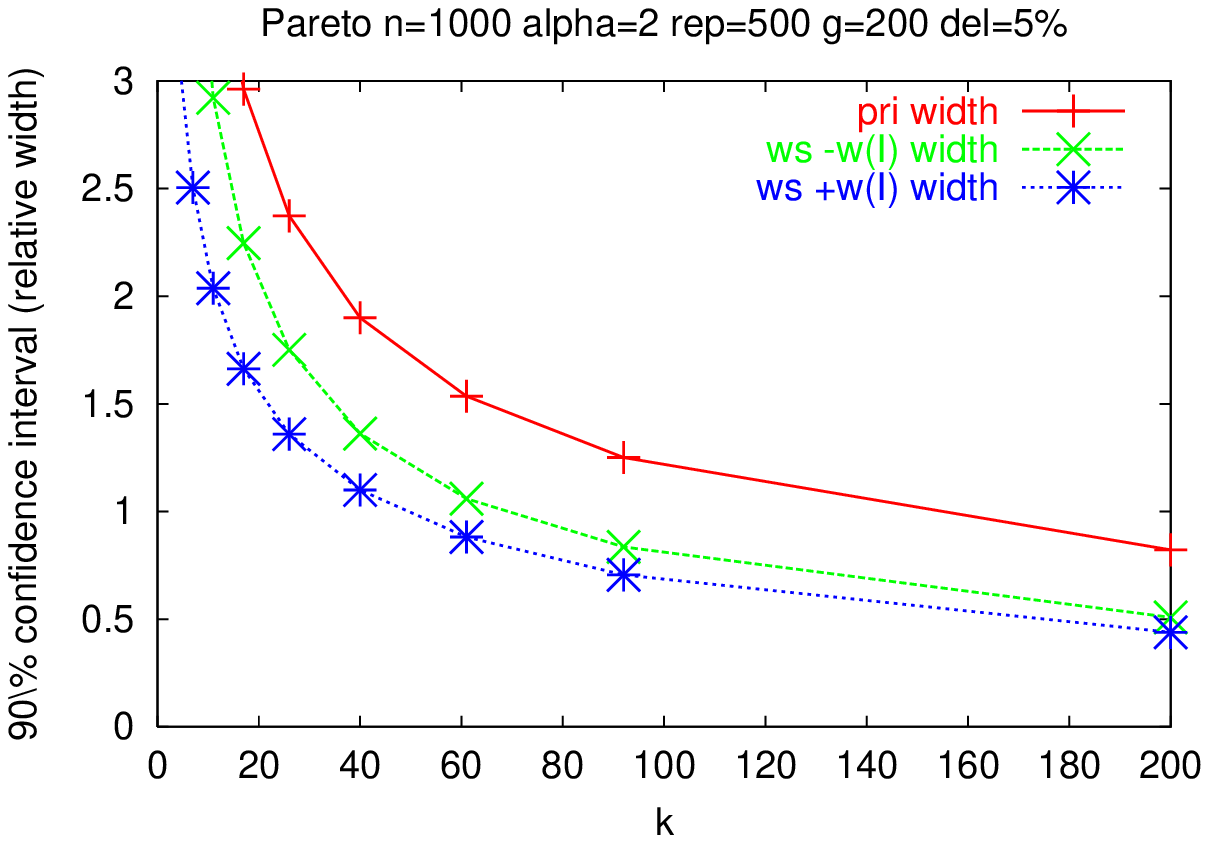,width=0.32\textwidth} \\
 \epsfig{figure=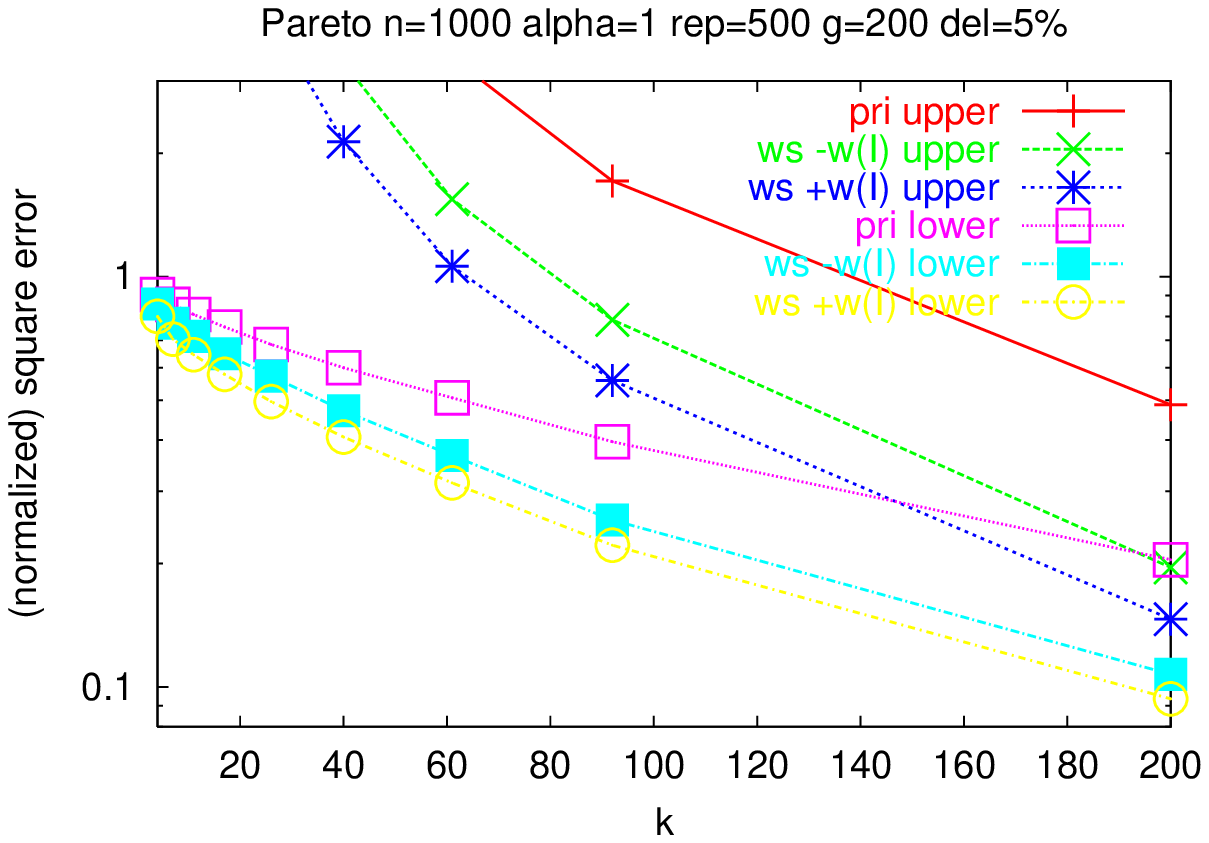,width=0.32\textwidth} &
 \epsfig{figure=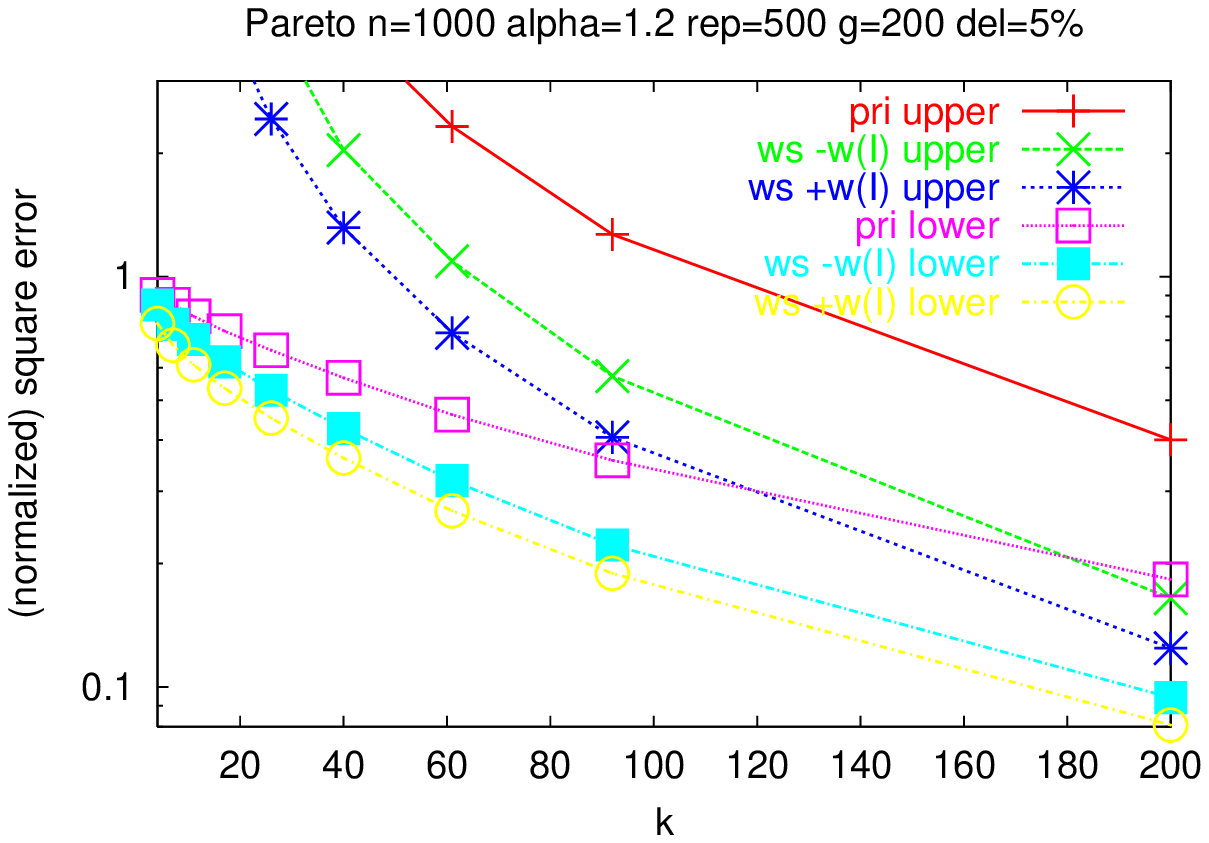,width=0.32\textwidth} &
 \epsfig{figure=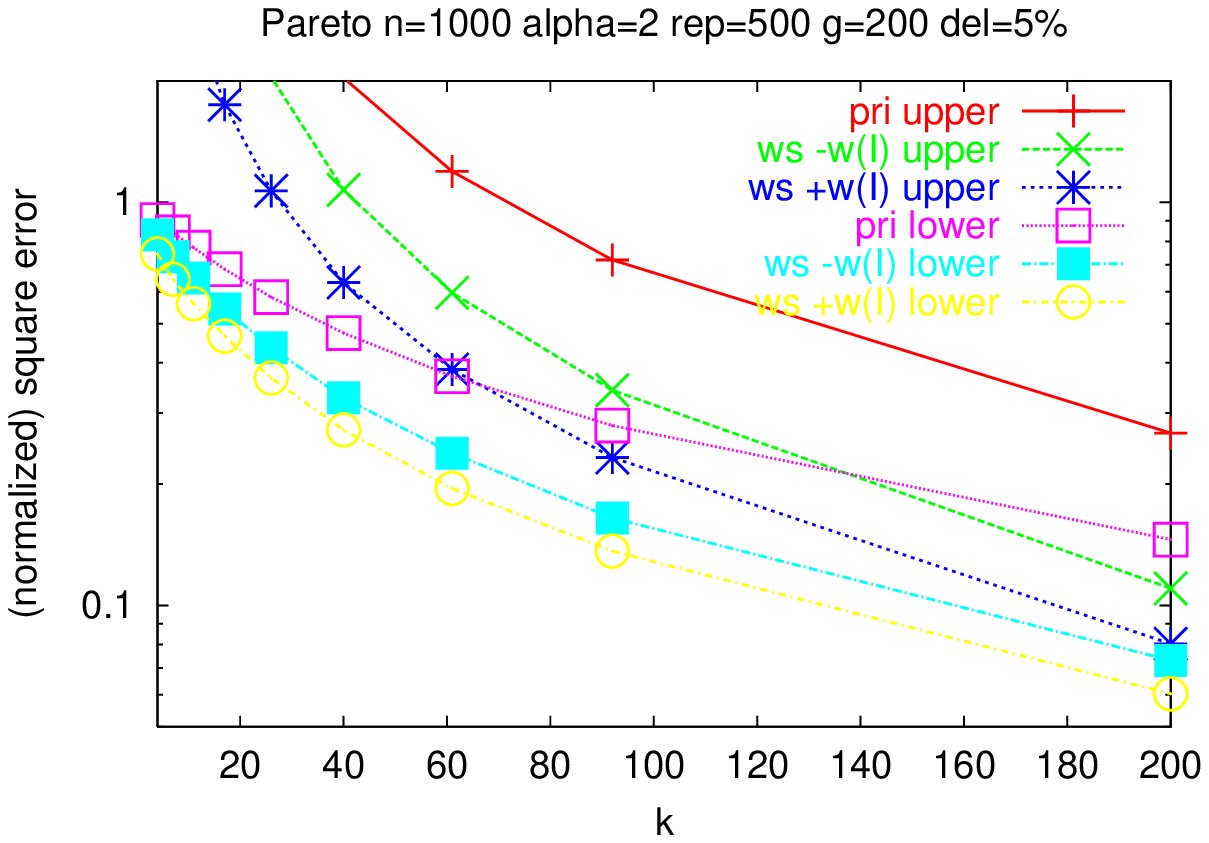,width=0.32\textwidth} \\
Pareto $n=1000$, $\alpha=1$ & Pareto $n=1000$, $\alpha=1.2$ & Pareto $n=1000$,$\alpha=2$ 
\end{tabular}
}
\caption{Subpopulation 95\% confidence bounds (top), 90\% confidence intervals (middle), and  (normalized) squared error of 
the 95\% confidence bounds (bottom) for  $g=200$.}
\label{conf_subpop200:fig}
\end{figure*}
}
\onlyinproc{
\begin{figure}[ht]
\centerline{\begin{tabular}{cc}
 \epsfig{figure=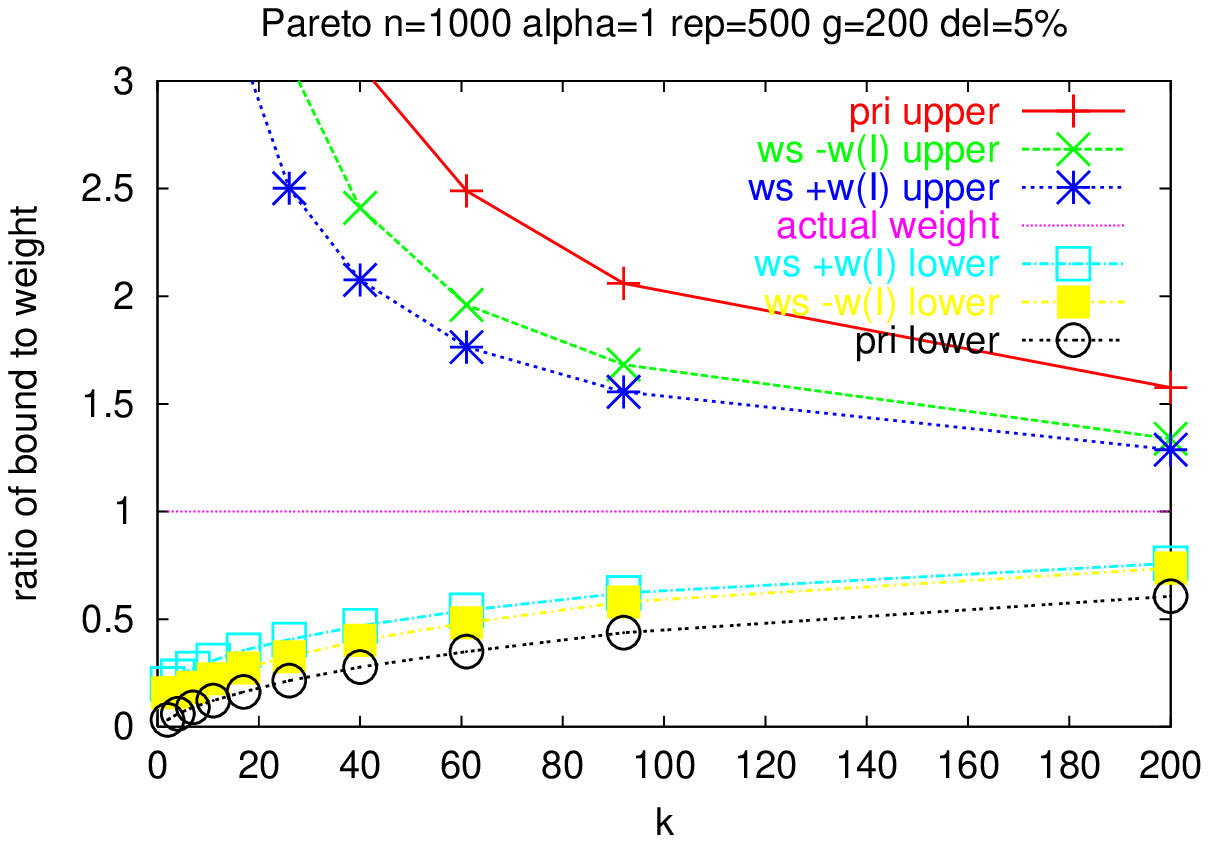,width=0.23\textwidth} &
 \epsfig{figure=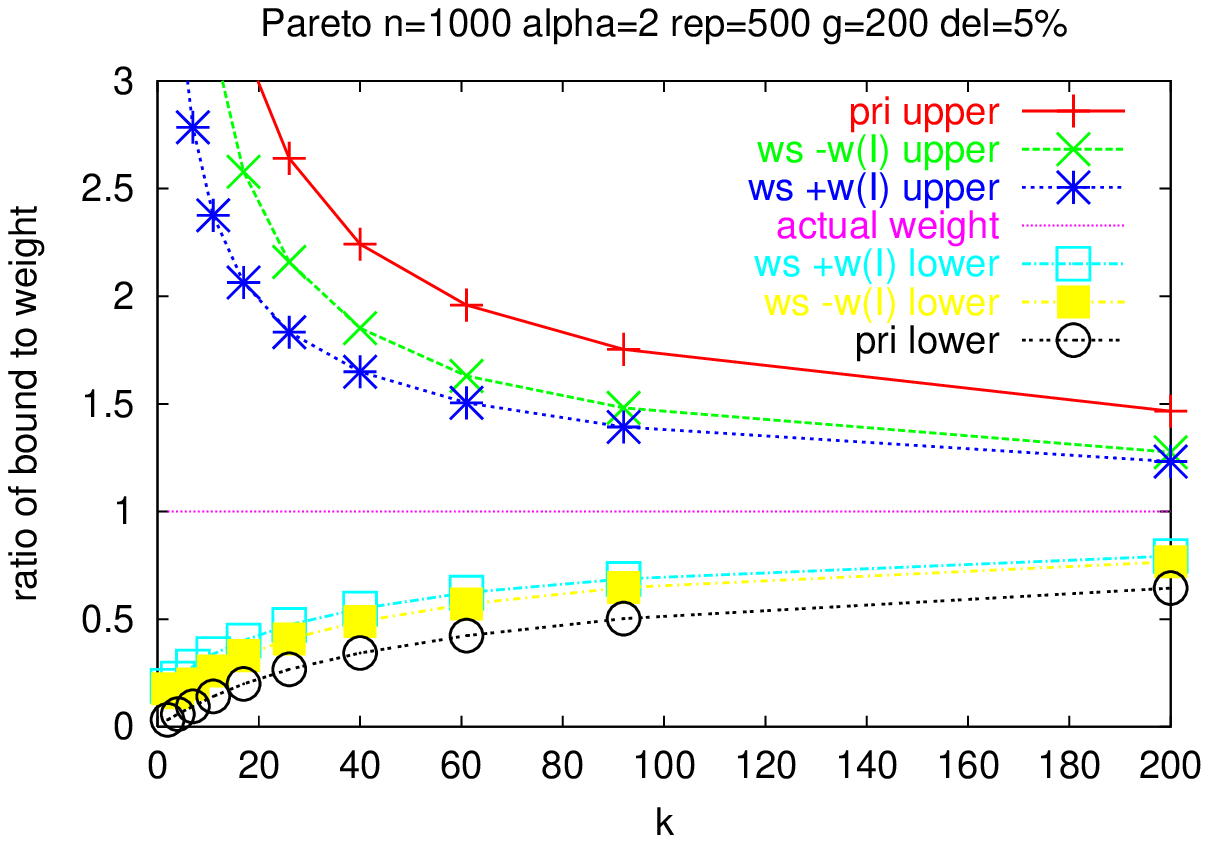,width=0.23\textwidth} \\
 \epsfig{figure=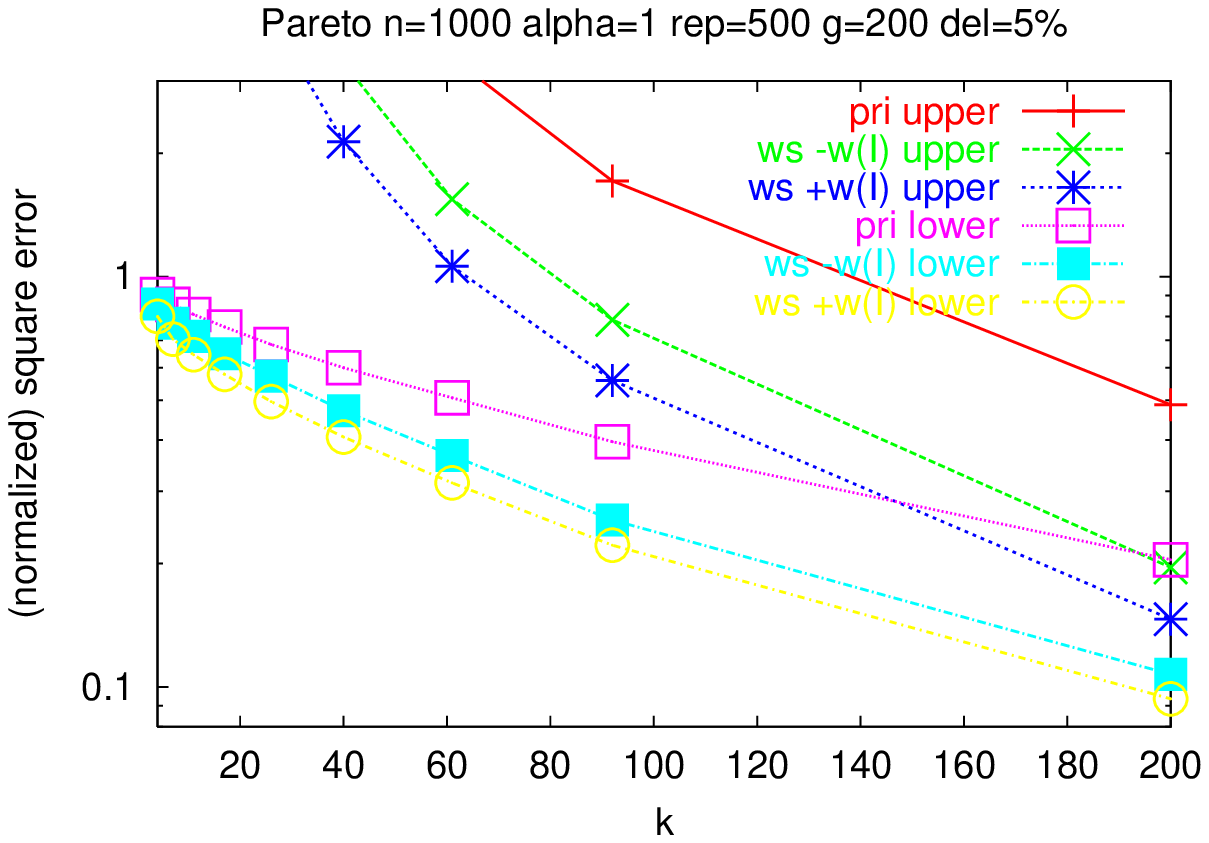,width=0.23\textwidth} &
 \epsfig{figure=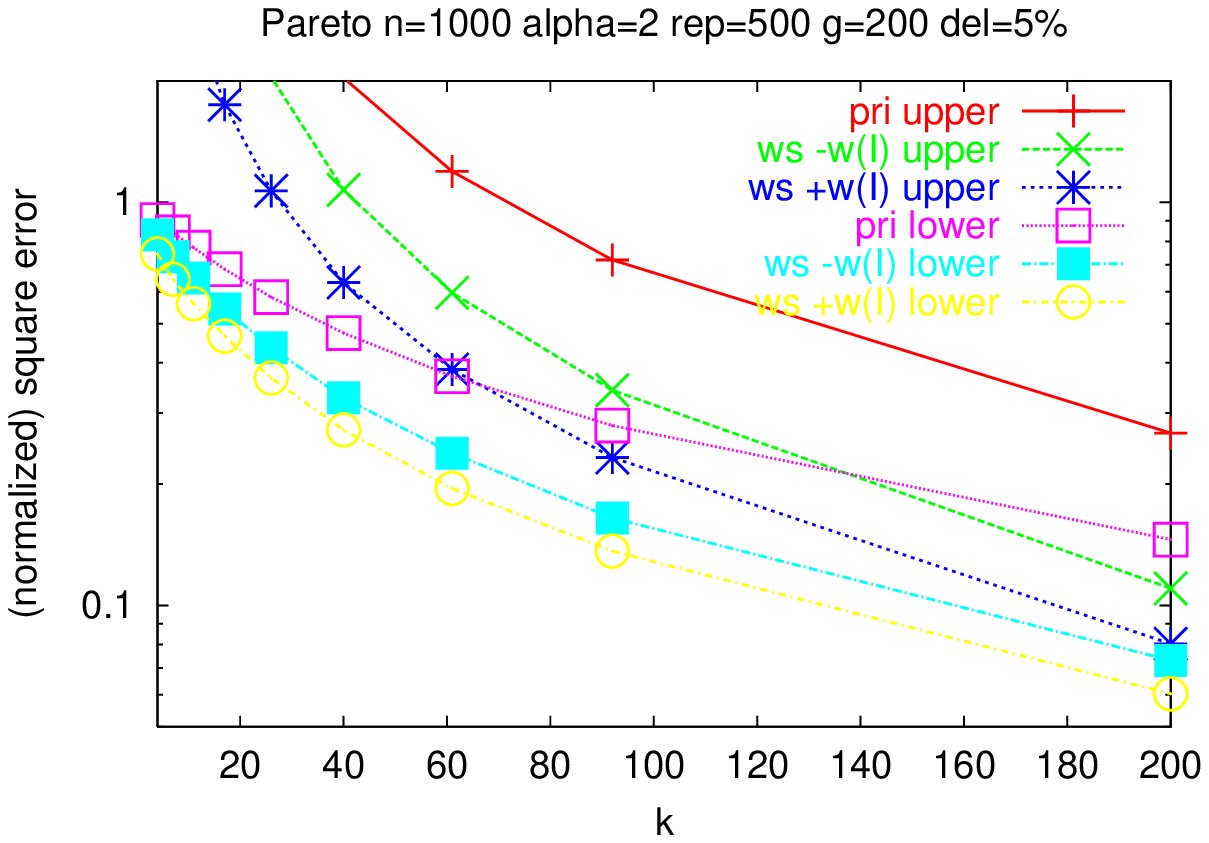,width=0.23\textwidth} \\
Pareto $n=1000$, $\alpha=1$ &  Pareto $n=1000$,$\alpha=2$ 
\end{tabular}
}
\caption{Subpopulation 95\% confidence bounds (top) and  (normalized) squared error of 
the 95\% confidence bounds (bottom) for  $g=200$.}
\label{conf_subpop200:fig}
\end{figure}

}

\ignore{
\begin{figure*}[ht]
\centerline{\begin{tabular}{ccc}
 \epsfig{figure=code2/results_conf/confb_pareto_1000_1_r_500_dl_0.05_q_200_g_500.eps,width=0.32\textwidth} &
 \epsfig{figure=code2/results_conf/confb_pareto_1000_1.2_r_500_dl_0.05_q_200_g_500.eps,width=0.32\textwidth} &
 \epsfig{figure=code2/results_conf/confb_pareto_1000_2_r_500_dl_0.05_q_200_g_500.eps,width=0.32\textwidth} 
\\
 \epsfig{figure=code2/results_conf/confw_pareto_1000_1_r_500_dl_0.05_q_200_g_500.eps,width=0.32\textwidth} &
 \epsfig{figure=code2/results_conf/confw_pareto_1000_1.2_r_500_dl_0.05_q_200_g_500.eps,width=0.32\textwidth} &
 \epsfig{figure=code2/results_conf/confw_pareto_1000_2_r_500_dl_0.05_q_200_g_500.eps,width=0.32\textwidth} 
\\
 \epsfig{figure=code2/results_conf/confbv_pareto_1000_1_r_500_dl_0.05_q_200_g_500.eps,width=0.32\textwidth} &
 \epsfig{figure=code2/results_conf/confbv_pareto_1000_1.2_r_500_dl_0.05_q_200_g_500.eps,width=0.32\textwidth} &
 \epsfig{figure=code2/results_conf/confbv_pareto_1000_2_r_500_dl_0.05_q_200_g_500.eps,width=0.32\textwidth} \\
Pareto $n=1000$, $\alpha=1$ & Pareto $n=1000$, $\alpha=1.2$ & Pareto $n=1000$,$\alpha=2$ 
\end{tabular}
}
\caption{Subpopulation 95\% confidence bounds (top), 90\% confidence intervals (middle), and (normalized) 
squared error of 95\% confidence bounds (bottom), for $g=500$.}
\label{conf_subpop500:fig}
\end{figure*}

\begin{figure*}[ht]
\centerline{\begin{tabular}{cc}
 \epsfig{figure=code2/results_conf/confb_pareto_20000_1.2_r_500_dl_0.05_q_200_g_4000.eps,width=0.4\textwidth} &
 \epsfig{figure=code2/results_conf/confb_pareto_20000_2_r_500_dl_0.05_q_200_g_4000.eps,width=0.4\textwidth} 
\\
 \epsfig{figure=code2/results_conf/confw_pareto_20000_1.2_r_500_dl_0.05_q_200_g_4000.eps,width=0.4\textwidth} &
 \epsfig{figure=code2/results_conf/confw_pareto_20000_2_r_500_dl_0.05_q_200_g_4000.eps,width=0.4\textwidth} 
\\
 Pareto $n=20000$, $\alpha=1.2$ & Pareto $n=20000$,$\alpha=2$ 
\end{tabular}
}
\caption{Subpopulation 95\% confidence bounds, and 90\% confidence intervals
for $20000$ items drawn from Pareto distribution with $\alpha=1.2$ and $\alpha =2$, and  $g=4000$.}
\label{conf_subpop4000:fig}
\end{figure*}
}

\section{Conclusion}

 We consider the fundamental problem of processing approximate
subpopulation weight queries over summaries of a set of weighted
records.  Summarization
methods supporting such queries include the $k$-mins 
format, which includes weighted sampling with
replacement (\wsr\ or PPSWR Probability Proportional to Size With
Replacement) and the bottom-$k$ format which includes weighted
sampling without replacement (\ws, also known as PPSWOR - PPS WithOut
Replacement) and priority sampling (\pri)~\cite{DLT:sigmetrics04}
which is related to IPPS (Inclusion Probability Proportion to
Size)~\cite{Szegedy:stoc06}.

\ignore{
.  \ws\ sketches are of particular interest
among bottom-$k$ sketches because they can be computed much more
efficiently in some common settings.  \pri\ sketches are of interest
because they have estimators with nearly-minimal sum of per-item
variances
}

  Surprisingly perhaps, the vast literature on survey sampling and PPS
and IPPS estimators (e.g.~\cite{Sampath:book,surveysampling:book}) is
mostly not applicable to our common database setting:
subpopulation-weight estimation, skewed (Zipf-like) weight
distributions, and summaries that can be computed efficiently over
massive datasets (such as data streams or distributed data).  Existing
unbiased estimators are the \HT\ and ratio estimators for PPSWR, the
\pri\ estimator~\cite{DLT:sigmetrics04,Thorup:sigmetrics06}, and a
\ws\ estimator based on mimicking \wsr\ sketches~\cite{bottomk07:ds}.

 We derive novel and significantly tighter estimators and confidence
bounds on subpopulation weight: better estimators for the classic \ws\
sampling method; better estimators than all known
estimators/summarizations (including \pri{}) for many data
representations including data streams; and tighter confidence bounds
across summarization formats.  Our derivations are complemented with
the design of interesting and efficient computation methods, including
a Markov chain based method to approximate the \ws\ \ssc\ estimator,
and the quantile method to compute the confidence bounds.

\ignore{   
 Using the total $w(I)$, refine performance metrics and optimality.
Obtain tighter estimators than pri for many important applications.

  Obtain better estimators for \ws\ and \wsr.
probability proportional to weight sampling is interesting,
applicable to unaggregated data, 
\ws\ sketches have some special properties that distinguishes them from other
bottom-$k$ sketches.   They allow for unbiased estimators for selectivity
that are tighter than those for \wsr\ sketches (see Appendix and~\cite{bottomk07:ds}); they support
estimators that cancel out variance for larger subsets (the subset
conditioning estimator in Section~\ref{WI:sec})
defacto better estimators overall than \pri\ \rc; tighter confidence
bounds for subpopulation weight, much tighter than all known bounds 
including those for \pri\ sketches.
 
 Interesting, current confidence bounds do not even capture \pri\ \rc\ optimality
in cases when it it nearly optimal.
}

  Our work reveals basic principles and
our techniques and methodology
 are a stand alone contribution with wide applicability to sketch-based
estimation.
\ignore{We list some further applications of our methodology.
\begin{itemize}
\item
 Tighter estimators for subset relations such as the weight
of the union, intersection, and resemblance based on the combined sketches
of supporting subsets.
Previously, these quantities were computed by applying an estimator to
the sketch of the union.
By using the sketch of the union with bottom-$k$ sketches, however,
we effectively use only $k$ out of the (up to) $ck$ samples, if $c$ is the
number of subsets. 
\item
 Over records with multiple weighted attributes,
(Eg, for IP flows we have bytes, packets, and unity for estimating the
number of flows) we are interested in subpopulation weight queries with
respect to different weight functions.  While
sketches derived for one set of weights can trivially be used to
obtain unbiased estimates with respect to another set of weights, 
the resulting estimates have
higher variance than sketches tailored to the weight function.
The variance is larger when there is smaller correlation between
the weight functions.
On the other extreme, computing a different sketch for each set
of weights provides  guarantees for each weight function but
is very wasteful, and does not benefit from correlation between
weight functions.
We derive a summarization algorithm and estimators
that provide guarantees for all weight functions and
perform better when the weight functions are correlated.
\item
 Our \HTp\ and \rc\ techniques guided the development and analysis of
summarization algorithms and estimators for unaggregated data 
streams~\cite{CDKLT:pods07,CDKLT:IMC07}.
\end{itemize}
}

\bibliographystyle{abbrv}
 \vspace*{1mm}
 \scriptsize
\bibliography{cycle,replace,data_structures}

\end{document}